\documentclass[aps,prd,twocolumn,groupedaddress,nofootinbib]{revtex4-1}
\pdfoutput=1

\usepackage{mathtools}
\usepackage{ulem}
\usepackage{color}
\usepackage{amsmath}

\begin{document}

\title{Cosmic variance of $H_0$ in light of forthcoming high-redshift surveys}

\author{Giuseppe Fanizza$^1$}
\email{gfanizza@fc.ul.pt}
\author{Bartolomeo Fiorini$^2$}
 \email{bartolomeo.fiorini@port.ac.uk}
\author{Giovanni Marozzi$^{3,4}$}
\email{giovanni.marozzi@unipi.it}
\affiliation{$^1$Instituto de Astrofis\'ica e Ci\^encias do Espa\c{c}o,
Faculdade de Ci\^encias da Universidade de Lisboa,
Edificio C8, Campo Grande, P-1740-016, Lisbon, Portugal}
\affiliation{$^2$Institute of Cosmology \& Gravitation, University of Portsmouth, Dennis Sciama Building, Burnaby Road, Portsmouth, PO1 3FX United Kingdom}
\affiliation{$^3$Dipartimento di Fisica, Universit\`a di Pisa, Largo B. Pontecorvo 3, 56127 Pisa, Italy}
\affiliation{$^4$Istituto Nazionale di Fisica Nucleare, Sezione di Pisa, Pisa, Italy}

\date{\today}

\begin{abstract} 
Forthcoming surveys will extend the understanding of cosmological large scale structures up to unprecedented redshift. According to this perspective, we present a fully relativistic framework to evaluate the impact of stochastic inhomogeneities on the determination of the Hubble constant. To this aim, we work within linear perturbation theory and relate the fluctuations of the luminosity distance-redshift relation, in the Cosmic Concordance model, to the intrinsic uncertainty associated to the measurement of $H_0$ from high-redshift surveys ($0.15\le z\le3.85$). We first present the detailed derivation of the luminosity distance-redshift relation 2-point correlation function and then provide analytical results for all the involved relativistic effects, such as peculiar velocity, lensing, time delay and (integrated) Sachs-Wolfe, and their angular spectra. Hence, we apply our analytical results to the study of high-redshift Hubble diagram, according to what has been recently claimed in literature. Following the specific of Euclid Deep Survey and LSST, we conclude that the cosmic variance associated with the measurement of the Hubble constant is at most of 0.1 \%. Our work extends the analysis already done in literature for closer sources, where only peculiar velocity has been taken into account. We then conclude that deep surveys will provide an estimation of the $H_0$ which will be more precise than the one obtained from local sources, at least in regard of the intrinsic uncertainty related to a stochastic distribution of inhomogeneities.
\end{abstract}

\pacs{Valid PACS appear here}

\maketitle


\section*{Introduction}
The last 30 years have witnessed the evolution of cosmology from an order of magnitude description of the Universe into an area of the hard sciences where precision measurements are achievable. This change has its cornerstone in the estimation of cosmological parameters through the detection of Cosmic Microwave Background (CMB) spectra which provide so far the most precise measurements of these parameters. Given this astonishing success, forthcoming missions aim to push forward this result in order to infer measurements of cosmological parameters also from late-time dataset, such as Large Scale Structure (LSS) surveys (see, for example, Euclid \cite{Laureijs:2011gra} and Vera C. Rubin Observatory Legacy Survey of Space and Time (LSST) \cite{Abell:2009aa}) and Intensity Mapping (IM) (for instance, SKA \cite{Mellema:2012ht} and HERA \cite{DeBoer:2016tnn}). One of the common goals shared within the community is to study late time probes in order to estimate cosmological parameters at a comparable level of precision with respect to the one achieved from CMB dataset. This program is highly motivated by several reasons. To mention a few, late time surveys provide 3-dimensional catalogs about the observed Universe, hence they may help understanding the evolution of the Universe rather than just giving a picture of it, as 2-dimensional datasets (CMB indeed) do\footnote{Here we underline that non-linearities in CMB lensing due to the bispectrum of the gravitational potential might turn CMB spectra into 3 dimensional datasets as well \cite{Marozzi:2016uob,Marozzi:2016und,Marozzi:2016qxl,DiDio:2019rfy}.}. Secondly, the non-linear evolution of the structures gets enhanced at later times. As a consequence, this allows to explore better non-linear scales in the distribution of matter in the Universe, distribution that contains useful information about the energy content of the Universe itself (for instance dark matter candidates and massive neutrinos). Last but not least, the measurements of late time probes furnish estimations for the cosmological parameters which are (almost) independent of the adopted model, differently from the CMB ones.

For what concerns the comparison between different reconstructions of cosmological parameters, last years have shown a discrepancy between the estimation of the present Hubble rate $H_0$ from CMB and late time probes. Indeed, CMB measurements provide a value for $H_0=67.36\pm0.54$ km s$^{-1}$ Mpc$^{-1}$ \cite{Aghanim:2018eyx}, whereas late time estimations based on local probes, such as Supernovae Ia (SnIa),
returns $H_0=73.2\pm1.3$ km s$^{-1}$ Mpc$^{-1}$ \cite{Riess:2020fzl} (see \cite{Freedman_2019,Freedman_2020,Wong_2019,Yuan_2019,Pesce_2020} for other estimations): a discrepancy of almost $5\sigma$ emerges. A priori, both measurements might be questionable because of the following reasons. CMB estimation provides nowadays the most precise measurement of $H_0$ but, to this end, requires (as mentioned) that a cosmological model is chosen in order to analyze the spectra. On the other hand, SnIa catalogs for close sources do not need any cosmological model to infer $H_0$ but require to calibrate the relative magnitude of the observed \textit{standard candles} wrt some sources whose luminosity is known (for instance the Cepheid host).

This tension seems to survive also when different late time probes are investigated \cite{Verde_2019}. This has raised the interest of the community into the following question: is there any need for new unknown physics to cancel (or at least mild) this discrepancy? In this regard, it might be that the sound horizon scale $r_s$ could be lower than the one predicted by our current knowledge of pre-recombination physics in the framework of $\Lambda$CDM model \cite{Bernal:2016gxb}. In fact, since the acoustic angular peak measured by the CMB is proportional to $r_s H_0$, this would automatically raise the value of $H_0$ detected from the CMB towards the one inferred from local measurements. However, as discussed in \cite{Jedamzik:2020zmd}, the modification of $r_s$ might be not enough to completely reabsorb the tension.
In general, the search for new physics which could explain the discrepancy questions several aspects of the current Cosmic Concordance model. In particular, in \cite{Beenakker:2021vff} a set of 7 key assumptions which might be broken to explain the tension has been identified. All these reasons have motivated an intense research activity during the last years (see \cite{Knox:2019rjx,DiValentino:2020zio} for overviews, and \cite{DiValentino:2016hlg,DiValentino:2017iww,Vagnozzi:2019ezj,Ballardini:2020iws,Ye:2021nej}, and references therein, for a partial coverage of the related literature). We mention also that a very interesting attempt to measure $H_0$ from CMB lensing, without invoking the knowledge of $r_s$, has been done in \cite{Baxter_2020}. This method returns a $r_s$-independent constraint of $H_0=73.5\pm5.3$ km s$^{-1}$ Mpc$^{-1}$, which seems to reabsorb the tension.

Given this state-of-art, we will adopt an agnostic approach to the problem. Indeed, within the conservative framework of $\Lambda$CDM model\footnote{See \cite{Perivolaropoulos:2021jda} for a detailed review about the new challenges for $\Lambda$CDM.}, the question we aim to answer is the following: in view of the forthcoming LSS surveys, is there any theoretical bias which might increase the standard deviation in order to alleviate the tension? This question has already been partially faced in the past. In particular, in \cite{Ben-Dayan:2014swa} the effect of velocity dispersion of local SnIa (at redshift lower than 0.1) has been studied and it has been shown that it can introduce a further intrinsic error in the local estimation of $H_0$ of $\sim 1\%$. However, this effect can mitigate but not resolve the tension. In this work, we extend our analysis to forecast the estimated precision for catalog which will have deeper extension in redshift (up to redshift 3.85), just like Euclid Deep Survey (EDS) and LSST. To this aim, we will assume that the new generation of standard candles known as Superluminous Supernovae (SLSNe) will provide a suitable dataset for the analysis, following what has been recently claimed in \cite{Inserra:2020uki}.
In the same spirit but concerning the forecast for future CMB surveys (such as CMB-S4 \cite{Abazajian:2019eic}), in \cite{Baxter_2020} the forecasted error for the above-mentioned $r_s$-independent measurement of $H_0$ is about $\sigma_{H_0}=3$ km s$^{-1}$ Mpc$^{-1}$.

To perform the analysis, our starting point is to derive the analytical formula for the 2-point correlation function of the luminosity distance-redshift relation. Hence, we will discuss general aspects of this function, whose interests go beyond the ones concerning $H_0$. In particular, we will investigate numerically the cosmological information encrypted in the lowest angular multipoles of the 2-point correlation functions. To conclude, we will finalize our analysis by providing forecasted errors of $\sim 0.1\%$ for EDS and $\sim 0.01\%$ for LSST in regard of the measurement of $H_0$. These results are obtained by using linear power spectrum. However, as we will show, non-linear scales do not dramatically change this scenario. Their contribution enhances the forecasted errors to $\sim 0.1\%$ for both EDS and LSST. This renders the measurement for high-redshift standard candles ideally much more precise than the one so far discussed in \cite{Ben-Dayan:2014swa} for close sources.

The paper is organized as follows. In Sect. \ref{sec:cos_var} we describe the general method followed to infer the value of $H_0$, from higher LSS surveys, thanks to the knowledge of the linear luminosity distance-redshift relation. In Sect. \ref{sec:an_res} we compute in details the 2-point correlation function for all the relativistic effects involved in the linear luminosity distance-redshift relation, providing general expressions for them.
In Sect. \ref{sec:multipoles}, we further give numerical details about the spectrum of lower angular multipoles for lensing and doppler effect, and a general treatment about the monopole concerning all the effects. In Sect. \ref{sec:numerics}, we assume a numerical set of the cosmological parameters within the $\Lambda$CDM model and estimate the 2-point correlation function effect by effect. Furthermore, we discuss some technical approximations which speed up the numerical evaluations of the lensing effect. In Sect. \ref{sec:next_gen_sur} we
outline the consequences of our numerical analysis according to the specifics of EDS and LSST and discuss the impact of non-linear scales. Finally, in Sect. \ref{sec:conclusions} we summarize our main conclusions.
Moreover, in App. \ref{terms_derivation} we furnish technical details about the calculations of the 2-point correlation function. In App. \ref{app:expansion}, we report explicit expression for the lower multipoles of the lensing angular spectrum. App. \ref{app:bessel} contains useful properties of the spherical Bessel functions, whereas App. \ref{app:legendre} contains details about the multipole expansion of lensing 2-point correlation function.


\section{Cosmic variance}
\label{sec:cos_var}
The starting point for our analysis is the well-known expression for the luminosity distance-redshift relation $d_L(z)$ in the homogenous and isotropic Cosmic Concordance model
\begin{equation}
d_L(z)=  \frac{1+z}{H_0}  \int _{0} ^z \frac{dz'}{\sqrt{\Omega_{0m}(1+z')^3+\Omega_{0\Lambda}} },
\label{d_L(H)}
\end{equation}
where $H_0$ is indeed the Hubble constant and $\Omega_{0m}$ and $\Omega_{0\Lambda}$ are respectively the energy density for the Cold Matter and Cosmological Constant today. Within the $\Lambda$CDM model assumptions, it is well known that $\Omega_{0\Lambda}=1-\Omega_{0m}$ such that Eq. \eqref{d_L(H)} contains only two free parameters. Eq. \eqref{d_L(H)} can then be inverted and provides a relation which can be used to infer the value of $H_0$ as
\begin{equation}
H_0 = \frac{1+z}{d_L(z)}  \int _{0} ^z \frac{dz'}{\sqrt{\Omega_{0m}(1+z')^3+1-\Omega_{0m}} } \, .
\label{eq:H0_back}
\end{equation}
In this way, given an \textit{a priori} knowledge of $\Omega_{0m}$, independent estimations for redshift and luminosity distance from a given sample of sources can provide a measurement of $H_0$.

Within this framework, the question that we want to address is then the following: how precise can in principle be this estimation if we consider the inhomogeneities all around our observed Universe? In other words, the observed luminosity distance-redshift relation $d_L(z)$ is affected by the inhomogeneities and this provides an intrinsic dispersion for $d_L(z)$ which is governed by the way the cosmological structures are distributed and evolve. Hence, in this regard, we consider the observed inhomogeneous luminosity distance-redshift relation $\widetilde{d_L}(z)$
\begin{equation}
\widetilde{d_L}(z,{\bf n})=d_L(z) \left[1 + \delta^{(1)}(z,{\bf n})+ \delta^{(2)}(z,{\bf n})\right] \, ,
\label{eq:dL_pert}
\end{equation}
where ${\bf n}$ is the observed direction for the given source and $\delta^{(1)}$ and $\delta^{(2)}$ are linear and second order corrections to the luminosity distance-redshift relation. Let us underline that here we do not have to consider perturbations in the redshift $z$. Indeed, the inhomogeneous $\widetilde{d_L}(z)$ is evaluated by construction at constant observed redshift hypersurfaces. This means that redshift corrections are already taken into account in the $\widetilde{d_L}(z,{\bf n})$. From the geometrical viewpoint, this choice coincides with slicing the space-time on constant observed redshift time-like hypersurfaces and then the time-like gauge mode sets redshift perturbations null by construction. Since $\widetilde{d_L}(z)$ is an observable, this choice is completely allowed and does not affect the result.

At this point, following \cite{Ben-Dayan:2014swa},we define the inhomogeneous value of the Hubble constant $\widetilde{H_0}$ as
\begin{align}
\widetilde{H_0} \equiv& \frac{1+z}{\widetilde{d_L}(z)}  \int _{0} ^z \frac{dz'}{\sqrt{\Omega_{0m}(1+z')^3+1-\Omega_{0m}} }\nonumber\\
=& H_0\frac{d_L(z)}{\widetilde{d_L}(z)}\nonumber\\
=&H_0\left[1 - \delta^{(1)} - \delta^{(2)} + \left(\delta^{(1)}\right)^2 \right](z,{\bf n}) \,.
\label{eq:H0_pert}
\end{align}
Eq. \eqref{eq:H0_pert} contains also pure second order perturbations of the luminosity distance-redshift relation. Because of that, $\widetilde{H_0}$ inferred from the observation of a single source is expected to deviate from $H_0$ (see \cite{Ben-Dayan:2014swa} for the case of small redshift surveys). To estimate this deviation, we should select a prescription of the {\it light-cone average} taken all around the observed sky at fixed redshift which is well-suited for our observables (see \cite{Gasperini:2011us,Bonvin:2015kea,Heinesen:2018vjp,Fanizza:2019pfp} for the general classification of the viable well-posed prescriptions for the light-cone averages). However, if we consider only two dimensional spheres at constant redshift, the impact of such a measure is null on the estimation of the variance at the leading order \cite{BenDayan:2012pp}. In addition, also the estimation of three dimensional light-cone averages over a redshift bin reduces to a two dimensional average for the limit case of small redshift bin (see \cite{Fanizza:2019pfp} for the detailed discussion of this limit and also \cite{Fleury:2016fda,Yoo:2017svj}). Hence, also in this case the prescription for the average is irrelevant for the leading order term of the variance\footnote{This result is powerful enough to ensure the variance at leading order is not affected by any bias between the distribution of matter and the one of the sources. Indeed, following \cite{Fanizza:2019pfp}, a weight involving the matter density in the measure for the light-cone average can be added to the exact expression. However, since it is involved in the measure, its effect on the variance is null at leading order, again according to \cite{BenDayan:2012pp}.}. Because of that, we simply skip the measure in our formalism and define the {\it light-cone average} as
\begin{equation}\label{sol_ang_ave}
\left \langle \ldots \right\rangle \equiv \frac{1}{4 \pi} \int d \Omega (\ldots)\,,
\end{equation}
having in mind that this is no longer valid neither for the evaluation of next-to-leading order contribution nor for the finite-size redshift bin average. On top of that, we also denote with $\overline{\cdots}$ the {\it ensemble average} over all the possible configuration of cosmological perturbations, provided that linear perturbations have a gaussian distribution with null mean value\footnote{We will provide analytic version for these assumptions later, in Eqs. \eqref{eq:delta}.}. In this way, the variance related to the estimation of $H_0$ is given by
\begin{equation}
\sigma^2_{H_0}\equiv \overline{\langle \widetilde{H_0}^2 \rangle}-\overline{\langle \widetilde{H_0} \rangle}^{\,2}\,.
\label{eq:variance}
\end{equation}
Hence, from Eq. \eqref{eq:H0_pert} we get at second order
\begin{align}
\overline{\langle\widetilde{H_0}^2\rangle}=&H^2_0\left[1 - 2\,\overline{\langle\delta^{(2)}\rangle} + 3\,\overline{\langle\left(\delta^{(1)}\right)^2\rangle} \right]\nonumber\\
\overline{\langle \widetilde{H_0} \rangle}^{\,2}=&H^2_0\left[1 - 2\,\overline{\langle\delta^{(2)}\rangle}
+ 2\,\overline{\langle\left(\delta^{(1)}\right)^2\rangle}\right]\,.
\end{align}
It then follows from Eq. \eqref{eq:variance}
\begin{equation}
\sigma^2_{H_0}=H^2_0\,\overline{\langle \left(\delta^{(1)}\right)^2 \rangle}\,.
\label{eq:th_variance}
\end{equation}
As above-mentioned, the leading order of $\sigma^2_{H_0}$ is entirely given by linear perturbation theory. This is in agreement with \cite{BenDayan:2012pp} and is a quite general result, independent of the chosen observable (see also \cite{Fanizza:2015gdn} for the application of this result related to the estimation of cosmological parameters to other cosmological observables).

Eq. \eqref{eq:th_variance} provides then the intrinsic uncertainty to the estimation of $H_0$ given by the presence of inhomogeneities all around our observed Universe and it is the lowest theoretical uncertainty we can reach, according to the sample of sources we have access to. This quantity is usually named {\it cosmic variance} and quantifies the error for the estimation of $H_0$ from a single source placed in an ideal survey of a large number of sources for each constant redshift hypersurface, uniformly distributed all over the sky. In practice, however, all the surveys contain a finite number of sources $N$ which can cover only a partial window of the sky. Given that, following \cite{Ben-Dayan:2014swa}, a more observationally oriented definition for the estimation of $\sigma^2_{H_0}$ is provided by
\begin{equation}
\overline{\left\langle \ldots \right\rangle} \to \frac{1}{N^2} \sum_{i,j} \overline{(\ldots)}\,,
\label{eq:sum}
\end{equation} 
where the indices $i,j$ run over all the pairs $(z_i,{\bf n}_i)$, respectively labeling redshift and observed position of the $i$-th source in the survey. Eq. \eqref{eq:sum} then adopt the following prescription for the uncertainty
\begin{equation}
\frac{\sigma^2_{H_0}}{H_0^2}=\frac{1}{N^2} \sum_{i,j} \overline{\delta^{(1)}(z_i,{\bf n}_i)\,\delta^{(1)}(z_j,{\bf n}_j)}\,,
\label{eq:cov_matrix}
\end{equation}
rather than Eq. \eqref{eq:th_variance}. Eq. \eqref{eq:cov_matrix} is the variance associated to the average value of $H_0$ inferred from a finite survey of $N$ sources and corresponds to the locally measured Hubble parameter $H_0$ from the covariance matrix of the $\widetilde{d_L}(z)$, given an arbitrarily distributed sample of $N$ observed sources at positions $(z_i, {\bf n}_i)$.
Indeed, the variance associated to the average value of $H_0$ inferred from a finite survey of $N$ sources is
\begin{align}
\sigma^2_{H_0}=&\overline{\left( \sum_{i}\frac{\widetilde{H_{0}}(z_i,{\bf n}_i)}{N}-H_0 \right)\left( \sum_{j}\frac{\widetilde{H_{0}}(z_j,{\bf n}_j)}{N}-H_0 \right)}
\nonumber\\
=&\frac{1}{N^2}\sum_{i,j}\overline{\left( \widetilde{H_{0}}(z_i,{\bf n}_i) \widetilde{H_{0}}(z_j,{\bf n}_j)-H_0^2 \right)}
\nonumber\\
=&\frac{H^2_0}{N^2} \sum_{i,j} \overline{\delta^{(1)}(z_i,{\bf n}_i)\,\delta^{(1)}(z_j,{\bf n}_j)}\,,
\end{align}
which precisely corresponds to Eq. \eqref{eq:cov_matrix}.

In the next section, these general preliminaries will be applied to the case of linear perturbations of the luminosity distance. This will provide the explicit expression for $\sigma^2_{H_0}$ due to all the linear relativistic corrections.


\section{Analytic results}
\label{sec:an_res}
In the previous section, we have shown in complete generality that the cosmic variance $\sigma^2_{H_0}$ is sourced at the leading order only by linear perturbations. To make the explicit evaluation of all the terms needed for its estimation, we first need to consider all the linear relativistic corrections involved in the $\delta^{(1)}$. To this aim, we only consider linear scalar perturbations in the Longitudinal Gauge without anisotropic stress\footnote{This assumption might look too restrictive. However in the following sections, we will take into account sources located after the decoupling, where our assumption works well. The reader interested in the general expression in presence of anisotropic stress can have look at \cite{Marozzi:2014kua} for the general non-linear expression of luminosity distance-redshift relation and \cite{Fanizza:2018qux} for the non-linear expression of redshift containing also the observer terms.}
\begin{align}
ds^2=a^2(\eta)\left\{ -\left( 1+2\psi \right)d\eta^2+\left( 1-2\psi \right)\left[dr^2+r^2d\Omega^2\right] \right\}\,,
\nonumber\\
\end{align}
and then formally write the linear perturbation of $\delta^{(1)}$ \cite{Bonvin:2005ps,BenDayan:2012wi,Fanizza:2015swa,Umeh:2014ana} as
\begin{equation}
\delta^{(1)}(z,{\bf n})= \sum_{E} \hat{O}_E\, \psi(\eta_E,r_E\,{\bf n})\,,
\label{eq:formal_sum}
\end{equation}
where the index $E$ denotes the sum over the linear relativistic effects, the linear operators\footnote{Here the subscripts stand for Peculiar Velocity (PV), Sachs-Wolfe (SW), Integrated Sachs-Wolfe (ISW), Time Delay (TD) and Lensing (L). Hence the label $E$ runs in the set (PV, SW, ISW, TD, L). We omit relativistic corrections due to the gravitational potential at the observer position.} are
\begin{align}
\hat{O}_{PV}=& -\Xi_s \int_{\eta_{in}}^{\eta_{s}}  d\eta \frac {a(\eta)} {a(\eta_o)} \partial_r (\ldots) \nonumber\\
\hat{O}_{SW}=&  - (1 + \Xi_s) (\ldots) \nonumber\\
\hat{O}_{ISW}=&  -2 \, \Xi_s \int_{\eta_{s}}^{\eta_o}  d\eta\, \partial_\eta (\ldots) \nonumber\\
\hat{O}_{TD}=& \frac{2}{\Delta \eta_s} \int_{\eta_{s}}^{\eta_o}  d\eta (\ldots) \nonumber\\
\hat{O}_{L}=& - \frac{1}{\Delta \eta_s} \int_{\eta_{s}}^{\eta_o}  d\eta \frac{ \eta - \eta_{s}}{ \eta_o - \eta } \Delta_2 (\ldots) \label{eq:Operators}\,,
\end{align}
where $\eta_o$ is the present conformal time, $\eta_s$ is the conformal time of the source, $\eta_{in}$ is an initial time when perturbations were negligible (or, more precisely, the integrands of the related operators were negligible), $\Delta \eta_x= \eta_o -\eta_x$ (where $x$ can be either $s$ or $i$), $\Delta_2$ is the angular Laplacian and
\begin{equation}
\Xi_s = \left(1-\frac{1}{\mathcal{H}_s \Delta \eta_s} \right)\,,
\end{equation}
where $\mathcal{H}_s=\partial_\eta a(\eta_s)/a(\eta_s)$ is the conformal Hubble function. In Eq. \eqref{eq:formal_sum}, $\eta_E$ and $r_E$ in $\psi$ depends on which operator acts on $\psi$. Indeed, for the relativistic effects integrated along the observer's past light-cone, i.e. ISW, TD and L, we have that $\eta_E=\eta$ and $r_E=\eta_o-\eta$, so both of these variables are integrated. On the other hand, for what concerns the PV, the $\psi$ is integrated along the source world-line and then $\eta_E=\eta$ whereas $r_E=\eta_o-\eta_s$. This means that for the PV only time is integrated when the $\hat{O}_{PV}$ acts on $\psi$. Finally, SW is a local relativistic effect and then, in this case $\eta_E=\eta_s$ and $r_E=\eta_o-\eta_s$.

At this point, for a practical evaluation of the ensemble average, we move from real to $k$-space. We then Fourier transform the gravitational potential as
\begin{equation}
\psi \left( \eta_E,r_E {\bf n}\right)=\frac{1}{(2\pi)^{3/2}} \int d^3 k \,  e^{i{\bf k}\cdot{\bf n }\,r_E } \frac{g(\eta_E)}{g(\eta_o)} \widetilde{\psi} ( {\bf k} )  \, ,
\label{eq:fourier}
\end{equation}
where $g(\eta)$ is the standard approximated expression of the growth function of scalar perturbations in terms of the current values of the critical density parameters $\Omega_{m0}$ and $\Omega_\Lambda$  (see e.g. \cite{Peter:2013avv}), namely 
\begin{align}
g(\eta)=&\frac{5}{2}\,g_\infty\frac{\Omega_m}{\Omega^{4/7}_m-\Omega_\Lambda+\left( 1+\frac{\Omega_m}{2} \right)\left( 1+\frac{\Omega_\Lambda}{70}\right)}\nonumber\\
\Omega_m=&\frac{\Omega_{m0}(1+z)^3}{\Omega_{m0}(1+z)^3+\Omega_{\Lambda 0}},
\nonumber\\
\Omega_\Lambda=&\frac{\Omega_{\Lambda 0}}{\Omega_{m0}(1+z)^3+\Omega_{\Lambda 0}},
\label{eq:g}
\end{align}
where $\Omega_{m0}+\Omega_{\Lambda 0}=1$, and where $g_\infty$ is a normalization constant fixed such that $g(\eta_o)=1$. Moreover $\widetilde{\psi}({\bf k})$ are delta-correlated functions
\begin{equation}
\overline{ \widetilde{\psi} ({\bf k}) }= 0 \qquad,\qquad  \overline{\widetilde{\psi}( {\bf k} )\,\widetilde{\psi}( {\bf p} )}= | \psi_k |^2 \delta({\bf k} +{\bf p}) \, .
\label{eq:delta}
\end{equation}
In terms of this expansion, linear perturbations in Eq. \eqref{eq:formal_sum} can be written as
\begin{equation}
\delta_s^{(1)}=\frac{1}{(2\pi)^{3/2}} \int d^3 k \,  \sum_E\,\hat{O}_E\left[e^{i{\bf k}\cdot{\bf n }\,r_E } \frac{g(\eta_E)}{g(\eta_o)} \widetilde{\psi} ( {\bf k} )\right]\, .
\label{eq:formal_sum_k}
\end{equation}
In this way, the combination of Eq. \eqref{eq:cov_matrix} with Eq. \eqref{eq:Operators} gives
\begin{align}
\frac{\sigma^2_{H_0}}{H_0^2}&=\frac{ 1 }{ N^2 }\sum_{i,j}\overline{ \delta_i^{(1)} \delta_j^{(1)} } \nonumber\\
 &=\frac{ 1 }{ N^2 }\sum_{i,j}\sum_{E,E'}\hat{O}_{Ei}   \hat{O}_{E'j} \overline{ \psi(\eta_{Ei},r_{Ei}\,{\bf n}_i) \psi(\eta_{E'j},r_{E'j}\,{\bf n}_j) } \, ,
\label{sig_teo}
\end{align}
where the index $i,j$ run over the sources and the index $E,E'$ run over all the effects for each source. Hence, by inserting the expansion \eqref{eq:fourier} in Eq. \eqref{sig_teo} and using Eq. \eqref{eq:delta} we obtain
\begin{equation}
\frac{\sigma^2_{H_0}}{H_0^2} = \frac{ 1 }{ N^2 }\sum_{i,j} \sum_{E,E'} \int \frac{dk } {k} \mathcal{P} _\psi(k)\mathcal{W}_{Ei,E'j}\,,
\label{DH_(W)}
\end{equation}
where we have defined
\begin{align}
\mathcal{W}_{Ei, E'j} \equiv& \frac{ 1 } { 4 \pi } \int d\Omega_k\,\hat { O } _ { E i } \hat { O } _ { E' j }
\nonumber\\
&\times\left[\frac{g(\eta_{Ei})}{g(\eta_o)} \frac{g(\eta_{E'j})}{g(\eta_o)}e ^ { i {\bf k} \cdot (r_{Ei}{\bf n}_i-r_{E'j}{\bf n}_j )} \right]\,,
\label{eq:general_kernels}
\end{align}
with $\Omega_k$ to be meant as the solid angle in $k$-space and we used the so-called dimensionless power spectrum

\begin{equation}
\mathcal{P} _\psi(k) \equiv \frac{k^3}{2 \pi^2} |\psi_k|^2 \, .
\end{equation}

The variance in Eq. \eqref{DH_(W)} can be written as a sum of the contribution over different pairs of effects. Since we have 5 different effects we will find 15 different pairs of effects. We distinguish between the contribution of the effects of the same kind, which we refer to as \textit{pure terms} and of the effects of different kinds, which we refer to as \textit{mixed terms}
\begin{align}
\frac{\sigma^2_{H_0}}{H_0^2} =& \frac{ 1 }{ N^2 }\sum_{i,j}\sum_E \xi_E(z_i,z_j,{\bf n}_i\cdot{\bf n}_j)
\nonumber\\
&+\frac{ 1 }{ N^2 }\sum_{i,j} \sum_{E\ne E'} \xi_{EE'}(z_i,z_j,{\bf n}_i\cdot{\bf n}_j)\,,
\label{eq:correlation_sum}
\end{align}
where we have defined
\begin{align}
\xi_{EE'}(z_i,z_j,{\bf n}_i\cdot{\bf n}_j)=&\int \frac{dk } {k} \mathcal{P} _\psi(k)\mathcal{W}_{EE'ij}\,,
\nonumber\\
\xi_E(z_i,z_j, {\bf n}_i\cdot{\bf n}_j)=&\,\xi_{EE}(z_i,z_j, {\bf n}_i\cdot{\bf n}_j)\,.
\label{eq:corr_func}
\end{align}
$\xi_{EE'}(z_i,z_j,{\bf n}_i\cdot{\bf n}_j)$ are nothing but the 2-point correlation functions between the relativistic effects $E$ and $E'$ evaluated for two different sources with redshifts $z_i$ and $z_j$ along the observed directions ${\bf n}_i$ and ${\bf n}_j$. We conclude this section with the explicit expressions for all the $\mathcal{W}_{EE'}$. In particular, we define
\begin{align}
G_i=&\int_{\eta_{in}}^{\eta_i}d\eta \, \frac{a(\eta)}{a(\eta_i)} \frac{g(\eta)}{g(\eta_o)}\nonumber\\
R(\eta_x,\eta_y,\nu)=&\sqrt{\Delta\eta^2_x+\Delta\eta^2_x-2\Delta\eta_x\Delta\eta_y\nu}
\nonumber\\
L(\eta_x,\eta_y,\nu) =& \frac{\Delta\eta_x\,\Delta\eta_y \nu}{R\left( \eta_x,\eta_y,\nu \right)}\nonumber\\
H(\eta_x,\eta_y,\nu) =& \frac{\Delta\eta_x\,\Delta\eta_y \sqrt{1-\nu^2}}{R\left( \eta_x,\eta_y,\nu \right)}\,,
\label{eq:important_eqs}
\end{align}
where $R$ is the distance between two sources and $L$ and $H$ are respectively the normalized scalar and (modulo of the) vector products between the two directions of the sources.
We then have that the pure terms are
\begin{widetext}
\begingroup
\allowdisplaybreaks
\begin{align}
 \mathcal{W}_{PVij} &= \Xi _ { i } \Xi _ { j } G _ { i } G _ { j } k^2 \left\{\frac{\Delta\eta_i\,\Delta\eta_j(1-\nu^2)}{R^2}j_2 (kR)+    \frac{\nu}{3}\left[ j_0\left(kR\right)-2j_{2}\left(kR\right)\right]\right\}(\eta_i,\eta_j,\nu)
 \nonumber\\
\mathcal{W}_ { SW i j }=& (1+ \Xi_i) (1+ \Xi_j) \, \frac { g ( \eta _ { i } ) } { g ( \eta_o ) } \frac { g ( \eta _ { j } ) } { g ( \eta_o ) } j_0(kR(\eta_i,\eta_j,\nu))
\nonumber\\
\mathcal{W}_{ISWij} =&  4\,\Xi _ { i }\,\Xi _ { j } \int _ { \eta _ { i }} ^ { \eta_o } d \eta \int _ { \eta _ { j }} ^ { \eta_o } d \eta'  \frac { \partial _{\eta}  g ( \eta  ) } { g ( \eta_o ) } \frac {  \partial _{\eta'} g ( \eta ' ) } { g ( \eta_o ) }  j_0(kR(\eta,\eta',\nu))
\nonumber\\
\mathcal{ W } _ {TDij} =& \frac{4}{\Delta \eta_i\Delta \eta_j} \int_{\eta_{i}}^{\eta_o}  d\eta \int_{\eta_{j}}^{\eta_o}  d\eta' \frac {  g ( \eta ) } { g ( \eta_o ) } \frac { g ( \eta ^ { \prime } ) } { g ( \eta_o ) } j_0(kR(\eta,\eta',\nu))
\nonumber\\
\mathcal { W } _ { L ij } = & \frac { 1 } { \Delta \eta _ { i } } \frac{1}{\Delta \eta _ { j } }  \int _ { \eta _ { i }} ^ { \eta_o } d \eta \frac { \eta - \eta _ { i }} { \eta_o - \eta } \int _ { \eta _ { j } } ^ { \eta_o } d \eta' \frac { \eta' - \eta _ { j } } { \eta_o - \eta' } \frac{g (\eta )g (\eta')  }{g^2 (\eta_o) } \Big[  k^4 H^4 j_4(kR) - 8 k^3 H^2 L\,j_3(kR)\nonumber\\
&+ k^2 \left( 8 L^2 -6 H^2 \right) j_2(kR) + 4\,k\,L\,j_1(kR) \Big]\left( \eta,\eta',\nu \right)\,,
\label{eq:pure_kernels}
\end{align}
\endgroup
where $\nu\equiv {\bf n}_i\cdot {\bf n}_j$, $j_n$ are the spherical Bessel functions of $n$-th order and, in the same way, the mixed terms are
\begingroup
\allowdisplaybreaks
\begin{align}
\mathcal{W} _ { PVi, L j } =& \frac{\Xi_i}{\Delta \eta_j} G_i \int_{\eta_{j}}^{\eta_o}  d\eta \frac{ \eta - \eta_{j} }{ \eta_o - \eta }\frac{g(\eta)}{g(\eta_o)} \left\{-k^3 \Delta\eta^2 \frac{(\Delta\eta_i - \nu \Delta\eta) (\Delta\eta - \nu \Delta\eta_i)^2 }{ R^3 } j_3(kR)\right. \nonumber\\
&\left.+ k^2 \Delta\eta \left(3 \Delta\eta \frac{\Delta\eta_i - \nu\,\Delta\eta } { R^2 } - 2 \nu \right) j_2(kR) - \frac{k\Delta\eta}{R} \left[ k^2 \Delta\eta (\Delta\eta_i-\nu \Delta\eta)-2 \nu \right] j_1(kR)\right\}\left( \eta_i,\eta,\nu \right)\nonumber\\
\mathcal{ W } _ { PV i, SW j } =& \Xi_i (1+ \Xi_j) G_i \frac{g(\eta_j)}{g(\eta_o)} \, k \, \left( \nu \Delta\eta_j - \Delta\eta_i\right) \left(\frac{j_1(kR)}{R}\right)\left( \eta_i,\eta_j,\nu \right)
\nonumber\\ 
\mathcal{ W } _ { PV i, ISW j } =& 2 \, \Xi_i \Xi_j G_i \int_{\eta_{j}}^{\eta_o}  d\eta \frac{\partial _{\eta}g(\eta)}{g(\eta_o)}  \, k \,  \left(\nu \Delta\eta - \Delta\eta_i \right) \left(\frac{j_1(kR)}{R}\right)\left( \eta_i,\eta,\nu \right) \nonumber\\
\mathcal{ W } _ { PV i, TD j } =& -2 \,  \frac{\Xi_i} {\Delta \eta_j} G_i \int_{\eta_{j}}^{\eta_o}  d\eta \frac{g(\eta)}{g(\eta_o)}  \, k \, \left(\nu \Delta\eta - \Delta\eta_i \right) \left(\frac{j_1(kR)}{R}\right)\left( \eta_i,\eta,\nu \right)
\nonumber\\
\mathcal{ W } _ { L i, SW j } =& \frac{1+\Xi_j}{\Delta \eta_i} \frac{g(\eta_j)}{g(\eta_o)} \int_{\eta_i}^{\eta_o}  d\eta \frac{ \eta - \eta_i }{ \eta_o - \eta } \frac{g(\eta)}{g(\eta_o)} 
 \left[ k^2 H^2 j_0(kR) - k \left( \frac{3 H^2}{R} - 2 L \right) j_1(kR) \right ]\left( \eta,\eta_j,\nu \right) \nonumber\\
\mathcal{ W } _ { L i, ISW j } =& 2 \,  \frac{\Xi_j}{\Delta \eta_i}  \int_{\eta_i}^{\eta_o}  d\eta \frac{ \eta - \eta_i }{ \eta_o - \eta } \frac{g(\eta)}{g(\eta_o)}   \int_{\eta_j}^{\eta_o}  d\eta' \frac{\partial _{\eta'}g(\eta')}{g(\eta_o)}  \left[ k^2 H^2 j_0(kR) - k \left( \frac{3 H^2}{R} - 2 L \right) j_1(kR) \right ]\left( \eta,\eta',\nu \right) \nonumber\\
\mathcal{ W } _ { L i, TD j } =& - 2 \,  \frac{1}{\Delta \eta_i \Delta \eta_i}  \int_{\eta_i}^{\eta_o}  d\eta \frac{ \eta - \eta_i }{ \eta_o - \eta } \frac{g(\eta)}{g(\eta_o)}   \int_{\eta_j}^{\eta_o}  d\eta' \frac{g(\eta')}{g(\eta_o)} 
 \left[ k^2 H^2 j_0(kR) - k \left( \frac{3 H^2}{R} - 2 L \right) j_1(kR) \right ]\left( \eta,\eta',\nu \right) \nonumber\\
\mathcal{ W } _ { SW i, ISW j } =& 2 \, (1+\Xi_i) \Xi_j \frac{g(\eta_i)}{g(\eta_o)} \int_{\eta_j}^{\eta_o}  d\eta \frac{\partial _{\eta}g(\eta)}{g(\eta_o)}  \, j_0\left(kR(\eta_i,\eta,\nu)\right) \nonumber\\
\mathcal{ W } _ { SW i, TD j } =& -2 \, \frac{1+\Xi_i}{\Delta \eta_j}\frac{g(\eta_i)}{g(\eta_o)} \int_{\eta_j}^{\eta_o}  d\eta \frac{g(\eta)}{g(\eta_o)}  \, j_0\left(kR(\eta_i,\eta,\nu)\right) \nonumber\\
\mathcal{ W } _ { ISW i, TD j } =& - 4 \, \frac{\Xi_i}{\Delta \eta_j}\int_{\eta_i}^{\eta_o}  d\eta \frac{\partial _{\eta}g(\eta)}{g(\eta_o)}  \int_{\eta_j}^{\eta_o}  d\eta' \frac{g(\eta')}{g(\eta_o)}  \, j_0\left(kR(\eta,\eta',\nu)\right) \,.
\label{eq:mixed_kernels}
\end{align}
\endgroup
\end{widetext}
The detailed analytic derivation for these 15 different contributions is reported in App. \ref{terms_derivation} and follows the derivation obtained in \cite{Fiorini2018}. We just remark that the kernel $\mathcal{W}_{PVij}$ is in agreement with the one found in \cite{Ben-Dayan:2014swa}.


\section{Multipoles analysis}
\label{sec:multipoles}
Let us now investigate the angular decomposition of the 2-point correlation function $\xi$ in Eqs. \eqref{eq:corr_func}. To this end, we expand the angular dependence in multipoles as
\begin{equation}
\xi_{EE'}(z_1,z_2,\nu)=\sum_{\ell=0}^\infty C^{EE'}_\ell(z_1,z_2)\,P_\ell(\nu)
\label{eq:multipoles}
\end{equation}
where $P_\ell(x)$ are the Legendre polynomials of order $\ell$ and
\begin{equation}
C^{EE'}_\ell(z_1,z_2)=\frac{2\ell+1}{2}\int_{-1}^{1}d\nu\,\xi(z_1,z_2,\nu)\,P_\ell(\nu)\,.
\label{eq:LegendreDecomposition}
\end{equation}
With this decomposition, we aim to investigate the behavior of low multipoles, in order to understand how fast the truncated version of Eq. \eqref{eq:multipoles} converges to the full numerical results. To the extent of this paper, we just limit our analysis to the lensing and doppler terms in the 2-point correlation function, but this analysis can be applied to all the effects.

\subsection{Lensing}
We start by analyzing in detail the 2-point correlation function for the lensing. Hence, we apply Eq. \eqref{eq:multipoles} to the term
\begin{equation}
\xi_L(z_1,z_2,\nu)=\int \frac { d k } { k } \mathcal{P} _ { \psi } ( k ) \mathcal { W } _ { L 12}(z_1,z_2,\nu)
\end{equation}
where $\mathcal{W}_{L12}$ is the lensing kernel for the 2-point correlation function as reported in Eqs. \eqref{eq:pure_kernels}. First of all, we notice that the monopole is exactly $C_0= 0$, regardless of the chosen redshifts (see Appendix \ref{app:legendre} for details). In the ideal case of infinite number of sources densely distributed in each redshift bin all over the sky, the statistical average tends to the monopole. Hence, in this case, lensing is not expected to affect the variance of $H_0$ at all. It is interesting to notice that this property about the monopole stands for all the cross-correlation terms between lensing and other effects, as we will prove in Eqs. \eqref{eq:commutation} and \eqref{monopole_terms}. This means that the above-mentioned ideal case is not affected at all by the expected leading correction due to lensing.

\begin{figure}[ht!]
\centering
\includegraphics[scale=0.9]{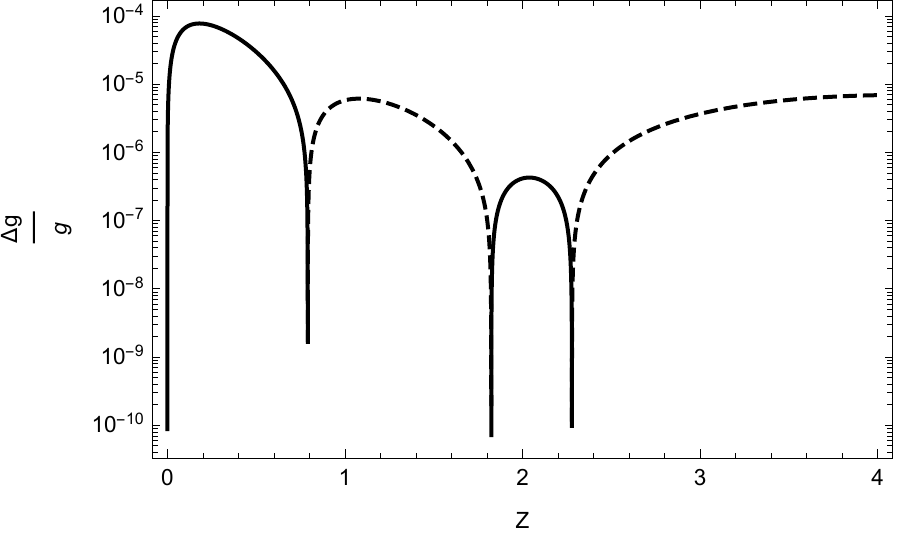}
\caption{Relative error between the exact numerical solution for the growth function in term $z$ and its approximated polynomial expression with coefficient $g_0=1,\,g_1=1.10015 \times 10^{-4},\,g_2=-1.81753\times 10^{-8},\,g_3=1.52535\times 10^{-12},\,g_4=-6.9544\times 10^{-17}\text{ and }g_5=1.1202\times 10^{-21}$. Solid lines refer to positive values and dashed lines stand for negative ones. As we can see from this plot, our polynomial approximation for $g(z)$ is precise at $0.01\%$ level.}
\label{fig:approx}
\end{figure}

However, realistic surveys deals with partial sky coverages. As discussed before, this sky coverage is very limited for realistic forthcoming surveys (see, for instance, Euclid Deep Survey and LSST). Hence, the effect of higher multipoles is expected to contribute to the variance for realistic surveys. This is indeed due to the fact that window function introduced by the partial sky-coverage is convolved in $\ell$-space with the higher multipoles and then an amount of power is transferred from higher multipoles to the monopole itself. In particular, each multipoles can be written as an integral in $k$-space as
\begin{equation}
C^L_{\ell}=\int \frac{dk}{k} \mathcal{P}_\psi (k)\mathcal{L}^L_\ell(z_1,z_2,k)\,,
\label{eq:CL}
\end{equation}
where the kernel of the integrand $\mathcal{L}^L_\ell$ is
\begin{equation}
\mathcal{L}^L_\ell=\frac{2\ell+1}{2}\int_{-1}^{1}d\nu\,\mathcal{W}_L(z_1,z_2,\nu)P_\ell(\nu)\,.
\label{eq:kernel}
\end{equation}
It is hard to solve Eq. \eqref{eq:kernel} analytically. However, we have outlined the following approximation scheme to deal with it. We first perform variable changes $x=\eta_o-\eta$ and $y=\eta_o-\eta'$ into the integrals in $\mathcal{W}_L$ in Eq. \eqref{eq:pure_kernels}. This leads to
\begin{widetext}
\begin{align}
\mathcal { W } _ { L ij } = & \frac { 1 } { \Delta \eta _ { i } } \frac{1}{\Delta \eta _ { j } }  \int_0^{\Delta\eta_i} dx \frac { \Delta\eta _ { i }-x} {x} \int_0^{\Delta\eta_j}d y \frac { \Delta\eta_j-y } {y} \frac{g (\eta_o-x )g (\eta_o-y)  }{g^2 (\eta_o) } \Big[  k^4 H^4 j_4(kR) - 8 k^3 H^2 L\,j_3(kR)\nonumber\\
&+ k^2 \left( 8 L^2 -6 H^2 \right) j_2(kR) + 4\,k\,L\,j_1(kR) \Big]\left(x,y,\nu \right)\,,
\label{eq:Lensing_Kernel_Changed}
\end{align}
\end{widetext}
where the geometrical functions $R$, $L$ and $H$ now simplify to
\begin{align}
R\left( x,y,\nu \right)=&\sqrt{x^2+y^2-2xy\nu}\nonumber\\
L(x,y,\nu) =& \frac{x\,y\,\nu}{R\left( x,y,\nu \right)}\nonumber\\
H(x,y,\nu) =& \frac{x\,y\,\sqrt{1-\nu^2}}{R\left( x,y,\nu \right)}\,.
\label{eq:important_eqs_2}
\end{align}
Hence we adopt a polynomial expansion of the growth function $g$ as $g(\eta_o-x)=\sum_{i=0}^5 g_i x^i$ , where the precision of the approximation is shown in Fig. \ref{fig:approx} and the coefficient $g_i$ are reported in the related caption. As a consequence, Eq. \eqref{eq:Lensing_Kernel_Changed} as well can be expanded as a sum of polynomial terms as
\begin{widetext}
\begin{align}
\mathcal { W } _ { L ij } = & \frac { 1 } { \Delta \eta _ { i } } \frac{1}{\Delta \eta _ { j } } \sum_{n,m=0}^5 \frac{g_n\,g_m}{g^2 (\eta_o) }\int_0^{\Delta\eta_i} dx \frac { \Delta\eta _ { i }-x} {x} \int_0^{\Delta\eta_j}d y \frac { \Delta\eta_j-y } {y} x^n\,y^m \Big[  k^4 H^4 j_4(kR) - 8 k^3 H^2 L\,j_3(kR)\nonumber\\
&+ k^2 \left( 8 L^2 -6 H^2 \right) j_2(kR) + 4\,k\,L\,j_1(kR) \Big]\left(x,y,\nu \right)\,,
\label{eq:Lensing_Kernel_Expanded}
\end{align}
\end{widetext}

Thanks to this trick, integrals over $x,\,y\text{ and }\nu$ in Eq. \eqref{eq:CL} can be done analytically and the computation of the $C^L_\ell$'s eventually requires a single integration in $k$-space left. On one side, this polynomial expansion allows to get analytic results for the integrand of \eqref{eq:kernel}. On the other side, this allows to lower the computational time for each multipoles and increase the numerical precision. Moreover, thanks to the angular integration, integrals over $x$ and $y$ factorize. We show this feature for the dipole $\ell=1$, since the monopole $\ell=0$ is null, as above-mentioned. Indeed, after a long but straightforward calculation, the angular integration in Eq. \eqref{eq:kernel} for $\ell=1$ gives
\begin{align}
\mathcal{L}^L_1=&\frac{12}{k^4\, \Delta \eta _ { i }\Delta \eta _ { j }}
 \sum_{n,m=0}^5 \frac{g_n\,g_m}{g^2 (\eta_o) }\int_0^{\Delta\eta_i} dx \frac { \Delta\eta _ { i }-x} {x}
 \nonumber\\
 &\times\int_0^{\Delta\eta_j}d y \frac { \Delta\eta_j-y } {y} x^{n-2}\,y^{m-2}
\left( kx\cos kx-\sin kx \right)\nonumber\\
&\times\left(  ky\cos ky-\sin ky  \right)\nonumber\\
=&\frac{12}{\Delta \eta _ { i }\Delta \eta _ { j }}
\sum_{n,m=0}^5\frac{g_n\,g_m}{g^2(\eta_o)}Q_{1n}(\Delta\eta_i)Q_{1m}(\Delta\eta_j)\,,
\label{eq:52}
\end{align}
where
\begin{equation}
Q_{1n}(z)=-\int_0^z dx \left(z-x\right)x^{n-1}j_1(kx)\,.
\label{eq:53a}
\end{equation}
$Q_n$ are nothing but integrals of the form $\int dx\,x^\alpha\sin(kx)$ and $\int dx\,x^\alpha\cos(kx)$ which are analytically solvable with multiple integrations by part. The same evaluation performed for first 6 multipoles shows that we can generalize Eq. \eqref{eq:52} for $\ell\le 6$ as
\begin{align}
\mathcal{L}^L_\ell=&\frac{(2\ell+1)\ell^2\left(\ell+1\right)^2}{ \Delta \eta _ { i }\Delta \eta _ { j }} 
\int_0^{\Delta\eta_i} dx \frac { \Delta\eta _ { i }-x} {x}
 \nonumber\\
 &\times
\int_0^{\Delta\eta_j}d y \frac { \Delta\eta_j-y } {y} \frac{g (\eta_o-x )g (\eta_o-y)  }{g^2 (\eta_o) }
j_\ell(kx)j_\ell(ky)\nonumber\\
=&\frac{(2\ell+1)\ell^2\left(\ell+1\right)^2}{\Delta \eta _ { i }\Delta \eta _ { j }}
\sum_{n,m=0}^5\frac{g_n\,g_m}{g^2(\eta_o)}Q_{\ell n}(\Delta\eta_i)Q_{\ell m}(\Delta\eta_j)\,,
\label{eq:kell}
\end{align}
where $Q_{\ell n}$ trivially generalizes Eq. \eqref{eq:53a} as
\begin{equation}
Q_{\ell n}(z)=-\int_0^z dx \left(z-x\right)x^{n-1}j_\ell(kx)\,.
\label{eq:53}
\end{equation}
In App. \ref{app:expansion} we explicitly evaluate $Q_{\ell n}(z)$ for the lower multipoles $\ell=1,2,3$. Eq. \eqref{eq:kell} suggests that it might be generalized for every $\ell$. So far, the general proof for every $\ell$ is still missing, since we have only checked $\ell$ by $\ell$ its validity for the first 6 multipoles, also plotted in Fig. \ref{fig:PowerSpectra}. However, we notice that first line of Eq. \eqref{eq:kell} is consistent with the evaluations done in \cite{Bonvin:2005ps} about the multipoles expansion of the luminosity distance 2-point correlation function.

To show this point, we first remark that Eq. \eqref{eq:kell} takes into account the whole contribution of the angular laplacian in the lensing kernel, which contains also subleading terms in the counting of power of $k$ (see Eqs. \eqref{D} and \eqref{lens_dOk}). On the contrary, in \cite{Bonvin:2005ps} the lensing terms contains only leading terms according to the weight in $k$-space. Hence only first two terms in our Eq. \eqref{D} are taken into account in the term $C^{(5)}_\ell$ of \cite{Bonvin:2005ps}, whereas the last term in our expansion \eqref{D} is accounted for in the term $C^{(3)}_\ell$ of \cite{Bonvin:2005ps}. Beside this different classification, this means that our Eq. \eqref{eq:kell} should be compared with the appropriate combination of kernels of Eqs. (70), (73) and (74) of \cite{Bonvin:2005ps}, according to our Eq. \eqref{D}, rather than just their expression for $C^{(5)}_\ell$. Once this subtle point is taken into account, our evaluations agree with \cite{Bonvin:2005ps}. Since the result of \cite{Bonvin:2005ps} are valid for any $\ell$, this supports the fact that our Eq. \eqref{eq:kell} can be considered for any $\ell$.

We remark the appearance of the prefactor $\ell^2\left( \ell+1 \right)^2$ in Eq. \eqref{eq:kell}. This is due to the presence of $\Delta_2$ in each $\hat{O}_L$ involved in the lensing 2-point correlation function. The fact that monopole vanishes then directly follows from the fact that the eigenvalue of $P_0$ for $\hat{O}_L$ is 0 (see Appendix \ref{app:legendre} for details).

Beside the technical aspects, it is easy to study the behavior of $\mathcal{L}_\ell$'s themselves, such that they can be interpreted as power spectra for each multipole. In Fig. \ref{fig:PowerSpectra} we plot these power spectra for the first six multipoles for the lensing-lensing correlation at the same redshift, which ranges from $z=0.15$ (bluer) to $z=1.55$ (redder).
\begin{widetext}

\begingroup
\allowdisplaybreaks
\begin{figure}[ht!]
\centering
\includegraphics[scale=0.9]{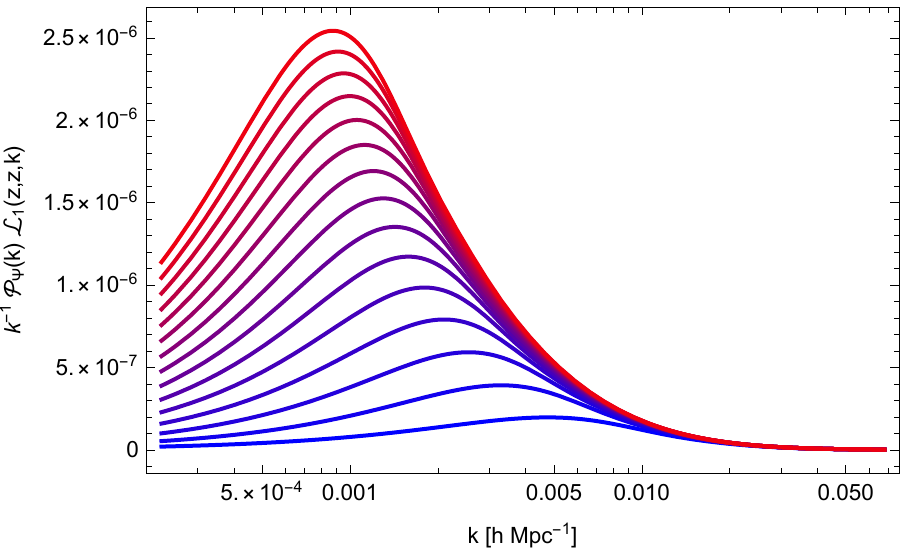}
\includegraphics[scale=0.9]{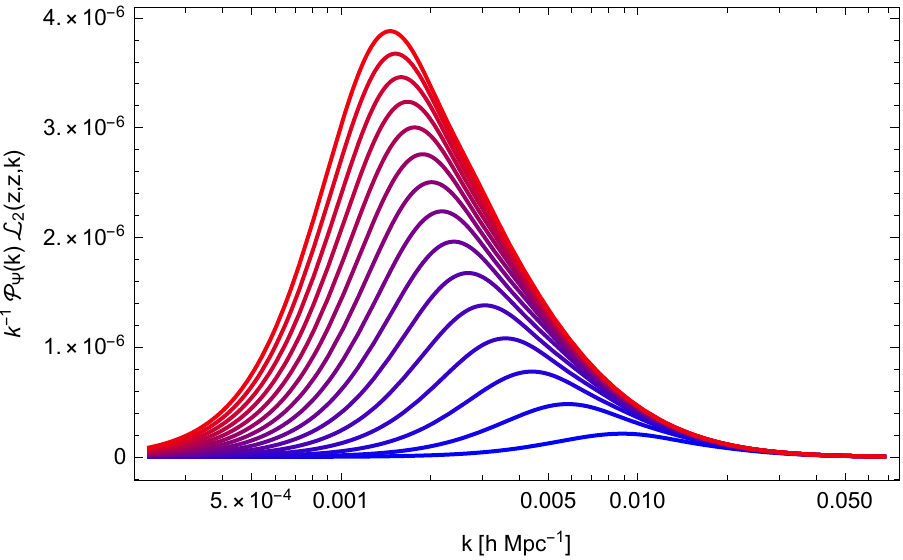}\\
\includegraphics[scale=0.9]{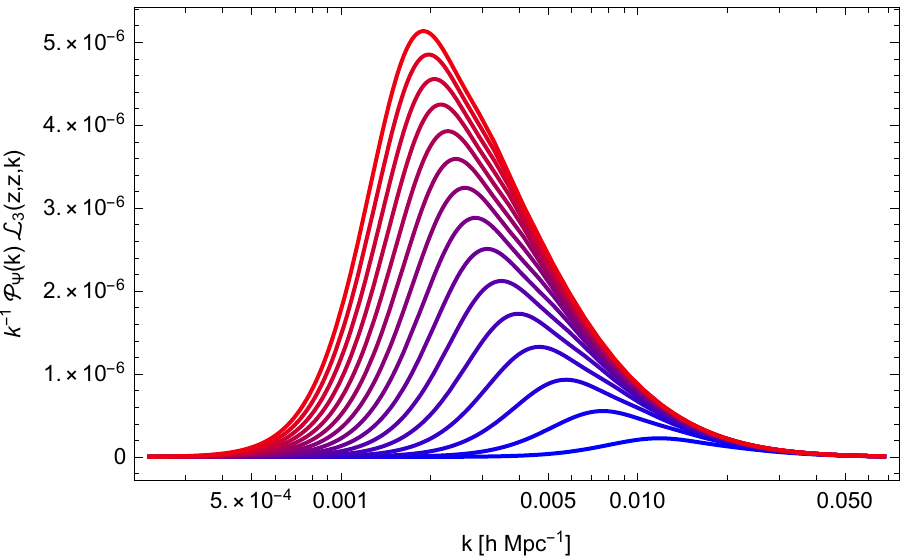}
\includegraphics[scale=0.9]{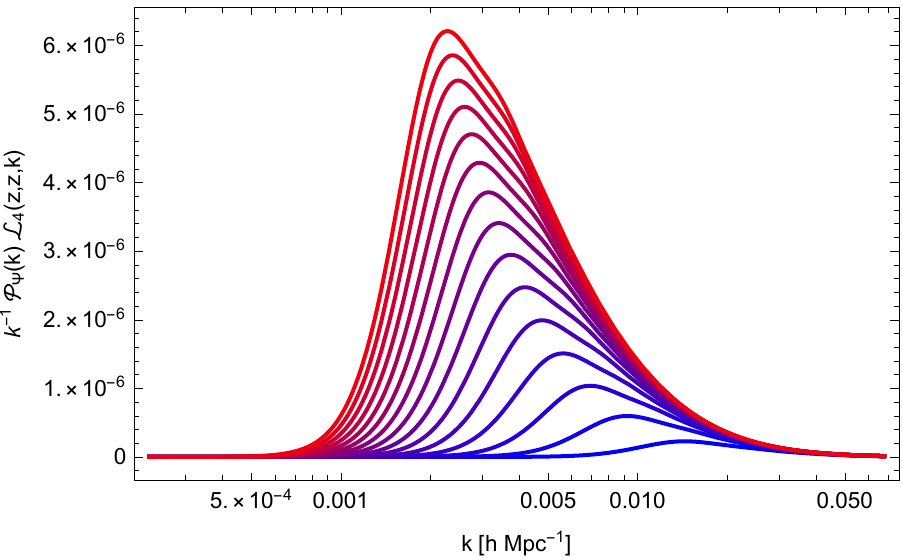}\\
\includegraphics[scale=0.9]{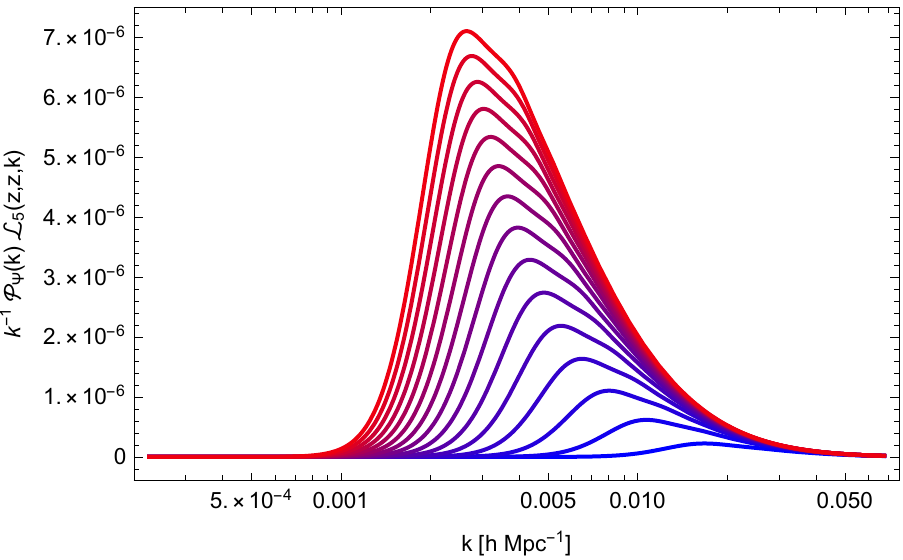}
\includegraphics[scale=0.9]{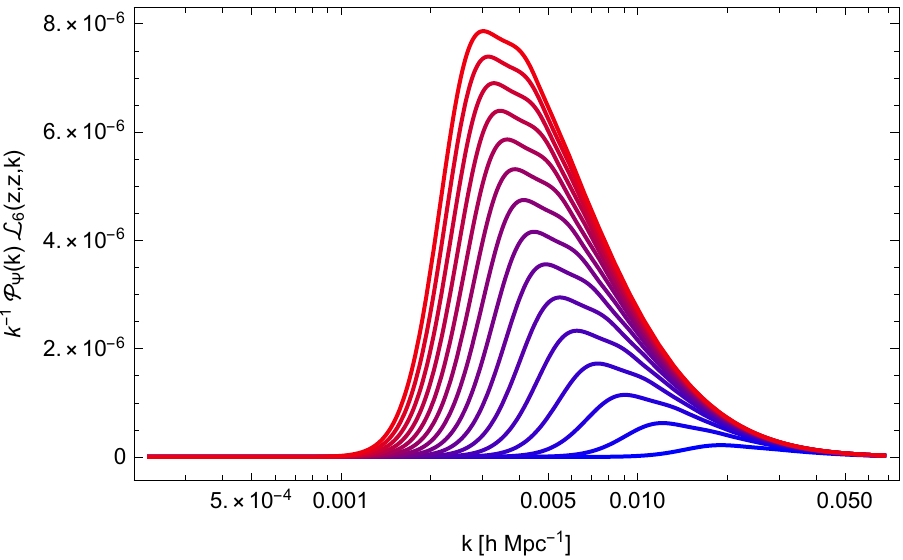}
\caption{Power spectrum for the different multipoles of the 2-point correlation function of the lensing term. The lowest multipoles are considered: $\ell=1$ (top-left), $\ell=2$ (top-right), $\ell=3$ (center-left), $\ell=4$ (center-right), $\ell=5$ (bottom-left) and $\ell=6$ (bottom-right). The monopole is not shown as it is null for the correlation function of the lensing correction.  All these lines refers to the same redshift in the correlation function, ranging from $z=0.15$ (bottom curves) to $z=1.55$ (top curves), with an interval step of 0.1. As we can notice, higher multipoles exhibit the presence of a second peak which becomes more important at higher redshifts. These plots show scales ranging from $k_{IR}=H_0=3\times 10^{-4}h\,\text{Mpc}^{-1}$ to $k_{UV}=0.1 h\,\text{Mpc}^{-1}$.}
\label{fig:PowerSpectra}
\end{figure}
\endgroup
\end{widetext}

\begin{figure}[ht!]
\centering
\includegraphics[scale=0.9]{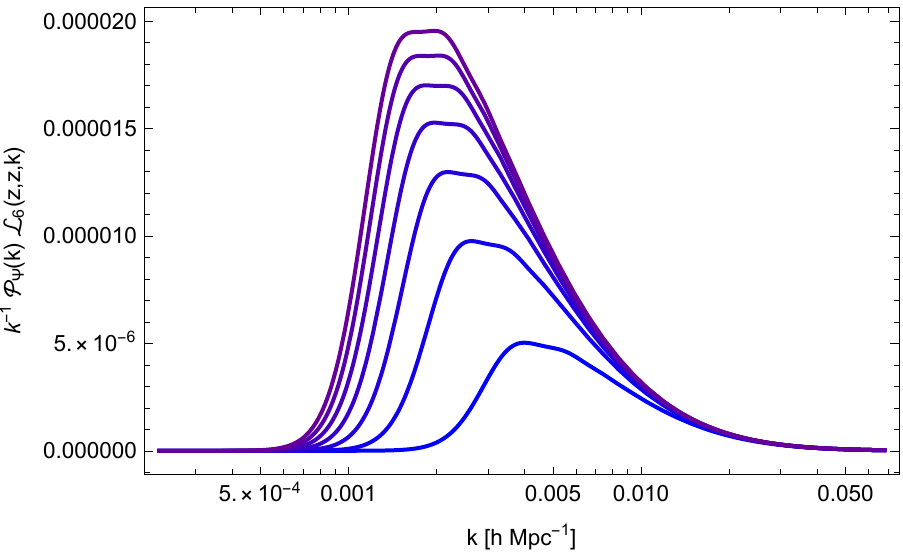}
\caption{The same as in Fig. \ref{fig:PowerSpectra} for $\ell=6$. Now the redshift ranges from $z=1$ (bottom curve) to $z=7$ (top curve) with a redshift gap of 1. As we can notice, at higher redshift the appearance of a second peak eventually flattens the curves at their maximum.}
\label{fig:PowerSpectraHigherZ}
\end{figure}

\begin{figure}[ht!]
\centering
\includegraphics[scale=0.9]{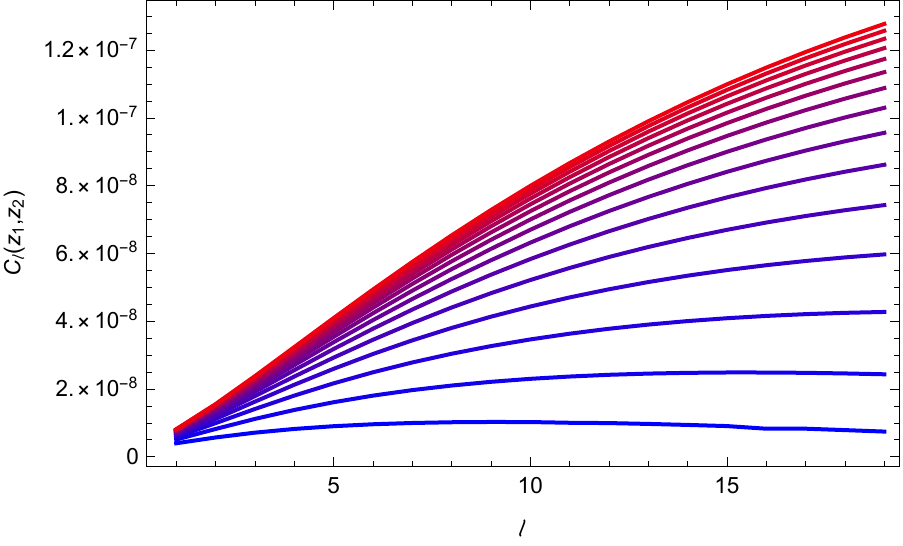}
\includegraphics[scale=0.9]{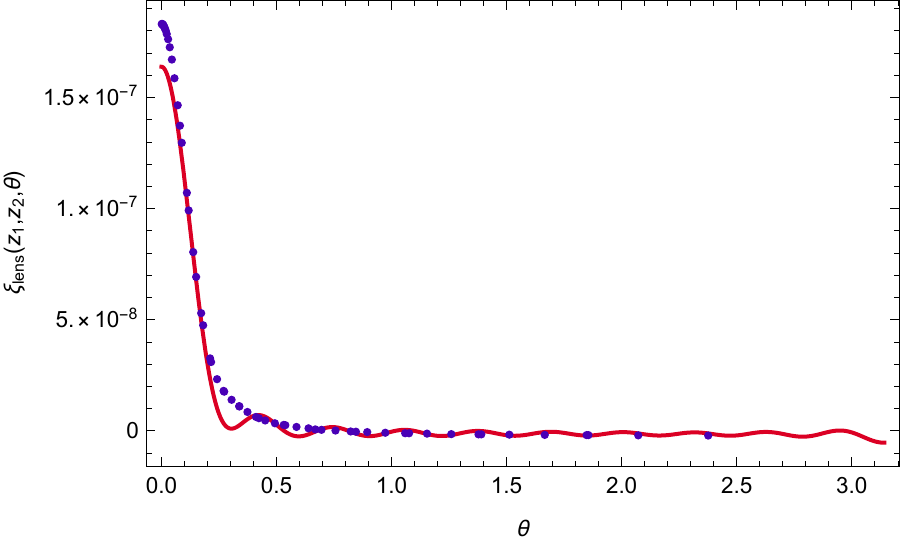}
\caption{Top-panel: 19 lowest multipoles of the 2-point angular correlation function of lensing. Here the first redshift is fixed at $z_1=0.75$ whereas the second one $z_2$ varies from $0.15$ (bottom curve) to $1.55$ (top curve), with a redshift step of $0.1$. Bottom-panel: comparison between the numerical evaluation of the 2-point angular correlation function for the lensing (blue points) with the semi-analytical estimation given by the sum of the multipoles truncated at $\ell=19$ (red curve). Here $z_1=0.75$ and $z_2=0.15$. The bottom panel corresponds to the re-summation of multipoles of the bottom curve in the top panel.
The comparison between the two figures makes clear that the multipoles for the considered redshift slowly decreases and the cutoff to be chosen in order to truncate the multipole expansion is not yet reached at $\ell=19$. This behavior is due to the steepness of the peak at $\nu=1$. The convergence becomes even slower at higher redshift, as shown in the other curves in the top panel.}
\label{fig:Multipoles}
\end{figure}

Along these plots, we notice several interesting features:
\begin{itemize}
\item{First of all, higher redshifts exhibit higher peaks than closer sources. This is somehow expected, since this behavior just reflects the fact that lensing is an integrated effect, such that the farer the source, the bigger the effect.}
\item{Secondly, for higher multipoles the peak is drifted towards the smaller scales. This is also expected since higher multipoles investigates the correlation on smaller angular scales. To be mentioned is the behavior for the dipole. Here at larger redshift we notice that a considerable amount of power lies on super-Hubble scales. This leads to a degeneracy between super-Hubble fluctuations and the choice of the background value. This is due to the fact that we have not taken into account corrections due to the observer's peculiar motion. Indeed, the latter contributes to the dipole and this renders the total 2-point correlation function independent of the physics beyond super-Hubbles scales, restoring causality as shown in \cite{Scaccabarozzi:2018vux}. However, as it emerges from Figs. \ref{fig:PowerSpectra}, the lensing dipole is comparable to other multipoles only for very low redshifts. Hence, truncation of the angular spectra at $\ell=2$ introduces a negligible error to the lensing contribution to the 2-point correlation function for high redshifts (see also Fig.  \ref{fig:Multipoles}). This method is in line with what suggested in \cite{Tansella:2018sld} and has the practical advantage of getting rid of the dependence on super-Hubble physics and lower the number of terms involved in the analysis.}
\item{Moreover, we notice the appearance of a secondary peak at higher multipoles. The relative importance of the second peak becomes higher for farer sources. This is shown in Fig. \ref{fig:PowerSpectraHigherZ}, where the same plot as in Fig. \ref{fig:PowerSpectra} is done for $\ell=6$ and where the redshift ranges from $z=1$ (bottom) to $z=7$ (top). It is evident how at higher redshift the peaks are such competitive and close that it sounds fairer to refer to a range of scales as the dominant ones rather than just a single peak. According to our understanding, this behavior reflects the fact that on larger multipoles and at higher redshifts, at the given angular scale, there is room for "resonances" in the $k$-space on the transverse plane wrt to the line-of sight.}
\end{itemize}

Figs. \ref{fig:PowerSpectra} also indicate that the $\ell$-expansion converges quite slowly to the full angular correlation function. Indeed, until $\ell=6$ the order of magnitude of the spectra tends to increase or stays constant for the same redshift. This behavior is more evident in the top panel of Fig. \ref{fig:Multipoles}. Here we plot the first 19 multipoles for the lensing and consider the cross correlation between the sources at redshift $z_1=0.75$ and other sources with possibly different redshift, letting $z_2$ range from $0.15$ (bluer, bottom) to $1.55$ (redder, top). As we can appreciate from the figure, the order of magnitude for the multipoles at the same redshift tends to remain the same or increase. For what concerns the pair $(z_1,z_2)=(0.15,0.75)$ (bottom curve in Fig. \ref{fig:Multipoles}), the multipoles decrease in amplitude after $\ell\sim 7$. However, this decreasing is not rapid enough to allow convergence to the full result for the chosen first 19 multipoles.
To make this point clearer, in the bottom panel of Fig. \ref{fig:Multipoles} we compare the full numerical estimation of the the angular correlation function at $z_1=0.15$ and $z_2=0.75$ (blue points) with its truncated $\ell$-expansion until $\ell=19$ (with coefficient given in the bottom curve of the top panel). We realize from this plot that intermediate angular ranges are in an acceptable agreement (modulo an oscillation due to the truncation in the Legendre expansion), whereas the maximum still admits a $\sim 15 \%$ missing correction. Considering that the numerical value of the peak is $3 \times 10^{-7}$ and the order of magnitude of the $C_\ell$ for this bin pair is $10^{-8}$, we can estimate the convergence $\ell$ scale thanks to the argument which follows. The value of the angular correlation at its peak corresponds to the full resummed series of all the $C_\ell$'s for a given redshift pair. This can be understood by looking at the Eq. \eqref{eq:LegendreDecomposition} since Legendre's polynomials all equal 1 when evaluated at $\nu=1$. Thanks to a simple analytical estimation, we can then infer that a $1\%$ error on the peak's resolution for the chosen case will be reached around at $\ell\sim 50$. The situation is even worse for the other redshift pairs.

Finally, the high value of these convergent angular scales means that the angular correlation function for the lensing is extremely peaked around its maximum and then tends to correlate between light-signals emitted from sources separated by a very narrow angular size.

\subsection{Doppler}
The same analysis can be done for the Doppler angular correlation function. The analogous of $\mathcal{L}^L_\ell$ for the Doppler effect can be written exactly for each multipoles without requiring the polynomial interpolation for $g$. This is due to the fact that Doppler effect is not integrated along the line-of-sight and then the integrations of the growth functions factorizes in the ultimate expression in the functions $G_i$ in $\mathcal{W}_{PV}$ in Eq. \eqref{eq:pure_kernels}. This leads to an important difference when compared with lensing. Indeed, for the case of Doppler, $\mathcal{L}^{PV}_\ell$ can be written as a product of combination of spherical Bessel functions for each $\ell$, and this product is symmetric under the exchange of $z_i\leftrightarrow z_j$. Moreover, the dependences on the two redshifts are decoupled. For a matter of clarity, we report the simplest cases of the monopole ($\ell=0$) and dipole ($\ell=1$) for the Doppler
\begin{align}
\mathcal{L}^{PV}_0(z_i,z_j)=&\,j_1\left(k \Delta\eta_i \right)j_1\left(k \Delta\eta_j \right)\nonumber\\
\mathcal{L}^{PV}_1(z_i,z_j)=&\,3\left[ \frac{j_1\left( k\Delta\eta_i \right)}{k\Delta\eta_i}-j_2\left( k\Delta\eta_i \right) \right]\nonumber\\
&\times\left[ \frac{j_1\left( k\Delta\eta_j \right)}{k\Delta \eta_j}-j_2\left( k\Delta\eta_j \right) \right]\,.
\label{eq:dopp_mon_kern}
\end{align}
It is worth to remark that Eq. \eqref{eq:dopp_mon_kern} can be written as $\mathcal{L}^{PV}_\ell=\left( 2\ell+1 \right)j'_\ell(k\Delta\eta_i)j'_\ell(k\Delta\eta_j)$. This is in agreement with results of \cite{Bonvin:2005ps}.
The same structure occurs for higher multipoles, but with a more involved combination of higher order $j_n$. Since the product of $j_n$'s shows a constructive interference only when $z_1\approx z_2$, $\mathcal{L}^{PV}_\ell$ is mostly positive only when the two redshift are almost the same. For larger separation, $\mathcal{L}^{PV}_\ell$ exhibits an oscillating behavior which suppresses the integration in $k$-space and reduces the amplitude of multipoles. This explain why Doppler angular correlation is relevant only for sources placed within the same redshift bin. Finally, we comment on the fact that this is not the case for lensing, when the $\mathcal{L}^L_\ell$'s are positive defined also for large angular separation. This is due to the fact that lensing is an integrated effect along the line-of-sight and then also two sources with a separation in redshift space contribute to the amplitude in a non-negligible way.


\subsection{Monopoles for the general 2-point correlation function}
The above-mentioned features for the spectral decomposition of the 2-point correlation functions can be generalized by evaluating the angular integral over $\Omega_k$ in Eq. \eqref{eq:general_kernels}. In fact, since the operators $\hat{O}_E$ do not act on the angular dependences in $d\Omega_k$ in the kernels $\mathcal{W}_{Ei,E'j}$, we can perform this integration independently by aligning the $z$ axis of ${\bf k}$ with the vector $r_{Ei}{\bf n}_i-r_{E'j}{\bf n}_j$. We then get 
\begin{align}
\int{d\Omega_k} e^{i {\bf k} \cdot  (r_{Ei}{\bf n}_i-r_{E'j}{\bf n}_j )}=& 2 \pi \int_{-1}^{1}d\left(\cos \theta_k\right) e^{i k R \cos \theta_k}\nonumber\\
=& 4 \pi j_0(kR)\,,
\end{align}
where $R=R(\eta_E,\eta_{E'},\nu)$. In this way, Eq. \eqref{eq:general_kernels} gets the simple expression
\begin{equation}
\mathcal{W}_{Ei, E'j} =\hat{O}_{Ei} \hat{O}_{E'j} \left[\frac{g(\eta_{Ei})}{g(\eta_o)} \frac{g(\eta_{E'j})}{g(\eta_o)}j_0(kR) \right]\,.
\label{eq:cornerstone}
\end{equation}
This exact expression is as simple as powerful. Indeed, thanks to Eq. \eqref{eq:cornerstone}, we can generate the kernel for any relativistic effect $E$ simply as the action of the operator $\hat{O}_E$ on the function $g(\eta_E)\,j_0(kR)$. Eq. \eqref{eq:cornerstone} is also useful since it provides a generating function for the monopole of the $\xi_{EE'}$. Indeed, the kernel for the monopole is given by the action of the angular average on the Eq. \eqref{eq:cornerstone}
\begin{align}
\mathcal{W}_{Ei,E'j}^0=&\frac{1}{2}\int_{-1}^{1}d\nu\,\mathcal{W}_{Ei,E'j}\nonumber\\
=&\frac{1}{2}\int_{-1}^{1}d\nu\,\hat{O}_{Ei} \hat{O}_{E'j} \left[\frac{g(\eta_{Ei})}{g(\eta_o)} \frac{g(\eta_{E'j})}{g(\eta_o)}j_0(kR) \right]\nonumber\\
=&\frac{1}{2}\hat{O}_{Ei} \hat{O}_{E'j} \left[\frac{g(\eta_{Ei})}{g(\eta_o)} \frac{g(\eta_{E'j})}{g(\eta_o)}\int_{-1}^{1}d\nu\,j_0(kR) \right]\nonumber\\
=&\hat{O}_{Ei}\left[\frac{g(\eta_{Ei})}{g(\eta_o)} j_0\left(k\Delta\eta_{Ei}\right)\right]\nonumber\\
&\times \hat{O}_{E'j} \left[\frac{g(\eta_{E'j})}{g(\eta_o)}j_0\left(k\Delta\eta_{E'j}\right)\right] \, ,
\label{eq:commutation}
\end{align}
where we used $\int_{-1}^1d\nu j_0(k R(x,y,\nu))=2 j_0(kx)j_0(ky)$. The second-last equality in Eq. \eqref{eq:commutation} is due to the fact that all the $\hat{O}_E$'s commute with the angular average. Indeed, for what concerns the lensing $\hat{O}_L=\Delta_2$, on one hand we recall that the angular average of the angular Laplacian is null. On the other hand, the angular average of a given function does not depend on the angle, then its Laplacian is null. Provided that, we can commute $\hat{O}_L$ with the angular average. In regard of all other operators, they do not depend on the angular coordinates, so they can be exchanged as well with the angular average.

The crucial aspect of Eq. \eqref{eq:commutation} is the complete factorization of the effects. This also allows to analytically evaluate the kernels for the monopole contribution due to all the 15 auto and cross 2-point correlation functions thanks to the action of only 5 different operators. In particular, we get
\begin{widetext}
\begin{align}
\hat{O}_{PV} \left[\frac{g(\eta_s)}{g(\eta_o)} \, j_0(k \Delta\eta_s)\right] =& k\,\Xi_s G_s \, j_1(k \Delta\eta_s)
\nonumber\\
\hat{O}_{SW} \left[\frac{g(\eta_s)}{g(\eta_o)} \, j_0(k \Delta\eta_s)\right] =& -(1+\Xi_s) \frac{g(\eta_s)}{g(\eta_o)} \, j_0(k \Delta\eta_s)
\nonumber\\
\hat{O}_{ISW} \left[\frac{g(\eta)}{g(\eta_o)} \, j_0(k \Delta\eta)\right] =& - 2\,\Xi_s \int_{\eta_{s}}^{\eta_o}  d\eta  \left(\frac{g'(\eta)}{g(\eta_o)} \, j_0(k \Delta\eta) + \frac{g(\eta)}{g(\eta_o)} \, j_1(k \Delta\eta) \right)
\nonumber\\
\hat{O}_{TD} \left[\frac{g(\eta)}{g(\eta_o)} \, j_0(k \Delta\eta)\right] =& \frac{2}{\Delta \eta_s} \int_{\eta_{s}}^{\eta_o}  d\eta \frac{g(\eta)}{g(\eta_o)} \, j_0(k \Delta\eta)
\nonumber\\
\hat{O}_{L} \left[\frac{g(\eta)}{g(\eta_o)} \, j_0(k \Delta\eta) \right]=& 0 \, .
\label{monopole_terms}
\end{align}
\end{widetext}
These relations are enough to evaluate the monopole for the whole $\xi$. We remark that lensing has no monopole. Hence, from Eqs. \eqref{monopole_terms} and the factorization in Eq. \eqref{eq:commutation} we infer that also all the cross-correlations between lensing and any other relativistic effect in the luminosity distance have vanishing monopole. This has consequences for what concerns the observation of large number of sources distributed all over the sky. Indeed, in the limit of large number of sources, the spatial average tends to the angular ones and then $\sigma^2_{H_0}$ is completely given by the sum of the monopoles for each combination of $EE'$. In this limit, then, any cross-correlation with lensing cannot contribute to the total effect. This independence on lensing is an interesting point, since this is usually one of the most important effect in the analysis of Large Scale Structure surveys.

Total monopole is then expected to be dominated by the auto correlation of Doppler. This indeed is due to the fact that Eqs. \eqref{monopole_terms} exhibits a further $k$ in the first line. The latter then contributes with a $k^2$ amplitude in the integration over $k$-space which then contributes more than the other pairs.

We underline that this features of monopoles follows from  the fact that $P_0$ is constant. The same analysis can be applied to higher multipoles. However, since $P_\ell$'s in general depend on $\nu$, they do not commute with all the operators $\hat{O}_E$ (in particular when $E=L$) and this aspect must be taken into account for higher multipoles. The cosmological information within LSS contained in higher multipoles has been discussed in recent papers as \cite{DiDio:2018zmk,Beutler:2020evf}, where the dipolar structure of galaxy number counts has been investigated. We postpone the investigation of these effects multipole by multipole within forthcoming surveys to future works. We stress, however, that the analysis performed in the next sections captures the presence of all the multipoles for the 2-point correlation function.

We then conclude this section by remarking that typical surveys have no access to the full sky coverage. This practical limitation makes lensing contribution no longer vanishing and then dominating the other effects. In the following sections, we will show this point and forecast for some cases of interest for forthcoming surveys the expected values for $\sigma^2_{H_0}$.


\section{Numerical evaluations}
\label{sec:numerics}
In this section, we will provide numerical evaluation of the 2-point correlation function $\xi$. To this aim, we will focus our analysis only on lensing and doppler terms. Indeed, from Eqs. \eqref{eq:pure_kernels} we have that the kernels regarding lensing and doppler effects, namely $\mathcal{W}_L$ and $\mathcal{W}_{PV}$, have the higher number of powers in terms of $k$, respectively $k^4$ and $k^2$. Moreover, $\mathcal{W}_{PV}$ contains the pre-factors $\Xi$ and these amplify the effect on small redshift. Motivated by the same argument, in Eqs. \eqref{eq:mixed_kernels} the only term of interest for us is $\mathcal{W}_{PV\,L}$ which indeed contains the highest number of power in $k$ and is amplified by the prefactor $\Xi$. To this aim, we consider the linear dimensionless power spectrum today
\begin{equation}
\mathcal{P}_\psi=A\left( \frac{k}{k_0} \right)^{n_s-1} {9\over 25}
\left[\frac{g(\eta_o)}{g_\infty} \right]^2T^2\left(\frac{k}{13.41\,k_\text{eq}}\right)\,,
\label{eq:PS}
\end{equation}
where $T(k)$ is the so-called transfer function which takes into account the sub-horizon evolution of modes re-entering the horizon during the radiation era. We have expressed $T(k)$ in the Hu-Eisenstein parametrization \cite{Eisenstein:1997ik}, given by
\begin{align}
T(q)=&\frac{L_0(q)}{L_0(q)+q^2\,C_0(q)}\,,\nonumber\\
L_0(q)=&\log(2\,e+1.8\,q)\,,\nonumber\\
C_0(q)=&14.2+\frac{731}{1+62.5\,q}\,.
\label{eq:HE}
\end{align}
We have integrated over the spectral distribution of frequency modes using the following infrared (IR) and ultraviolet (UV) cutoff values:
\begin{equation}
k_{\rm IR}= 3 \times 10^{-4} \,h \,{\rm Mpc}^{-1}, ~~~~~~~~
k_{\rm UV}= 0.1 \times h \,{\rm Mpc}^{-1}.
\label{eq:cutoffs}
\end{equation}
They roughly correspond to the present horizon scale and to the limiting scale of the linear spectral regime, respectively. The numerical values of the parameters appearing in Eqs. \eqref{eq:g}, \eqref{eq:PS} and \eqref{eq:HE} have been chosen, according to recent cosmological observations \cite{Aghanim:2018eyx}, as follows
\begin{align}
A=&\,2.2\times10^{-9},\quad n_s=0.96,\quad k_0=0.05\,\text{Mpc}^{-1},\nonumber\\
 k_\text{eq}=&0.07\,h^2\,\Omega_{m0},\quad h=\,0.68,\quad \Omega_{m0}=0.315\,.
\label{eq:par}
\end{align}
With these numerical specifications, we first want to plot the values for aligned ($\nu=1$) and antipodes ($\nu=-1$) correlation regarding $\xi^L$, $\xi^{PV}$ and $\xi^{PV,L}$.
To this aim, we underline that these explicit limits $\nu=\pm 1$ show a huge analytic simplification for $\mathcal{W}_L$, $\mathcal{W}_{PV}$ and $\mathcal{W}_{PV,\,L}$ in Eqs. \eqref{eq:pure_kernels}, \eqref{eq:mixed_kernels} and \eqref{eq:Lensing_Kernel_Changed}. We find indeed that those kernels become
\begin{widetext}
\begin{align}
 \mathcal{W}^\pm_{PVij} &= \pm\,\frac{1}{3}\Xi _ { i } \Xi _ { j } G _ { i } G _ { j } k^2\left[ j_0-2j_{2}\right]\left(k(\Delta\eta_i\mp\Delta\eta_j)\right)
 \nonumber\\
\mathcal { W }^\pm _ { L ij } = & \frac { 1 } { \Delta \eta _ { i } } \frac{1}{\Delta \eta _ { j } }  \int_0^{\Delta\eta_i} dx \frac { \Delta\eta _ { i }-x} {x} \int_0^{\Delta\eta_j}d y \frac { \Delta\eta_j-y } {y} \frac{g (\eta_o-x )g (\eta_o-y)  }{g^2 (\eta_o) } \Big[  
k^2 8 \frac{x^2\,y^2}{(x\mp y)^2}  j_2 \pm 4\,k\,\frac{x\,y}{x\mp y}\,j_1 \Big]\left(k(x\mp y) \right)
\nonumber\\
{ W }^\pm _ { PVi, L j } =& \frac{\Xi_i}{\Delta \eta_j} G_i \int_0^{\Delta\eta_j}  dx \frac{ \Delta\eta_j-x }{ x }\frac{g(\eta_o-x)}{g(\eta_o)} \Bigg\{-k^3 x^2  j_3
\nonumber\\
&
+ k^2 x \left(3 \frac{x} { \Delta\eta_i \mp x } \mp 2 \right) j_2
- \frac{k\,x}{\Delta\eta_i \mp x} \left[ k^2 x (\Delta\eta_i\mp x)\mp 2 \right] j_1\Bigg\}\left( k(\Delta\eta_i\mp x) \right)\,,
\label{eq:limit_kernels}
\end{align}
\end{widetext}
where $\pm$ stands for $\nu=\pm 1$ and, in the last lines of Eqs. \eqref{eq:limit_kernels}, we have performed the change $x=\eta_o-\eta$ in the integration variable. A further simplification can be done for both $\mathcal{W}^\pm_{L}$ and $\mathcal{W}^\pm_{PV,L}$ about the growth function $g$. Just as done for the multipoles in the previous section, we can analytically perform the integrals along the line-of-sight thanks to the polynomial expansion of $g(\eta_o-x)$. This decreases the computational time, since it reduces the evaluation of the function to only 1-dimensional numerical integration in $k$-space. Results are shown in Fig. \ref{fig:extremes}.

Numerical evaluation in the range of redshift of interest for forthcoming surveys ($0.15\le z\le 3.85$) show that lensing and doppler are competitive effects for redshift smaller than 1. Moreover, cross-correlations between these two effects are always negligible in the explored ranges of $z$. According to this, one might conclude that the contributions due to doppler correction are important. However, doppler terms are counterbalanced by the change in sign which occurs around $z=1.6$ and this leads to a suppression of the total contribution of doppler in deeper surveys. This cancellation can be analytically understood by looking at the explicit expressions involving PV in Eqs. \eqref{eq:pure_kernels} and \eqref{eq:mixed_kernels}. Indeed, whenever doppler appears in the 2-point correlation function, it shows a factor $\Xi=1-\frac{1}{\mathcal{H}\Delta\eta}$ in the expression of $\xi^{EE'}$. This coefficient is null precisely when $\Delta\eta =\mathcal{H}^{-1}$ and changes sign. According to our chosen cosmology in Eq. \eqref{eq:par}, this switch happens exactly at $z=1.6$ as it is shown in Fig. \ref{fig:extremes}.
\begin{widetext}

\begingroup
\allowdisplaybreaks
\begin{figure}[ht!]
\centering
\includegraphics[scale=0.7]{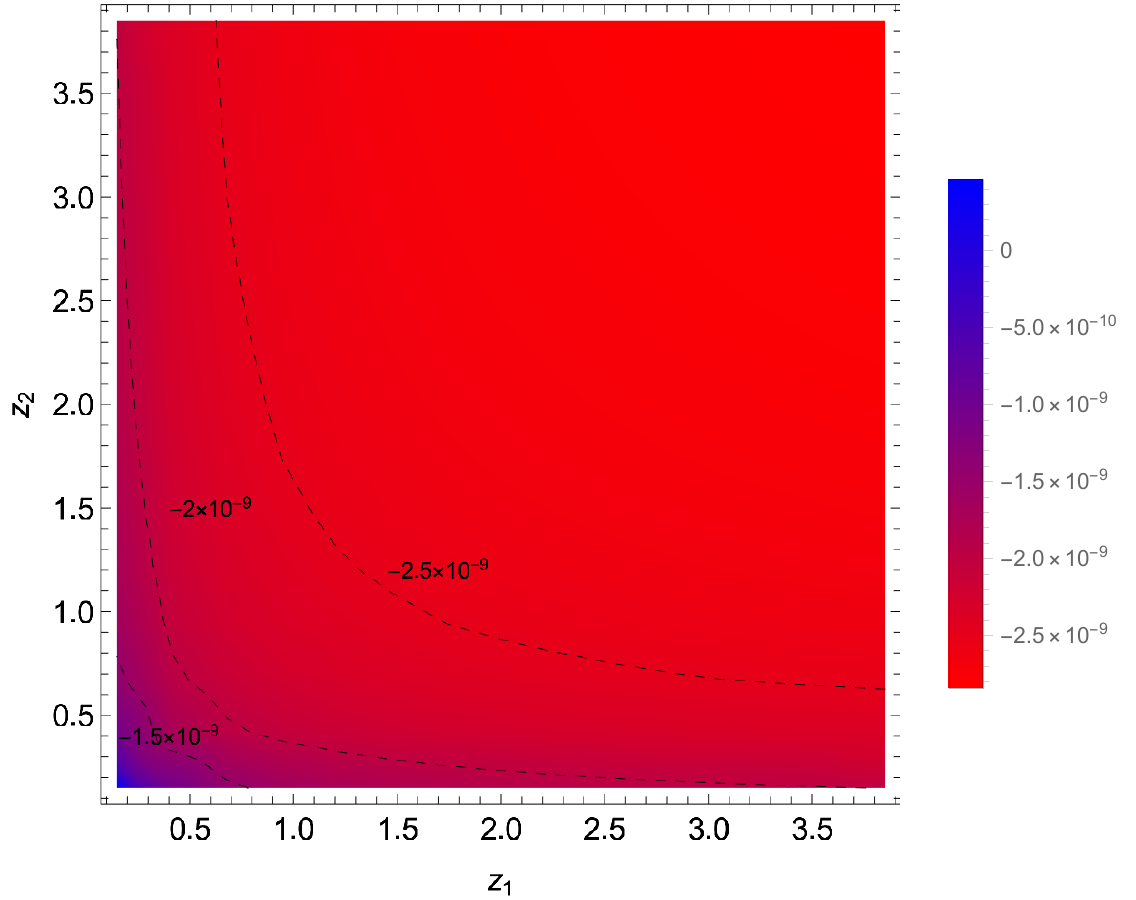}
\includegraphics[scale=0.7]{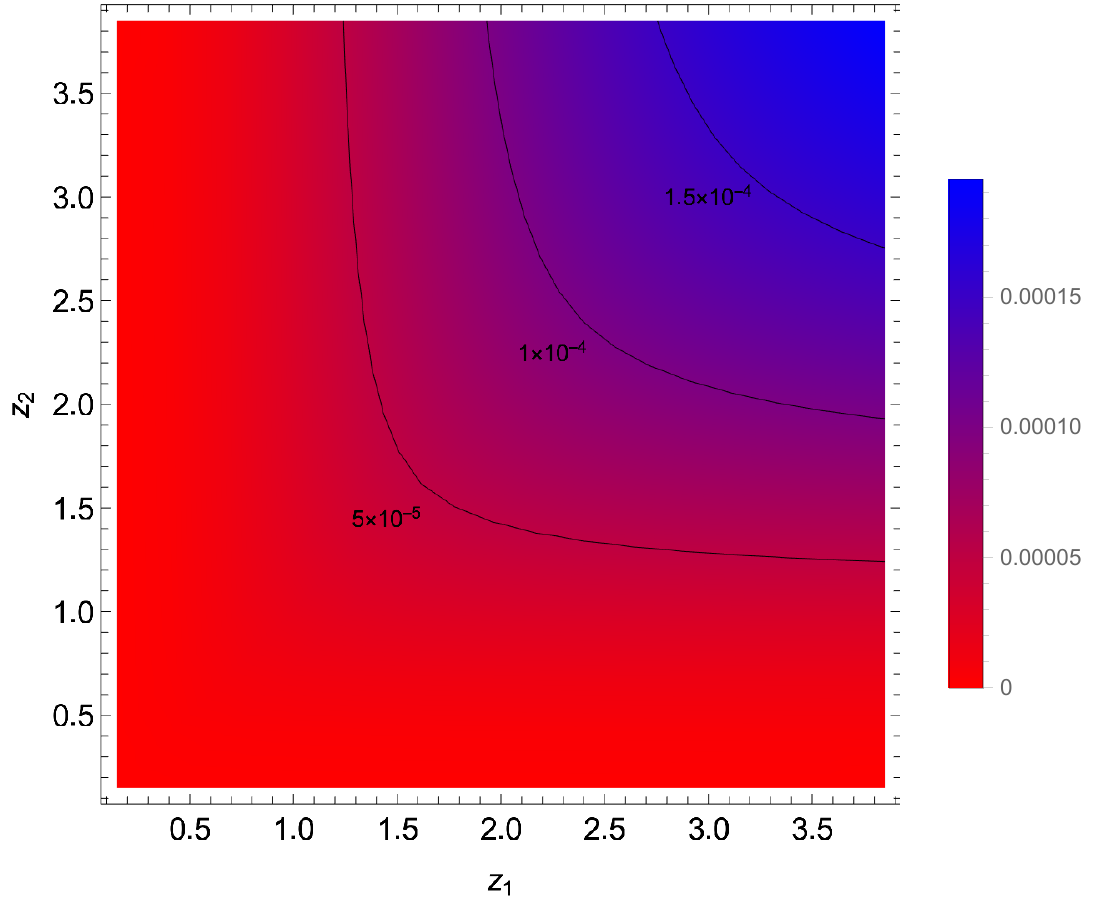}\\
\includegraphics[scale=0.7]{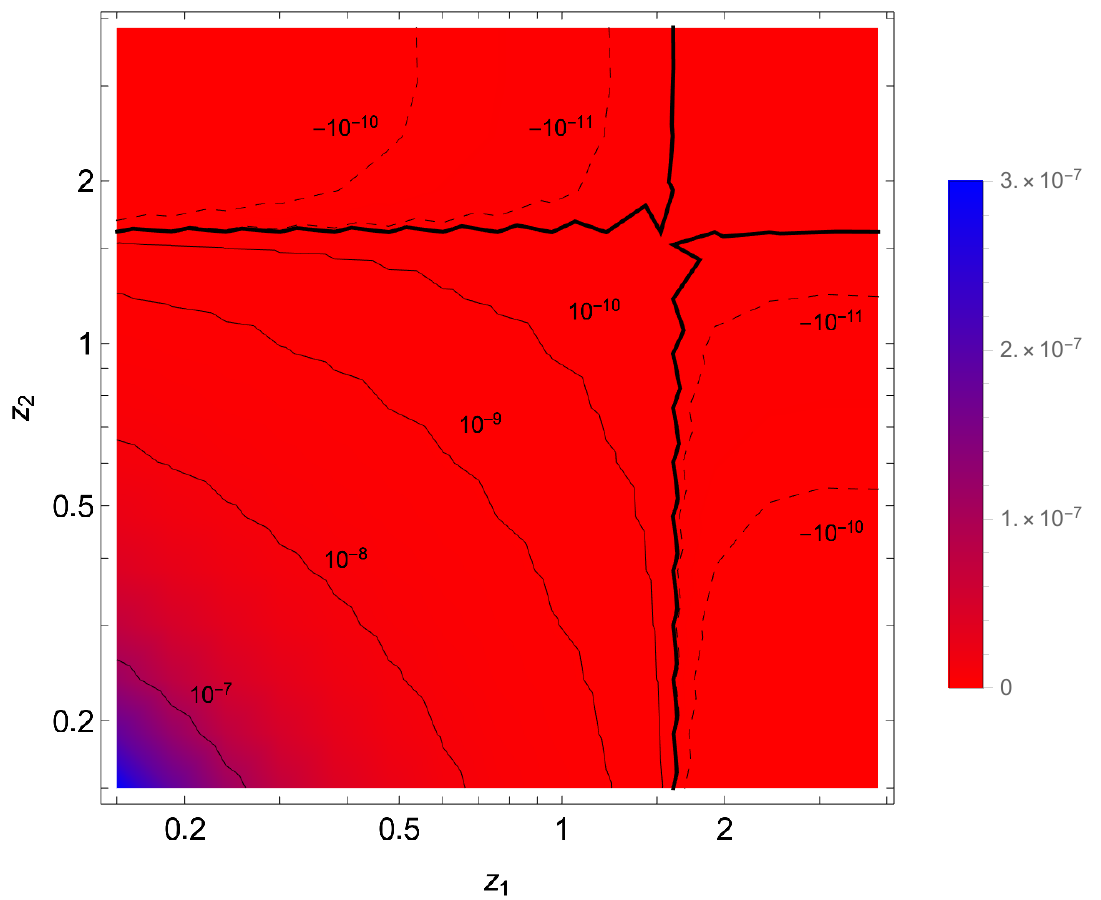}
\includegraphics[scale=0.7]{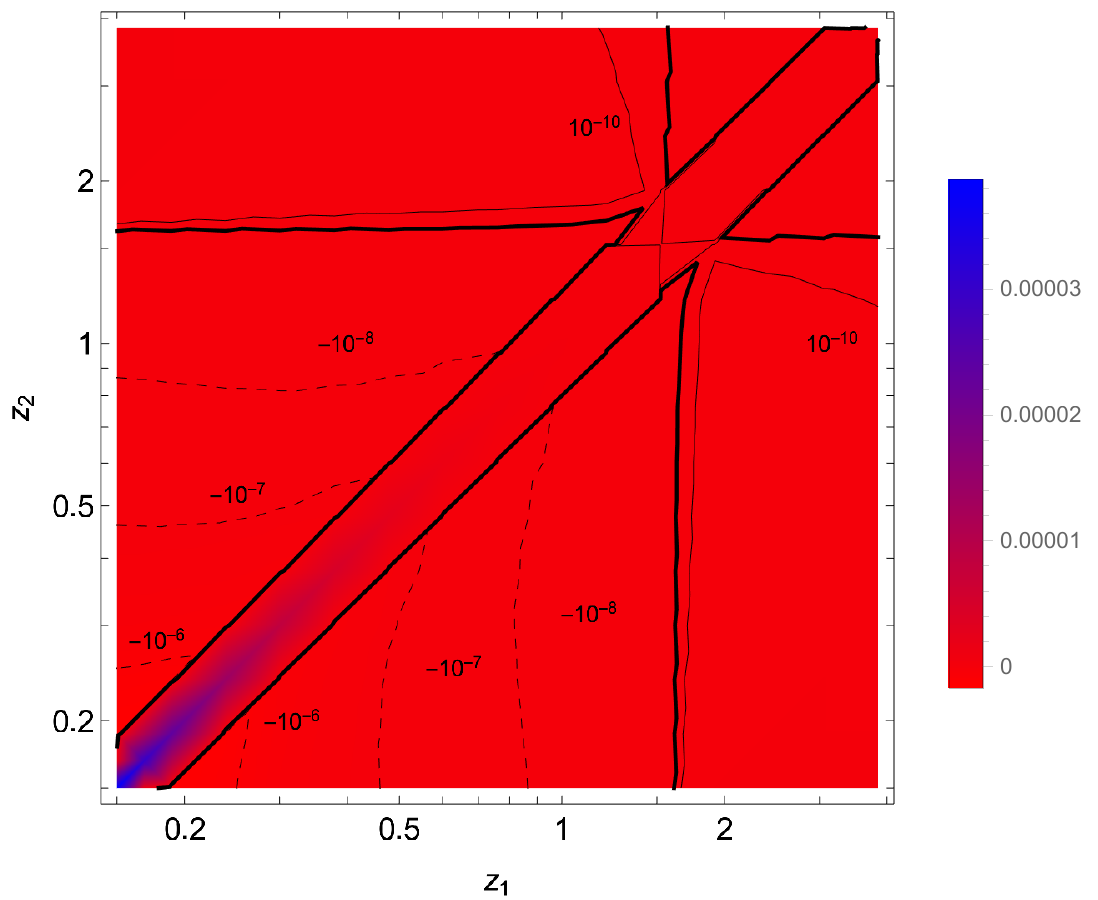}\\
\includegraphics[scale=0.7]{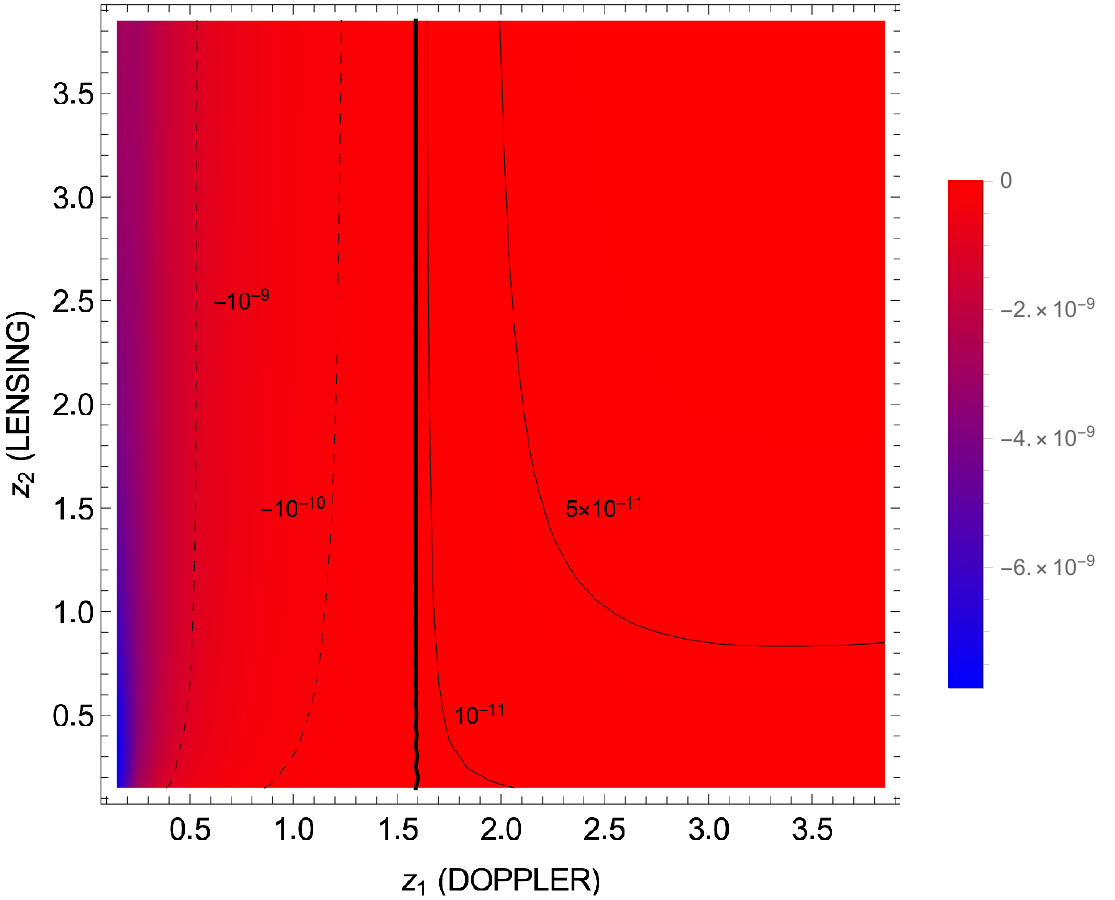}
\includegraphics[scale=0.7]{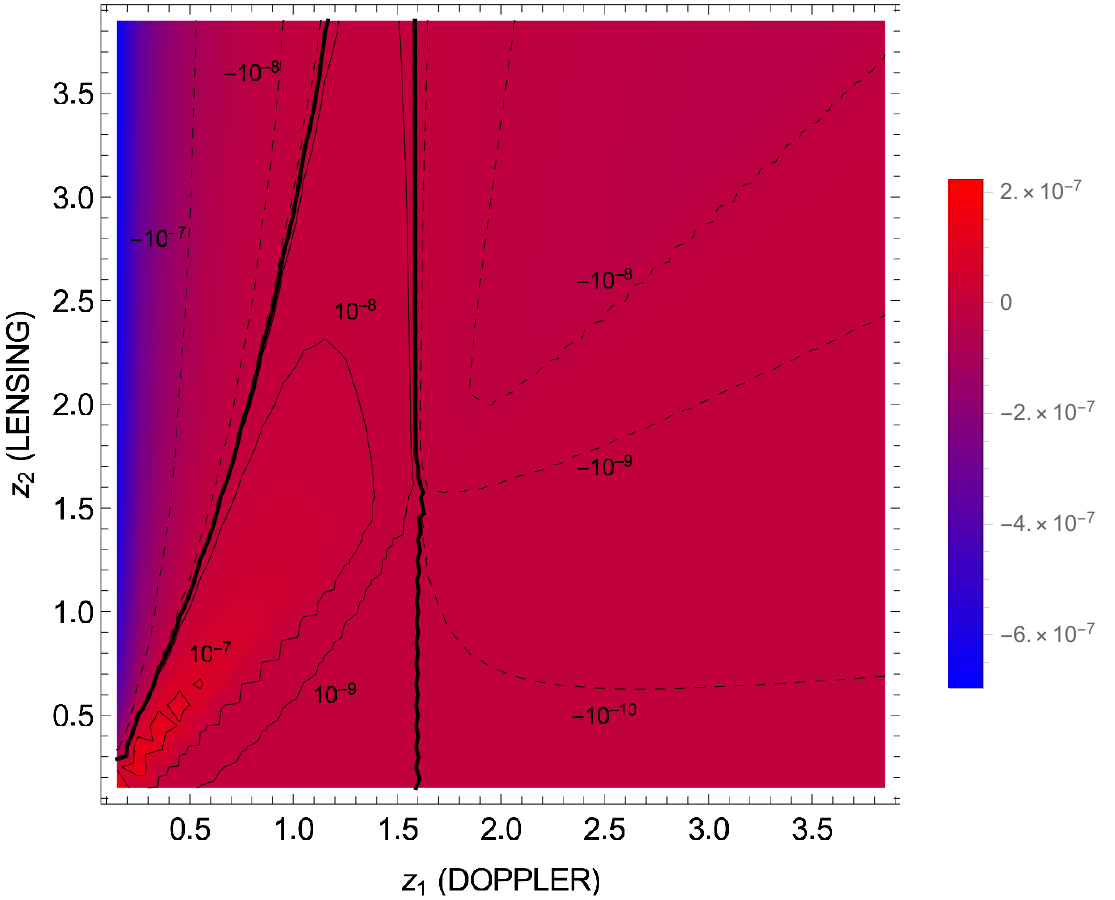}
\caption{Aligned-correlations ($\nu=1$, right panels) and antipodes-correlations ($\nu=-1$, left panels) as function of redshift. Thin solid lines refers to positive values as specified by the relative label. In the same way, dashed lines refers to negative values. Thick solid lines refers to the redshifts where correlations are null. We notice that the effect involving doppler terms (middle and bottom ones) always exhibit a null value when the redshift for the doppler effect is $z\approx 1.6$. This value corresponds to the scale where $\Xi_s=0$, namely $\Delta\eta_s = \mathcal{H}^{-1}_s$. Figures refer to lensing (top panels), doppler (middle panels) and cross-correlation between lensing and doppler (bottom panels). In the regime of interest for us, lensing is always leading with respect to the other terms for the aligned-correlations, whereas doppler turns out to be the leading effect for the antipodes-correlation at redshift smaller than 1. Plots for doppler effect terms are shown in logarithmic scales in order to show better the scaling of their values.}
\label{fig:extremes}
\end{figure}
\endgroup
\end{widetext}

\begin{figure}[ht!]
\centering
\includegraphics[scale=0.8]{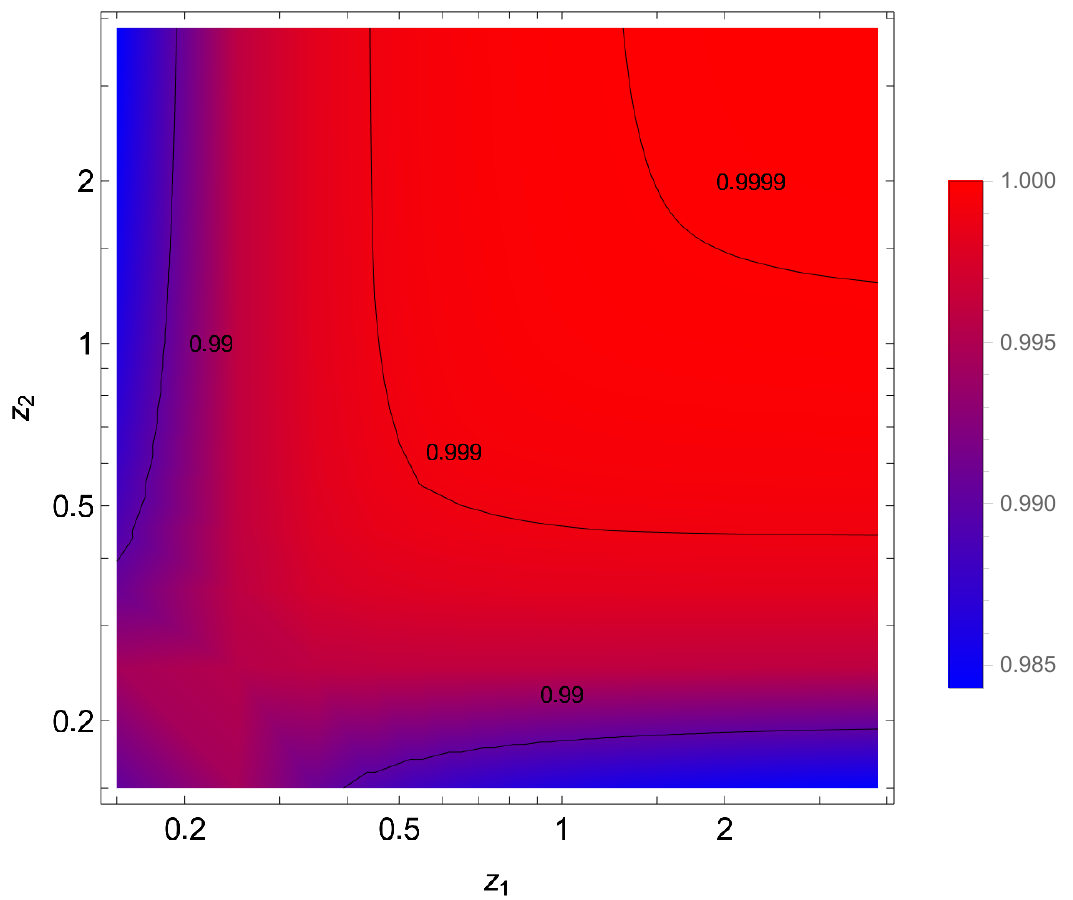}
\caption{Plot of $\nu^{ij}_{th}$ for the lensing 2-point correlation function. As we can see, $\xi^{Lij}$ rapidly peaks at values close to $\nu=1$.}
\label{fig:peak}
\end{figure}

Another important feature that comes from the numerical estimations is that close line-of-sights are more correlated than distant ones for what concerns lensing effect. This behavior is more evident for higher redshift. This then means that $\xi^{Lij}$ is highly peaked when $\nu\approx 1$ and rapidly decays to the constant value taken at $\nu=-1$. The angular scales $\nu^{ij}_{th}$ at which this decay occurs can be estimated by the following approximation method. We build an approximated $\xi^{Lij}_{app}$ as
\begin{equation}
\xi^{Lij}_{app}(\nu)=\Theta(\nu^{ij}_{th} - \nu) \left|\xi^{Lij}_+\right| - \Theta(\nu - \nu^{ij}_{th}) \left|\xi^{Lij}_-\right|\,,
\label{eq:xi_app}
\end{equation}
where $\Theta(x)$ is the Heaviside step function and subscripts in $\xi_+$ and $\xi_-$ respectively indicate $\nu=+1$ and $\nu=-1$. Eq. \eqref{eq:xi_app} shares the same extremes of $\xi^{Lij}$ shown in Fig. \ref{fig:extremes}. Moreover, $\nu^{ij}_{th}$ is determined by the further requirement that
the angular average over the full sky of $\xi^{Lij}_{app}$ is null, just as what is analytically shown for $\xi^{Lij}$, namely
\begin{equation}
\int_{-1}^1 d\nu\, \xi^{Lij}_{app}(\nu)=0\,.
\label{eq:null_average}
\end{equation}
Indeed, by combining Eqs. \eqref{eq:xi_app} and \eqref{eq:null_average}, we have that
\begin{equation}
\left(1-\nu^{ij}_{th}\right)\left|\xi^{Lij}_+\right|=\left( \nu^{ij}_{th}+1 \right)\left|\xi^{Lij}_-\right|\,
\end{equation}
which then returns
\begin{equation}
\nu^{ij}_{th}=\frac{\left|\xi^{Lij}_+\right|-\left|\xi^{Lij}_-\right|}{\left|\xi^{Lij}_+\right|+\left|\xi^{Lij}_-\right|}\,.
\label{eq:nu_th}
\end{equation}
From top panels of Fig. \ref{fig:extremes}, we notice that $\left|\xi^{Lij}_+\right|\gg\left|\xi^{Lij}_-\right|$ so $\nu^{ij}_{th}$ is expected to be very close but lower than 1. In Fig. \ref{fig:peak} we show the numerical values for the range of redshift of our interest.
This gives us a threshold angular scale.
Redshift by redshift, beyond this angular scale the lensing 2-point correlation function for a given pair of sources is negligible.

The previous estimation is very powerful, since $\nu^{ij}_{th}$ can be directly compared with the angular sky coverage of a given survey and tell whether the lensing effect is important or not for the chosen catalog. As an instance, for the EDS configuration, the angular sky coverage is approximatively $20 \text{ deg}^2$ per observation patch \cite{Laureijs:2011gra}. This solid angle can be associated to a typical cosine as $20=\left(\frac{180}{\pi}\right)^2\int_0^{2\pi}d\phi \int_{\nu_{EDS}}^1d\nu=2\times180^2 (1-\nu_{EDS})/\pi$, which returns $\nu_{EDS}=0.999$. From Figs. \ref{fig:peak}, we notice that $\nu_{EDS}$ is lower than any angular scale at $z>0.5$. Hence, in this range the narrow sky coverage of EDS is larger than the angular scales where $\xi^{Lij}$ is at its maximum. On the other hand, when one of the two sources stands at redshift $z<0.5$, $\nu_{EDS}$ is always greater than $\nu^{ij}_{th}$. In this regard, the approximation of $\xi^{Lij}$ with $\xi^{Lij}_{app}$ is expected to work quite well when one of the source is at redshift closer than 0.5. For fainter sources, the fact that we deal with very narrow line-of-sight might show the limit of our approximation. We remark that for larger sky-coverage, just like LSST, our approximation is expected to work well at any redshift. We argue this since larger angular openings tend to full-sky coverages and in this limit case the lensing 2-point correlation function is expected to be null and the peculiar sources located at the transition scales become statistically less significant. We will quantify all these aspects in the next section.

Finally, we conclude this section by underlining the importance of the numerical approximation presented in Eq. \eqref{eq:xi_app}. The analytical estimation of $\xi^{Lij}_\pm$ with the polynomial expansion of the growth function is numerically very easy to implement. Indeed, it requires only 1-D numerical integration in $k$-space and nothing else. Hence, in spite of its simplicity, the fitting function in Eq. \eqref{eq:xi_app} is powerful since its three parameters can be evaluated quickly for different cosmological parameters and then easily implemented in the Montecarlo analysis, where several runs over different cosmologies need to be accessible in short time. With the exact analytical expression of $\xi^{Lij}$, the multiple line-of-sight integrals are a huge obstacle in this regard. We finally underline that this fitting method works well only for lensing 2-point correlation function. This is due to the particular features that lensing effect is very peaked for two narrow line-of-sights and its angular average on the full sky is zero. Neither of these properties simultaneously occurs for any other effect here considered. However, this method remains powerful since lensing is anyway the leading effect among all.

In the next section, we will provide some forecast for the $\sigma^2_{H_0}$ for the case of the forthcoming surveys EDS and LSST. Finally, we will show the goodness of our fitting function in Eq. \eqref{eq:xi_app} in the estimation of $\sigma^2_{H_0}$ itself.


\section{A First Estimation for Next Generation Surveys}
\label{sec:next_gen_sur}
\begin{figure}[htbp]
\centering
\includegraphics[scale=0.75]{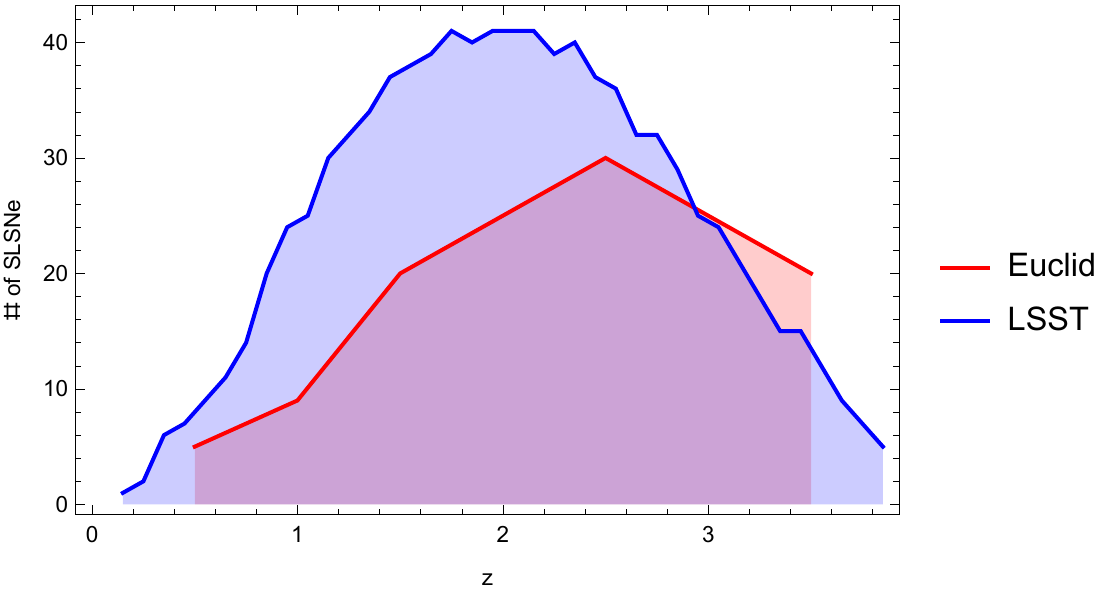}
\caption{Simulated SLSNe distributions for Euclid Deep Survey and LSST. From \cite{Inserra:2020uki}.}
\label{fig:hist}
\end{figure}

In this section, we want to apply the analytical and numerical results previously obtained to the case of Superluminous Supernovae (SLSNe). In particular, we will consider technical aspects for the forthcoming surveys of EDS \cite{Laureijs:2011gra} and LSST \cite{Abell:2009aa}. In this regard, we will follow the expected detection rate of SLSNe claimed in \cite{Inserra:2020uki} which we report in Fig. \ref{fig:hist}.
Having these histograms in mind, we generated two random surveys for the distribution of SLSNe with the following specifics
\begin{itemize}
\item 135 sources from EDS are generated in 7 redshift bins where $0.5\le z \le 3.5$ and the redshift bin width is assumed to be $\Delta z=0.5$. For this survey, the angular distribution covers two line-of-sights at North and South Poles, with angular opening of 20 deg$^2$ per line-of-sight,
\item 929 sources from LSST are generated in 38 redshift bins where $0.15\le z \le 3.85$ and the redshift bin width is assumed to be $\Delta z=0.1$. For this survey, the angular distribution spans a broad solid angle of 9000 deg$^2$.
\end{itemize}
Results are summarized in Table \ref{tab:results}.
\begin{table}[ht!]
\begin{tabular}{|c|c|c|}
\hline
$\sigma^2_{H_0}/H^2_0$&EDS&LSST\\
\hline
Lensing&$5.1\times 10^{-6}$&$7.6\times 10^{-8}$\\
Doppler&$2.1\times 10^{-9}$&$2.9\times 10^{-10}$\\
Approximated Lensing&$3\times 10^{-6}$&$7.1\times 10^{-8}$\\
\hline
\end{tabular}
\caption{Forecasts for the variance of $H_0$ in EDS and LSST. In the first line, exact 2-point correlation function for lensing is considered. In the second line, there are the contributions from 2-point correlation function of peculiar velocities. In the last line, the expected error from our approximated 2-point correlation function of lensing in Eq. \eqref{eq:xi_app} are shown. These values translate in the following values for the dispersion: for EDS, $\sigma_{H_0}/H_0=0.002$ for the exact estimation and $\sigma_{H_0}/H_0=0.002$ for the approximated estimation. For LSST,  $\sigma_{H_0}/H_0=0.0003$ for the exact estimation and $\sigma_{H_0}/H_0=0.0003$ for the approximated estimation.}
\label{tab:results}
\end{table}
Here we see that the dispersion associated to the measure of $H_0$, namely $\sigma_{H_0}\equiv \sqrt{\sigma^2_{H_0}}$ is of $\sim 0.2 \%$ for EDS but its value drops of almost 1 order of magnitude for LSST, where it contributes with a dispersion of $\sim 0.03\%$. This significant decrease can be understood by recalling that, for larger sky coverage and large number of sources, the total effect due to lensing must tend to 0. In fact, the specific of LSST are precisely along this direction. Indeed, the number of sources adopted in our forecast of LSST is almost 1 order of magnitude higher than the one of EDS. Furthermore, also the sky coverage is larger.

Another fact that we underline is that doppler effect is always subdominant. Hence we have that the total cosmic variance due to lensing and doppler is
\begin{align}
\sigma_{H_0}=&\sqrt{\sigma^2_{H_0 L}+\sigma^2_{H_0 PV}}
\nonumber\\
\approx&\sigma_{H_0 L}\left[ 1+\frac{1}{2}\frac{\sigma^2_{H_0 PV}}{\sigma^2_{H_0 L}}+\mathcal{O}\left(\left(\frac{\sigma^2_{H_0 PV}}{\sigma^2_{H_0 L}}\right)^2\right) \right]\,.
\end{align}
Again from Table \ref{tab:results}, we then get that the doppler effect corrects the total cosmic variance associated to $H_0$ by $0.2\%$ for LSST and by $0.02\%$ for EDS.

A further remark is about our approximation scheme for lensing proposed in Eq. \eqref{eq:xi_app}. From Table \ref{tab:results}, we see that the approximated method is in reasonable agreement with the exact evaluation done for LSST. The situation gets worse for EDS. We address this behavior to the specific sky coverage for the chosen surveys. Indeed, for EDS, the sky coverage is quite narrow and comparable for the redshift of our interest with the threshold scales estimated by $\nu^{ij}_{th}$. This implies that the particular details of the angular dependence in $\xi^L(\nu)$ are quite relevant. On the opposite case, LSST has a very broader angular opening. In this case, the sources are distributed with a larger angular opening and this renders the estimation of $\sigma_{H_0 L}$ less sensible to the specific values of $\nu^{ij}_{th}$.

\begin{figure}[ht!]
\includegraphics[scale=0.8]{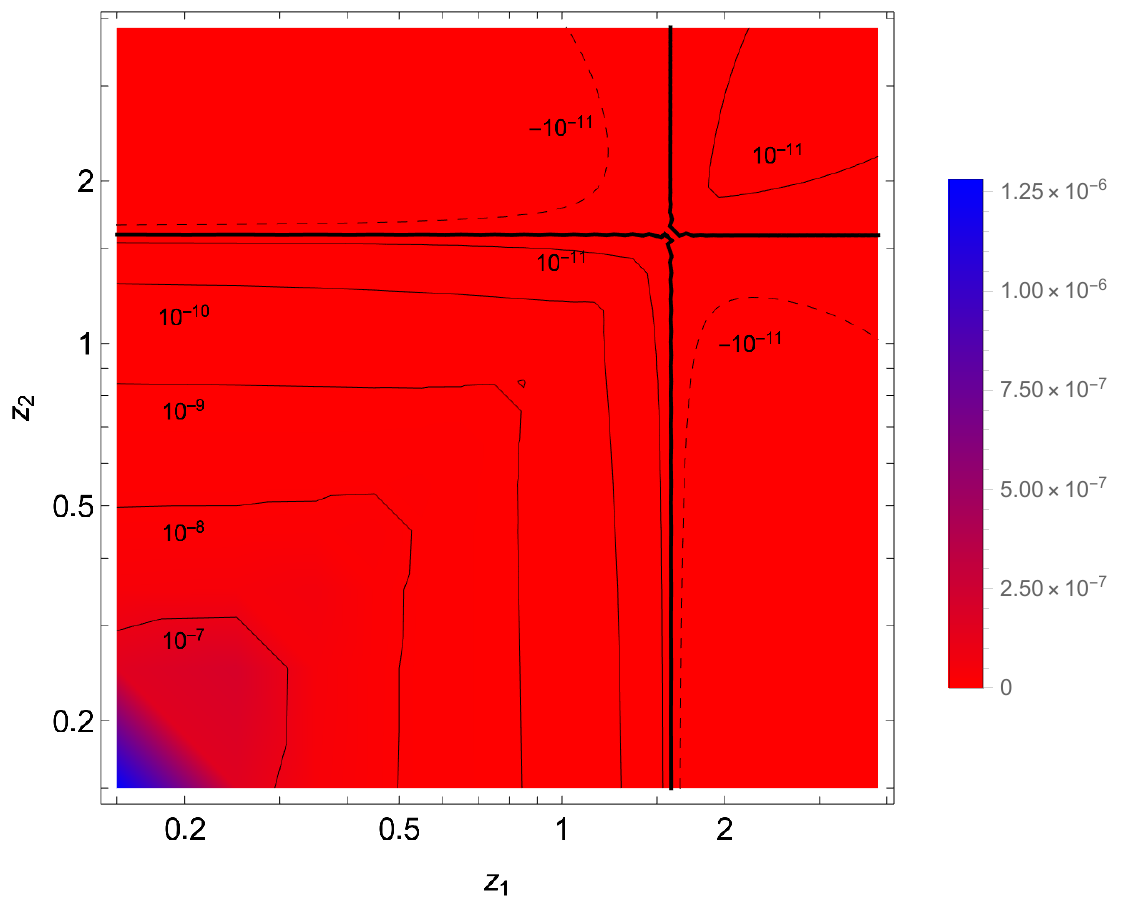}
\caption{Monopole of the doppler 2-point correlation function. Thick black lines indicate where function is 0. Dashed lines stand for negative values whereas continues lines refer to positive values.}
\label{fig:dopp_mon}
\end{figure}

Finally, we comment on the limit ideal case of full-sky coverage\footnote{See \cite{Yoo:2019skw} for a detailed discussion of this ideal case.}. For a large number of sources, lensing never contributes to the 2-point correlation function and then the leading correction is entirely addressed to the monopole of $\xi^{PV}$. This can be easily evaluated from the kernel in Eq. \eqref{eq:dopp_mon_kern} and results are shown in Fig. \ref{fig:dopp_mon}.
In this ideal case, $\sigma_{H_0}$ is entirely given by the doppler 2-point correlation function. According to the redshift distribution of LSST, we get that $\sigma_{H_0}/H_0=5.6\times 10^{-6}$.

\subsection{Non-linear scales}
So far we have taken into account only the linear power spectrum in our analysis. However, non-linear physics can affect significantly the amplitude of lensing (see, for instance \cite{BenDayan:2013gc}). Hence, it is important to understand how much our results are robust when non-linear scales in the power spectrum are taken into account. To this aim, we first adopt the approximation scheme outlined in Eq. \eqref{eq:xi_app}. In this case, since non-linear scales can enhance lensing up to 1 order of magnitude, we might naively expect that non-linear scales amplify our estimation for $\sigma_{H_0}$ by a factor $\sqrt{10}$. However, this estimation is too much conservative.

A more refined investigation about non-linear scales can be done just by looking at the kernels $\mathcal{W}^{\pm}_{Lij}$ in Eqs. \eqref{eq:limit_kernels}. Indeed, these are controlled by the functions $j_1(z)/z$ and $j_2(z)/z^2$. Both these functions have their maxima in $z=0$ and then oscillate around 0. Hence, they mostly contribute to the integrand when their argument is $\mathcal{O}(1)$. For the antipodes-correlation, from Eq. \eqref{eq:xi_app}, we get that the argument of the $j_n$'s is $k(x+y)$. Hence, the most relevant part of the integrand occurs when $k\sim (x+y)^{-1}$. It follows then that small scales contribute to $\xi^{Lij}_-$ only at very low redshifts, when $x+y$ approaches 0. On top of that, both $j_n$'s in $\mathcal{W}^-_{Lij}$ are multiplied by power of $xy$. This means that a competitive behavior occurs at low redshifts along the line-of-sight integrations and small scales tend to be suppressed in the ultimate evaluation. Thanks to this argument, we can infer that the value of $\xi^{Lij}_-$ is quite insensitive to the small scales and then $\xi^{Lij}_-$ does not change dramatically when non-linearities are taken into account.

On the contrary, the $j_n$'s in $\mathcal{W}^+_{Lij}$ depend on $k(x-y)$. This means that the integrand of $\xi^{Lij}_+$ is heavily sourced when $x\approx y$. If we call $d=x-y$, we get that the $j_n$'s gives a non-negligible contribution to the total $\xi^{Lij}_+$ whenever $k\sim d^{-1}$. This implies that, along the line-of-sight integrations, the smaller the $d$, the higher the contribution from higher $k$. Hence, $\xi^{Lij}_+$ is strongly dependent on the smaller scales. This is in line with the fact that non-linearities are expected to heavily source lensing. From our analysis, it turns out that this is true only for $\nu = 1$, namely when two sources are aligned.

The fact that $\xi^{Lij}_+$ is highly sensitive to non-linear scales whereas $\xi^{Lij}_-$ is almost independent of them has relevant implications on the angular correlation scales $\nu^{ij}_{th}$. Indeed, since the fact that lensing vanishes when integrated all around the observed sky is a pure geometrical effect, hence independent of the investigated scales, Eq. \eqref{eq:nu_th} is still a viable approximation. Then we get that the contribution from non-linear scales constrains $\nu^{ij}_{th}$ to be closer to one. In fact, when $|\xi^{Lij}_-|\ll|\xi^{Lij}_+|$, Eq. \eqref{eq:nu_th} can be expanded as
\begin{equation}
\nu^{ij}_{th}\approx 1-2\frac{|\xi^{Lij}_-|}{|\xi^{Lij}_+|}\,.
\end{equation}
In this way, if the only effect of non-linearities is to increase $|\xi^{Lij}_+|$, for instance, by a factor $C$, the angular scales conversely tends to 1 with a factor $C^{-1}$. We then have two competitive effects: on one hand, the maximum of the 2-point correlation function increases. On the other hand, the angular scales involved in this enhancement are less. Moreover, the product $(1-\nu_{th})|\xi^{Lij}_+|\approx 2 |\xi^{Lij}_-|$ is insensitive to the UV scales. As a consequence, due to non-linear effects in the power spectrum, only sources located almost along the same line-of sight show a non-negligible correction.

To test our arguments, we have then adopted the approximation method outlined in Eqs. \eqref{eq:xi_app} and assumed that $\xi^{Lij}_-$ remains the same as for the linear spectrum, whereas the net effect of non-linearities is to amplify $\xi^{Lij}_+$ by an overall factor 10. With this approach, we find that $\sigma^2_{H_0\,NL}=7.8\times 10^{-6}\,H^2_0$ for EDS and $\sigma^2_{H_0\,NL}=6.4\times 10^{-7}\,H^2_0$ for LSST. Hence, within the specific of LSST $\sigma^2_{H_0}$ is indeed enhanced by almost one order of magnitude, whereas for EDS $\sigma^2_{H_0}$ increases by almost a factor 1.5.

Despite of its rudeness, our analytical estimation agrees very well with the exact estimation where we have chosen $k_{UV}=10\, h$ Mpc$^{-1}$ and set a non-linear power spectrum with the HaloFit model \cite{Smith_2003,Takahashi_2012}. In this case we obtain $\sigma^2_{H_0\,NL}=1.1\times 10^{-5}\,H^2_0$ for EDS and $\sigma^2_{H_0\,NL}=7.8\times 10^{-7}\,H^2_0$ for LSST. Moreover, we have also applied the approximated formula in Eq. \eqref{eq:xi_app} with the non-linear values for $\xi^{Lij}_+$ and $\xi^{Lij}_-$ and we find that $\sigma^2_{H_0\,NL}=9.4\times 10^{-6}\,H^2_0$ for EDS and $\sigma^2_{H_0\,NL}=7.8\times 10^{-7}\,H^2_0$ for LSST. We notice that all our estimations agree quite well for both surveys.

Our results then show that, for EDS, even if non-linear scales are expected to enhance the lensing correction by almost one order of magnitude, the intrinsic error associated to the measurement of $H_0$ is almost insensitive to the non-linear scales, since it becomes $\sigma_{H_0\,NL}/H_0=0.003$. On the contrary, non-linear scales increase by roughly a factor 3 the dispersion of $H_0$ within the specific of LSST, raising $\sigma_{H_0}$ to the value $\sigma_{H_0\,NL}/H_0=0.0009$.

Thanks to our analysis, we can verify the claim done in \cite{Ben-Dayan:2014swa} about small redshift surveys, where it has been stated that the analysis is insensitive to smaller scales fluctuations due to the incoherence of such contributions. Our results seem to indicate that this is a reasonable expectation only for EDS. We address this feature to the fact that EDS covers smaller regions in the sky with higher angular density of sources.


\section{Summary and conclusions}
\label{sec:conclusions}
In this work, we have studied the impact of cosmological inhomogeneities on the estimation of $H_0$ from the high redshift Hubble diagram. Our analysis considers the possibility, discussed in \cite{Inserra:2020uki}, that a statistically relevant number of Superluminous Supernovae could be detected in the next years by EDS \cite{Laureijs:2011gra} and LSST \cite{Abell:2009aa}. In this regards, less conservative studies about the Hubble diagram at high redshifts ($z\le 1.5$) have been also investigated by exploiting exact inhomogeneous models in general relativity \cite{Cosmai:2013iga,Romano:2016utn,Cosmai:2018nvx,Vallejo-Pena:2019agp} or by considering strongly inhomogeneous dynamical dark energy models \cite{Cai:2021wgv}. These attempts look interesting especially in light of a recent analysis of the SNe Ia Pantheon sample \cite{Dainotti:2021pqg} suggesting that $H_0$ could be a decreasing function of redshift already at late time.

On the contrary, along our study, we have adopted a conservative approach based on linear perturbations within the Cosmic Concordance model. In this framework, first of all we have derived the 2-point correlation function of luminosity distance-redshift relation and provided explicit fully relativistic analytic expressions for its angular spectra. Our derivations agree with those already obtained in literature \cite{Bonvin:2005ps}, modulo a different classification of involved terms. It turns out in our analysis that lensing is the leading effect at the considered redshift, as one may expect.

In particular, as shown in Fig. \ref{fig:Multipoles}, we get that the angular multipoles expansion of the lensing 2-point correlation function at high redshift converges quite slowly. Moreover, lensing dipole rapidly becomes negligible with respect to the other multipoles. This allows to safely neglect its contribution to the 2-point correlation function, since it is expected to be contaminated by the observer's peculiar motion.

Furthermore, the role of higher multipoles in the lensing spectra plays a relevant role for partial sky-coverage surveys. Indeed, in the ideal case of large number of sources distributed all over the sky, lensing does not contribute to the total cosmic variance since its monopole is null. We have shown this also for cross-correlations of lensing with the other relativistic effects. In this ideal case, then, the leading correction to the estimation of $H_0$ is due to the peculiar velocities of the sources. However, realistic surveys deal with limited sky coverage and this makes lensing contribution no longer vanishing. In fact, according to the specific of EDS and LSST and to what has been claimed in \cite{Inserra:2020uki}, we forecast that the intrinsic error from cosmic variance associated to $H_0$ is of $\sim0.03\,\%$ for LSST and $0.3\,\%$ for EDS for the linear power spectrum. Non-linear scales contribute marginally to this estimation within the specific of EDS. For what regard the specific of LSST, the situation is a way worse. Indeed, in this case, we get that non-linear scales enhance our forecast by almost a factor 3. This is a direct consequence of the fact that lensing 2-point correlation function strongly depends on small scales fluctuations for the diagonal entries of the covariance matrix.

A similar analysis has been already performed in \cite{Ben-Dayan:2014swa}, where only close Supernovae (up to $z=0.1$) have been considered. Here we extend the analysis of \cite{Ben-Dayan:2014swa} since there only the peculiar motion of the sources is taken into account. In fact, this is the leading correction expected at those redshifts \cite{BenDayan:2013gc}. The interesting result is that low redshift surveys discussed in \cite{Ben-Dayan:2014swa} admits a cosmic variance for $H_0$ of $\sim 1\,\%$. Our analysis points out that surveys have an intrinsic error for $H_0$ which tends to decrease when higher redshift sources are considered.

Finally, we remark that our results are not able to alleviate the tension between local and distant measurements of the Hubble constant. However, they indicate that the analysis of fainter sources does not increase the theoretical uncertainty on $H_0$. The price to pay stands in the fact that the Hubble diagram at higher redshift is no longer model independent.


\section*{ACKNOWLEDGMENTS}
The authors are thankful to Vincenzo Cardone, Enea Di Dio, Ruth Durrer and Kazuya Koyama for useful discussions. GF acknowledges support by FCT under the program {\it ``Stimulus"} with the grant no. CEECIND/04399/2017/CP1387/CT0026. BF is supported by the PhD program of the University of Portsmouth. GM is supported in part by INFN under the program TAsP ({\it Theoretical Astroparticle Physics}).


\appendix
\begin{widetext}
\section{Kernels derivation}
\label{terms_derivation}
In this appendix, we report some technical aspects of the derivation of Eqs. \eqref{eq:pure_kernels} and \eqref{eq:mixed_kernels}. We start with the derivation of $\xi_E$ in Eqs. \eqref{eq:corr_func}. There are 5 terms and the geometrical sets for the integration are shown in Fig. \ref{geometry} respectively for local effects (left panel), such as PV and SW, and integrated terms (right panel), like TD, L and ISW.
\begin{figure}[htbp]
\begin{center}
\includegraphics[width=0.4 \textwidth]{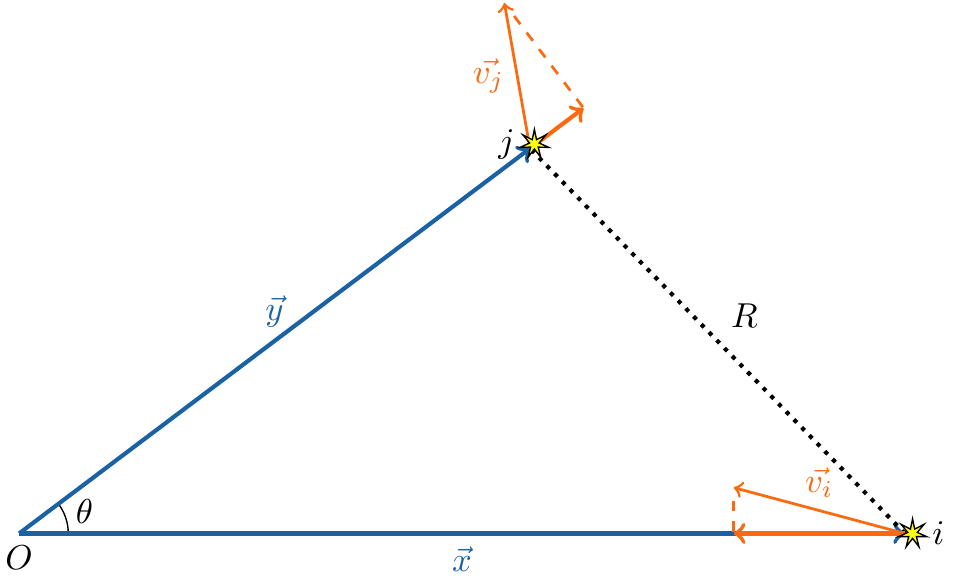}\qquad\qquad
\includegraphics[width=0.4 \textwidth]{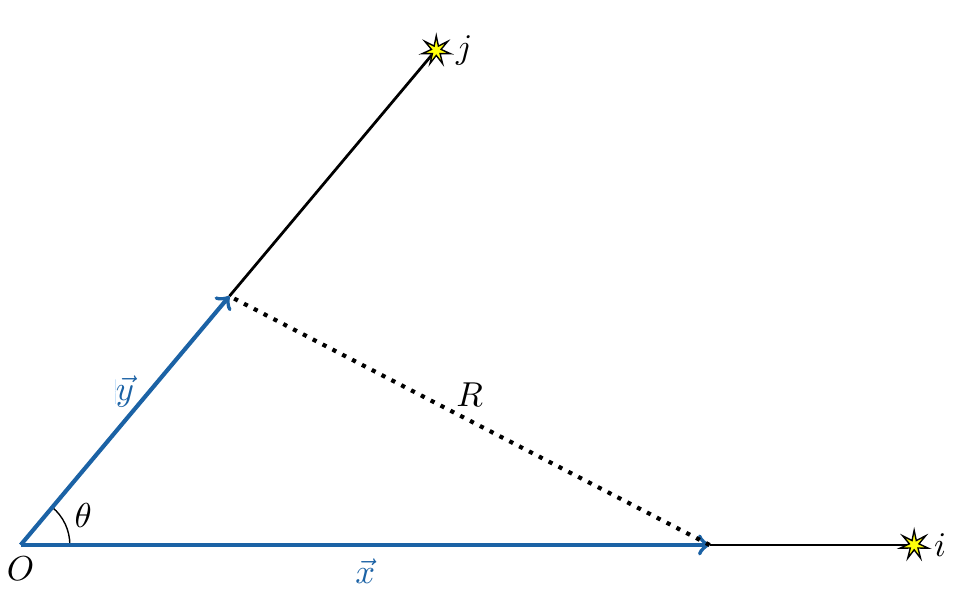}
\caption{Geometrical relations between line-of sight variables for local effects (left), i.e. PV and SW, and integrated ones (right), namely L, TD and ISW.}
\label{geometry}
\end{center}
\end{figure}

\vspace{1cm}

\paragraph{Peculiar Velocity}
As illustrated in Fig. \ref{geometry}, we define $x =\eta_o - \eta_i$ and $y= \eta_o - \eta_j$.
Combining Eqs. \eqref{eq:Operators} and \eqref{eq:general_kernels}, for peculiar velocity operator we obtain
\begin{equation}\label{W_PV}
\mathcal{W} _ { P V i j } =  \Xi_i \Xi_j    G_i G_j \partial_x  \partial_y \int \frac{d \Omega _ { k } } {4 \pi}  e ^ { i {\bf k} \cdot ( {\bf x} - {\bf y} )}\,,
\end{equation}
where we have used Eqs. \eqref{eq:important_eqs}.
Since ${\bf k}$ has rotational freedom, we align its component along the azimutal axis with ${\bf x}-{\bf y}$, so that the integral over the solid angle becomes
\begin{equation}
\int{d\Omega_k} e^{i {\bf k} \cdot ({\bf x}-{\bf y})}= 2 \pi \int_{-1}^{1}d\left(\cos \theta_k\right) e^{i k R \cos \theta_k}= 4 \pi j_0(kR)\,,
\label{int_dOk}
\end{equation}
where $j_0$ is the $0$-th order spherical Bessel function and $R=R(x,y,\nu)$ is taken as defined in Eqs. \eqref{eq:important_eqs_2}. Thus we are left with the evaluation of the derivatives of the spherical Bessel function
\begin{equation}
\partial_x\partial_y j _ { 0 } ( k R ) =\,k\,\partial_x \left[  j'_0 (kR)\partial_y R \right]
=k^2j''_0(kR)\partial_xR\,\partial_yR
+k\,j'_0(kR)\partial_x\partial_y R\,,
\end{equation}
where $j'_n(z)=\partial_z j_n(z)$, $j''_n(z)=\partial^2_z j_n(z)$ and so on. Finally, by exploiting the properties of the $j_n$ reported in Eq. \eqref{eq:jn_recursive}, we obtain
\begin{align}
\label{sig_pv2}
\mathcal { W } _ { P V i j } &= \Xi _ { i } \Xi _ { j } G _ { i } G _ { j } k^2 \left\{\frac{xy(1-\nu^2)}{R^2}j_2(kR) +    \frac{\nu}{3}\left[j_{0}(kR)-2j_{2}(kR)\right]\right\}\,,
\end{align}
in agreement with \cite{Ben-Dayan:2014swa}.

\vspace{1cm}

\paragraph{Lensing} From right panel of Fig. \ref{geometry}, we define $x=\eta_o - \eta$ and  $y =\eta_o - \eta'$.
From the definition of the lensing operator in Eqs. \eqref{eq:Operators}, we then obtain
\begin{equation}
\mathcal { W } _ { L  i j } =  \frac {1} { \Delta \eta _ { i } \Delta \eta _ { j }  } \int _ { \eta_i } ^ { \eta_o } d \eta \frac { \eta - \eta_i} { \eta_o - \eta } \int _ { \eta_ j } ^ { \eta_o } d \eta' \frac { \eta' - \eta_j } { \eta_o - \eta' }
 \frac{g (\eta )g (\eta')  }{g (\eta_o^2) }\int \frac{d \Omega _ { k } } {4 \pi}  \Delta _ { 2x }  \Delta _ { 2y } e ^ { i {\bf k} \cdot ( {\bf x} - {\bf y} ) }\,.
 \label{lens_int}
\end{equation}
We recall now that $ \Delta_{2r} = \partial _{\theta} ^2 + \cot { \theta } \, \partial _{ \theta } + 1 / \sin^2 { \theta } \, \partial_{\phi}^2$ is the angular Laplacian which can be either evaluated along the $r=x$ or $r=y$ direction. It can be written in terms of the 3-dimensional one as
\begin{equation}
\Delta_{2r}= r^2\left( \vec{\nabla}\cdot\vec{\nabla}-\nabla^2_r \right)
= r^2\left( \vec{\nabla}\cdot\vec{\nabla}-\partial_r^2 - \frac { 2 } { r } \partial_r\right)\,.
\label{D}
\end{equation}
We use Eq. \eqref{D} to replace the Laplacian in the angular coordinates and get easier evaluable derivatives. With such a substitution, the last integral in Eq. \eqref{lens_int} becomes
\begin{align}
\int \frac{d \Omega _ { k } } {4 \pi}  \Delta _ { 2 }  \Delta' _ { 2 } e ^ { i {\bf k} \cdot ( {\bf x} - {\bf y} ) }
=& \int \frac{d \Omega _ { k } } {4 \pi}   x^2\left(\vec{\nabla}_x\cdot\vec{\nabla}_x -\partial_x^2 -\frac{2}{x} \partial_x \right)  y^2\left(\vec{\nabla}_y\cdot\vec{\nabla}_y -\partial_y^2 -\frac{2}{y} \partial_y \right) e ^ { i {\bf k} \cdot ( {\bf x} - {\bf y} ) }\nonumber\\
=&\int \frac{d \Omega _ { k } } {4 \pi}   x^2\left(-k^2 -\partial_x^2 -\frac{2}{x} \partial_x \right)  y^2\left(-k^2 -\partial_y^2 -\frac{2}{y} \partial_y \right) e ^ { i {\bf k} \cdot ( {\bf x} - {\bf y} ) }\nonumber\\
=&x^2\left(-k^2 -\partial_x^2 -\frac{2}{x} \partial_x \right)  y^2\left(-k^2 -\partial_y^2 -\frac{2}{y} \partial_y \right)
j_0(kR)\,,
\label{lens_dOk}
\end{align}
where last equality has been obtained thanks to Eq. \eqref{int_dOk}.
Hence, by using the recursive relations of the $j_n$'s in Eq. \eqref{eq:jn_recursive}, after a bit of algebra, $\mathcal{W}_{Lij}$ can be rewritten as
\begin{align}
\mathcal{ W } _ { L  i j } =& \frac { 1 } { \Delta \eta _ { i } } \frac{1}{\Delta \eta _ { j } }  \int _ { \eta _i} ^ { \eta_o } d \eta \frac { \eta - \eta _i } { \eta_o - \eta } \int _ { \eta_j } ^ { \eta_o } d \eta' \frac { \eta' - \eta_j} { \eta_o - \eta' }  \frac{g (\eta )g (\eta')  }{g (\eta_o^2) } \left[  k^4 H^4 j_4(kR)
\right.\nonumber\\&\left.
- 8 k^3 H^2 L j_3(kR)
+ k^2 \left( 8 L^2 -6 H^2 \right) j_2(kR) + 4 k L j_1(kR) \right]
\end{align}
where $H$, L and $R$ are defined in Eqs. \eqref{eq:important_eqs_2}.

\vspace{1cm}

\paragraph{Sachs-Wolfe} In analogy with what has been done for the peculiar velocity, we refer to left panel of Fig. \ref{geometry}, we define $x=\eta_o-\eta_i$ and $y=\eta_o-\eta_j$ and consider the operator $\hat{\mathcal{O}}_{SW}$ in Eqs. \eqref{eq:Operators}. Since no spatial derivatives are considered, we simply get, though Eq. \eqref{int_dOk}
\begin{align}
\mathcal{W}_ { SW i  j }=& (1+ \Xi_i) (1+ \Xi_j)     \, \frac { g ( \eta _ { i } ) } { g ( \eta_o ) } \frac { g ( \eta _ { j } ) } { g ( \eta_o ) } \int \frac{d \Omega _ { k } } {4 \pi}  e ^ { i {\bf k} \cdot \left( {\bf x}- {\bf y}  \right) }
\nonumber\\
=& (1+ \Xi_i) (1+ \Xi_j) \, \frac { g ( \eta _ { i } ) } { g ( \eta_o ) } \frac { g ( \eta _ { j } ) } { g ( \eta_o ) } j_0(kR)\,,
\end{align}
where $R=R(\eta_i,\eta_j,\nu)$ is taken from Eqs. \eqref{eq:important_eqs}.

\vspace{1cm}

\paragraph{Integrated Sachs-Wolfe} In analogy with the geometrical set used for lensing (right panel in Fig. \ref{geometry}), we define
$x= \eta_o - \eta$ and $y= \chi' = \eta_o - \eta'$. Given that we simply obtain
\begin{align}
\mathcal { W } _ {ISW i j } =&  4 \,  \Xi_i \Xi_j    \int _ { \eta _i } ^ { \eta_o } d \eta \int _ { \eta _j  } ^{ \eta_o } d \eta'  \frac { \partial _{\eta}  g ( \eta  ) } { g ( \eta_o ) } \frac {  \partial _{\eta'} g ( \eta ' ) } { g ( \eta_o ) } \int \frac{d \Omega _ { k } } {4 \pi}  e ^ { i {\bf k} \cdot \left( {\bf x}- {\bf y}  \right) }
\nonumber\\
 =&  4 \Xi _ { i } \Xi _ { j } \int _ { \eta_i } ^ { \eta_o } d \eta \int _ { \eta_j} ^ { \eta_o } d \eta'  \frac { \partial _{\eta}  g ( \eta  ) } { g ( \eta_o ) } \frac {  \partial _{\eta'} g ( \eta ' ) } { g ( \eta_o ) }  j_0(kR)\,,
\end{align}
where we have used again Eq. \eqref{int_dOk}. We stress that now $R=R(\eta,\eta',\nu)$ from Eqs. \eqref{eq:important_eqs} is integrated along the two line-of-sights, differently from the case of PV and SW.

\vspace{1cm}

\paragraph{Time Delay} TD term can be evaluated in complete analogy with ISW. The only difference is that the growth functions are not derived. Following then the evaluation for ISW, we get
\begin{equation}
\mathcal{W} _ { TD i  j } = \frac{4}{\Delta \eta_i\Delta \eta_j} \int_{\eta_i}^{\eta_o}  d\eta \int_{\eta_j}^{\eta_o}  d\eta' \frac {  g ( \eta ) } { g ( \eta_o ) } \frac { g ( \eta ^ { \prime } ) } { g ( \eta_o ) } j_0(kR) \, .
\end{equation}

\vspace{1cm}

There are then 10 mixed terms that come out of the sum in Eq. \eqref{sig_teo} when $E\ne E'$.  As done for the pure terms, we will use the auxiliary variables $x,y$, whose definitions will be given for each term and differ whether the effect is local or integrated along the line-of-sight. To our aim, we first report the following useful expressions
\begin{align}
\int \frac{d \Omega _ { k } } {4 \pi}  \partial_x e ^ { i {\bf k} \cdot ( {\bf x} - {\bf y}  )}
=& k \frac{ \nu y - x }{R} j_1(kR)\,,\nonumber\\
\int \frac{d \Omega _ { k } } {4 \pi}  \Delta_{2x} e ^ { i {\bf k} \cdot ( {\bf x} - {\bf y}  )}
=& k^2 H^2 j_0(kR) - k \left( \frac{3 H^2}{R} - 2 L \right) j_1(kR) \,,\nonumber\\
\int \frac{d \Omega _ { k } } {4 \pi} \partial_x\Delta_{2y} e ^ { i {\bf k} \cdot ( {\bf x} - {\bf y}  )}
=& -k^3 y^2 \frac{(x - \nu y) (y - \nu x)^2 }{ R^3 } j_3(kR)
 + k^2 y \left[3 y \frac{(x - \nu y )} { R^2 } - 2 \nu \right] j_2(kR) \nonumber\\
&- \frac{ky}{R} \left[ k^2 y (x-\nu y)-2 \nu \right] j_1(kR)\, . 
\label{eq:important_mixed}
\end{align}
where $R$, $L$ and $H$ are taken from Eqs. \eqref{eq:important_eqs_2}. 

\vspace{1cm}

\paragraph{Peculiar Velocity - Lensing} From the definition of the operators in Eqs. \eqref{eq:Operators}, we define 
$x = \eta_o - \eta_i$ and $ y  = \eta_o - \eta'$, since PV is a local effect whereas L is integrated along the line-of-sight. With this, we get
\begin{equation}
\mathcal{W} _ { PVi ,L j } =\frac{\Xi_i}{\Delta \eta_j} \int_{\eta_{in}}^{\eta_{i}}  d\eta \frac {a(\eta)} {a(\eta_o)} \frac{g(\eta)}{g(\eta_o)} \int_{\eta_{j}}^{\eta_o}  d\eta' \frac{ \eta' - \eta_{j} }{ \eta_o - \eta'} 
 \int \frac{d \Omega _ { k } } {4 \pi}   \frac{g(\eta')}{g(\eta_o)} \partial_x  \Delta_{2y}   e ^ { i {\bf k} \cdot ( {\bf x} - {\bf y}  )} \, . 
\end{equation}
Hence, by using Eqs. \eqref{eq:important_mixed}, we obtain
\begin{align}
\mathcal{W}_ { PVi, L j } =& \frac{\Xi_i}{\Delta \eta_j} G_i \int_{\eta_{j}}^{\eta_o}  d\eta' \frac{ \eta' - \eta_{j} }{ \eta_o - \eta' } \left\{-k^3 y^2 \frac{(x - \nu y) (y - \nu x)^2 }{ R^3 } j_3 
+ k^2 y \left[3 y \frac{(x - \nu y )} { R^2 } - 2 \nu \right] j_2\right.\nonumber\\
&\left. - \frac{ky}{R} \left[ k^2 y (x-\nu y)-2 \nu \right] j_1\right\}\,,
\end{align}
where we have omitted the dependence on $k R(x,y,\nu)$ in the $j_n$'s.

\vspace{1cm}

\paragraph{Peculiar Velocity - Sachs-Wolfe} Since both effects are local effects, we define $x = \eta_o - \eta_i$ and $y= \eta_o - \eta_j$. From the definitions of the respective operators in Eqs. \eqref{eq:Operators}, we find
\begin{align}
\mathcal{ W } _ { PV i ,SW j } = &\Xi_i (1+ \Xi_j) \int_{\eta_{in}}^{\eta_i}  d\eta \frac {a(\eta)} {a(\eta_o)} \frac{g(\eta)}{g(\eta_o)} \frac{g(\eta_j)}{g(\eta_o)}  \int \frac{d \Omega _ { k } } {4 \pi}\partial_x   e ^ { i {\bf k} \cdot ( {\bf x} - {\bf y}  )} \nonumber\\
=& \Xi_i (1+ \Xi_j) G_i \frac{g(\eta_j)}{g(\eta_o)} \, k \, \frac{ \nu y - x }{R} j_1(kR)\,,
\end{align}
where we made use again of Eqs. \eqref{eq:important_mixed}.

\vspace{1cm}

\paragraph{Peculiar Velocity - Integrated Sachs-Wolfe} Since PV is a local effect whereas ISW is integrated along the line-of-sight, we define $x  = \eta_o - \eta_i$ and $y = \eta_o - \eta'$. Hence, from the definitions of the operators in Eqs. \eqref{eq:Operators}, we get
\begin{align}
\mathcal{ W } _ { PV i ,ISW  j } =& 2 \, \Xi_i \Xi_j \int_{\eta_{in}}^{\eta_i}  d\eta \frac {a(\eta)} {a(\eta_o)} \frac{g(\eta)}{g(\eta_o)}
\int_{\eta_{j}}^{\eta_o}  d\eta' \frac{\partial _{\eta'}g(\eta')}{g(\eta_o)}  \int \frac{d \Omega _ { k } } {4 \pi}    \partial_x   e ^ { i {\bf k} \cdot ( {\bf x} - {\bf y}  )} \nonumber\\
=& 2 \, \Xi_i \Xi_j G_i \int_{\eta_{j}}^{\eta_o}  d\eta' \frac{\partial _{\eta'}g(\eta')}{g(\eta_o)}  \, k \, \frac{ \nu y - x }{R} j_1(kR) \,,
\end{align}
where we have used as before Eqs. \eqref{eq:important_mixed}.

\vspace{1cm}

\paragraph{Peculiar Velocity - Time Delay}
 Since PV is a local effect whereas TD is integrated along the line-of-sight, we define $x  = \eta_o - \eta_i$ and $y = \eta_o - \eta'$. Hence, from the definitions of the operators in Eqs. \eqref{eq:Operators}, we obtain
\begin{align}
\mathcal{ W } _ { PV i ,TD j } =& - 2 \, \frac{\Xi_i} {\Delta \eta_j} \int_{\eta_{in}}^{\eta_i} d\eta \frac {a(\eta)} {a(\eta_o)} \frac{g(\eta)}{g(\eta_o)} 
 \int_{\eta_j}^{\eta_o}  d\eta' \frac{g(\eta')}{g(\eta_o)}  \int \frac{d \Omega _ { k } } {4 \pi}   \partial_x e ^ { i {\bf k} \cdot ( {\bf x} - {\bf y}  )} \nonumber\\
 =& -2 \,  \frac{\Xi_i} {\Delta \eta_j} G_i \int_{\eta_j}^{\eta_o}  d\eta' \frac{g(\eta')}{g(\eta_o)}  \, k \, \frac{ \nu y - x }{R} j_1(kR) 
\end{align}
where we have used as before Eqs. \eqref{eq:important_mixed}.

\vspace{1cm}

\paragraph{Lensing - Sachs-Wolfe} Since L is an integrated effect and SW is a local one, we set the geometrical variables as $x= \eta_o - \eta$ and $y = \eta_o - \eta_j$. Hence, the definition of the operators in Eqs. \eqref{eq:Operators} leads to
\begin{align}
\mathcal{ W } _ { L i ,SW j } = &\frac{1+\Xi_j}{\Delta \eta_i} \frac{g(\eta_j)}{g(\eta_o)} \int_{\eta_i}^{\eta_o}  d\eta \frac{ \eta - \eta_j }{ \eta_o - \eta } \frac{g(\eta)}{g(\eta_o)} \int \frac{d \Omega _ { k } } {4 \pi}   \Delta_{2y}   e ^ { i {\bf k} \cdot ( {\bf x} - {\bf y}  )}\nonumber\\
 =& \frac{1+\Xi_j}{\Delta \eta_i} \frac{g(\eta_j)}{g(\eta_o)} \int_{\eta_i}^{\eta_o}  d\eta \frac{ \eta - \eta_j }{ \eta_o - \eta } \frac{g(\eta)}{g(\eta_o)} 
 \left[ k^2 H^2 j_0(kR) - k \left( \frac{3 H^2}{R} - 2 L \right) j_1(kR) \right ] \,,
\end{align}
thanks to the Eqs. \eqref{eq:important_mixed}.

\vspace{1cm}

\paragraph{Lensing - Integral Sachs Wolfe} Since both effects are integrated along the line-of-sights, we define $x = \eta_o - \eta$ and $y =\eta_o - \eta'$. Hence, the action of the operators \eqref{eq:Operators}, combined with Eqs. \eqref{eq:important_mixed}, gives
\begin{align}
\mathcal{ W } _ { Li ,ISW  j } = &2 \, \frac{\Xi_j}{\Delta \eta_i}  \int_{\eta_i}^{\eta_o} d\eta \frac{ \eta - \eta_j }{ \eta_o - \eta } \frac{g(\eta)}{g(\eta_o)}
 \int_{\eta_j}^{\eta_o}  d\eta' \frac{\partial _{\eta'}g(\eta')}{g(\eta_o)} \int \frac{d \Omega _ { k } } {4 \pi}   \Delta_{2y}   e ^ { i {\bf k} \cdot ( {\bf x} - {\bf y}  )} \nonumber\\
=& 2 \,  \frac{\Xi_j}{\Delta \eta_i} \int_{\eta_i}^{\eta_o}  d\eta \frac{ \eta - \eta_j }{ \eta_o - \eta } \frac{g(\eta)}{g(\eta_o)}   \int_{\eta_j}^{\eta_o}  d\eta' \frac{\partial _{\eta'}g(\eta')}{g(\eta_o)}
\left[ k^2 H^2 j_0(kR) - k \left( \frac{3 H^2}{R} - 2 L \right) j_1(kR) \right ] \, .
\end{align}

\vspace{1cm}

\paragraph{Lensing - Time Delay} Since L and TD are both integrated along the line-of-sights, we define
$x= \eta_o - \eta$ and $y= \eta_o - \eta'$, so, with the definition \eqref{eq:Operators}, we simply evaluate
\begin{align}
\mathcal{ W } _ { Li ,TD j } = - 2 \, \frac{1}{\Delta \eta_i \Delta \eta_j}  \int_{\eta_i}^{\eta_o}  & d\eta \frac{ \eta - \eta_j }{ \eta_o - \eta } \frac{g(\eta)}{g(\eta_o)}
\int_{\eta_j}^{\eta_o}  d\eta' \frac{g(\eta')}{g(\eta_o)} \int \frac{d \Omega _ { k } } {4 \pi}   \Delta_{2y}   e ^ { i {\bf k} \cdot ( {\bf x} - {\bf y}  )}\nonumber\\
=& - 2 \,  \frac{1}{\Delta \eta_i \Delta \eta_i} \int_{\eta_i}^{\eta_o}  d\eta \frac{ \eta - \eta_j }{ \eta_o - \eta } \frac{g(\eta)}{g(\eta_o)}   \int_{\eta_j}^{\eta_o}  d\eta' \frac{g(\eta')}{g(\eta_o)}
\left[ k^2 H^2 j_0(kR) - k \left( \frac{3 H^2}{R} - 2 L \right) j_1(kR) \right ] \,,
\end{align}
thanks to the relations \eqref{eq:important_mixed}.

\vspace{1cm}

\paragraph{Time Delay - Integrated Sachs-Wolfe} Just as for the previous term, we define $x= \eta_o - \eta$ and $y= \eta_o - \eta'$ since both TD and ISW are integrated effects. Hence, from the operators defined in Eqs. \eqref{eq:Operators}, we obtain
\begin{align}
\mathcal{ W } _ { TDi, ISW j } =& 2 \, (1+\Xi_i) \Xi_j  \frac{g(\eta_i)}{g(\eta_o)}  \int_{\eta_j}^{\eta_o}  d\eta' \frac{\partial _{\eta'}g(\eta')}{g(\eta_o)}  \int \frac{d \Omega _ { k } } {4 \pi}      e ^ { i {\bf k} \cdot ( {\bf x} - {\bf y}  )}\nonumber\\
=& 2 \, (1+\Xi_i) \Xi_j \frac{g(\eta_i)}{g(\eta_o)} \int_{\eta_j}^{\eta_o}  d\eta' \frac{\partial _{\eta'}g(\eta')}{g(\eta_o)}  \, j_0(kR)\,,
\end{align}
where we have applied Eq. \eqref{int_dOk}.

\vspace{1cm}

\paragraph{Sachs-Wolfe - Time Delay} SW and TD are respectively local and integrated effects. Because of that, we define $x = \eta_o - \eta_i$ and $y = \eta_o - \eta'$ and then, from the operators \eqref{eq:Operators}, we obtain
\begin{equation}
\mathcal{ W } _ { SW i, TD j } = -2 \, \frac{1+\Xi_i}{\Delta \eta_j}\frac{g(\eta_i)}{g(\eta_o)}  \int_{\eta_j}^{\eta_o}  d\eta' \frac{g(\eta')}{g(\eta_o)}  \int \frac{d \Omega _ { k } } {4 \pi}      e ^ { i {\bf k} \cdot ( {\bf x} - {\bf y}  )}
= -2 \, \frac{1+\Xi_i}{\Delta \eta_j}\frac{g(\eta_i)}{g(\eta_o)} \int_{\eta_j}^{\eta_o}  d\eta' \frac{g(\eta')}{g(\eta_o)}  \, j_0(kR)\,,
\end{equation}
where we have applied Eq. \eqref{int_dOk}.

\vspace{1cm}

\paragraph{Integrated Sachs-Wolfe - Time Delay} ISW and TD are both integrated effects. Hence, we define $x = \eta_o - \eta$ and $y = \eta_o - \eta'$ and then, from the operators \eqref{eq:Operators}, we obtain
\begin{align}
\mathcal{ W } _ { ISW i,TD j } = &
 - 4  \, \frac{\Xi_i}{\Delta \eta_j} \int_{\eta_{i}}^{\eta_o}  d\eta \frac{\partial _{\eta}g(\eta)}{g(\eta_o)} \int_{\eta_j}^{\eta_o}  d\eta' \frac{g(\eta')}{g(\eta_o)}  \int \frac{d \Omega _ { k } } {4 \pi} e ^ { i {\bf k} \cdot ( {\bf x} - {\bf y}  )} 
 \nonumber\\
=& - 4 \, \frac{\Xi_i}{\Delta \eta_j}\int_{\eta_i}^{\eta_o}  d\eta \frac{\partial _{\eta}g(\eta)}{g(\eta_o)}  \int_{\eta_j}^{\eta_o}  d\eta' \frac{g(\eta')}{g(\eta_o)}  \, j_0(kR)\,,
\end{align}
where we have applied Eq. \eqref{int_dOk}.


\section{Useful Expressions}
\label{app:expansion}
Here we report the explicit value for the $Q_{\ell n}$ in Eq. \eqref{eq:53} for $\ell=1,2,3$, which allows to write analytically the kernel for the dipole of the lensing 2-point correlation function in Eq. \eqref{eq:kell}. For $\ell =1$, we obtain
\begin{align}
Q_{10}(z)=&\,-\frac{1}{2} z\, \text{SinInt}(zk)-\frac{\sin (zk)}{2\,k^2\,z}-\frac{1}{2k} \cos (zk)+\frac{1}{k}\,,\nonumber\\
Q_{11}(z)=&\,\frac{1}{k^2}\text{SinInt}(z k)-\frac{z}{k}\,,\nonumber\\
Q_{12}(z)=&\,-\frac{z}{k^2}\,\text{SinInt}(z k)-\frac{2 \cos (z k)}{k^2}+\frac{2}{k^3}\,,\nonumber\\
Q_{13}(z)=&\,\frac{3 \sin (z k)}{k^4}-\frac{2\,z}{k^3}-\frac{z \cos (z k)}{k^3}\,,\nonumber\\
Q_{14}(z)=&\,-\frac{z^2 \cos (z k)}{k^3}+\frac{5\,z \sin (z k)}{k^4}+\frac{8 \cos (z k)}{k^5}-\frac{8}{k^5}\,,\nonumber\\
Q_{15}(z)=&\,\frac{7\,z^2 \sin (z k)}{k^4}-\frac{z^3 \cos (z k)}{k^3}+\frac{8 z}{k^5}
-\frac{30 \sin (z k)}{k^6}+\frac{22 z \cos (z k)}{k^5}\,.
\end{align}
For $\ell=2$, we have
\begin{align}
Q_{20}(z)=&\,-\frac{\sin (k z)}{2 k^3 z^2}+\frac{\cos (k z)}{2 k^2 z}+\frac{\text{SinInt}(k z)}{2 k}-\frac{z}{3}\,,\nonumber\\
Q_{21}(z)=&\,-\frac{3 \sin (k z)}{2 k^3 z}-\frac{\cos (k z)}{2 k^2}+\frac{2}{k^2}-\frac{z \text{SinInt}(k z)}{2 k}\,,\nonumber\\
Q_{22}(z)=&\,\frac{3 \text{SinInt}(k z)}{k^3}-\frac{2 z}{k^2}-\frac{\sin (k z)}{k^3}\,,\nonumber\\
Q_{23}(z)=&\,-\frac{3 z \text{SinInt}(k z)}{k^3}-\frac{z \sin (k z)}{k^3}-\frac{8 \cos (k z)}{k^4}+\frac{8}{k^4}\,,\nonumber\\
Q_{24}(z)=&\,-\frac{z^2 \sin (k z)}{k^3}-\frac{8 z}{k^4}+\frac{15 \sin (k z)}{k^5}-\frac{7 z \cos (k z)}{k^4}\,,\nonumber\\
Q_{25}(z)=&\,-\frac{z^3 \sin (k z)}{k^3}-\frac{9 z^2 \cos (k z)}{k^4}+\frac{33 z \sin (k z)}{k^5}+\frac{48 \cos (k z)}{k^6}-\frac{48}{k^6}\,.
\end{align}
Finally, for $\ell=3$ we get
\begin{align}
Q_{30}(z)=&\,-\frac{5 \sin (k z)}{4 k^4 z^3}+\frac{5 \cos (k z)}{4 k^3 z^2}-\frac{\sin (k z)}{8 k^2 z}-\frac{z \text{SinInt}(k z)}{8}-\frac{\cos (k z)}{8 k}+\frac{2}{3 k}\,,\nonumber\\
Q_{31}(z)=&\,\frac{3 \text{SinInt}(k z)}{2 k^2}-\frac{5 \sin (k z)}{2 k^4 z^2}+\frac{5 \cos (k z)}{2 k^3 z}-\frac{2 z}{3 k}\,,\nonumber\\
Q_{32}(z)=&\,-\frac{3 z \text{SinInt}(k z)}{2 k^2}-\frac{15 \sin (k z)}{2 k^4 z}-\frac{\cos (k z)}{2 k^3}+\frac{8}{k^3}\,,\nonumber\\
Q_{33}(z)=&\,\frac{15 \text{SinInt}(k z)}{k^4}-\frac{8 z}{k^3}-\frac{8 \sin (k z)}{k^4}+\frac{z \cos (k z)}{k^3}\,,\nonumber\\
Q_{34}(z)=&\,-\frac{15 z \text{SinInt}(k z)}{k^4}+\frac{z^2 \cos (k z)}{k^3}-\frac{10 z \sin (k z)}{k^4}-\frac{48 \cos (k z)}{k^5}+\frac{48}{k^5}\,,\nonumber\\
Q_{35}(z)=&\,-\frac{12 z^2 \sin (k z)}{k^4}+\frac{z^3 \cos (k z)}{k^3}-\frac{48 z}{k^5}+\frac{105 \sin (k z)}{k^6}-\frac{57 z \cos (k z)}{k^5}\,.
\end{align}


\section{Spherical Bessel functions}
\label{app:bessel}
Spherical Bessel functions $j_n(z)$ are generated for any integer $n$ through the recursive formula
\begin{equation}
j_n(z)=(-1)^n\,z^n\,\left(\frac{1}{z}\frac{d}{dz}\right)^n\left(\frac{\sin z}{z}\right)\,,
\label{eq:jn}
\end{equation}
and are the fundamental solutions of the ODE
\begin{equation}
z^2\frac{d^2}{dz^2}j_n(z)+2\,z\frac{d}{dz}j_n(z)+\left[ z^2-n(n+1) \right]j_n(z)=0\,.
\label{eq:jn_eigen}
\end{equation}
The useful relations that we report for our purposes are the following
\begin{align}
j_{n-1}(z)=&\frac{2n+1}{z}j_n(z)-j_{n+1}(z)\,,\nonumber\\
j'_n(z)=&\frac{n}{2n+1}\,j_{n-1}(z)-\frac{n+1}{2n+1}\,j_{n+1}(z)\,,\nonumber\\
j'_n(z)=&j_{n-1}(z)-\frac{n+1}{z}j_n(z\,,)\nonumber\\
j'_n(z)=&-j_{n+1}(z)+\frac{n}{z}j_n(z)\,,
\label{eq:jn_recursive}
\end{align}
where, in particular, the first equation allows to extend the definition of the $j_n(z)$ also when $n$ is negative, whereas the remaining three equations allows to express also their derivatives in terms of the $j_n$'s themselves.


\section{Legendre polynomials and multipole expansion}
\label{app:legendre}
Let be $f(\cos\theta)$ a function which does not depend on the azimutal angle $\phi$. In terms of $\nu\equiv{\bf n}_x\cdot {\bf n}_y=\cos\theta$, we can write the angular laplacian $\Delta_2$ as
\begin{equation}
\Delta_2 f(\nu)=\partial_\nu\left[\left( 1-\nu^2 \right)\partial_\nu f(\nu)\right]\,.
\end{equation}
In this way, the multipole expansion of $\Delta_2f$ in terms of the Legendre polynomials $P_\ell(\nu)$ can be integrated by part twice. This leads to
\begin{align}
\int_{-1}^{1}d\nu\,P_\ell(\nu)\Delta_2 f(\nu)
=&
\int_{-1}^{1}d\nu\,P_\ell(\nu)\partial_\nu\left[\left( 1-\nu^2 \right)\partial_\nu f(\nu)\right]
\nonumber\\
=&
-\int_{-1}^{1}d\nu\,\partial_\nu P_\ell(\nu)\left[\left( 1-\nu^2 \right)\partial_\nu f(\nu)\right]
\nonumber\\
=&
\int_{-1}^{1}d\nu\,\Delta_2P_\ell(\nu)\,f(\nu)
=
-\ell\left(\ell+1\right)\int_{-1}^{1}d\nu\,P_\ell(\nu)\,f(\nu)\,,
\label{eq:D2}
\end{align}
where in the last line we have used the fact that $P_\ell(\nu)$ are eigenfunctions of $\Delta_2$. This simple result enlightens the case of lensing multipoles in Eq. \eqref{eq:kell}. Indeed, those coefficients can be formally written in the following form
\begin{equation}
\int_{-1}^1d\nu\,P_\ell(\nu)\Delta_{2x}\Delta_{2y} f(R(\nu))\,.
\label{eq:D3}
\end{equation}
In fact, since \eqref{eq:D3} is invariant under rotation of the coordinate system in the plane spanned by ${\bf x}$ and ${\bf y}$, we can apply Eq. \eqref{eq:D2} as follows: we first rotate the coordinate system in the integral in order to align them with the direction ${\bf x}$, hence we apply Eq. \eqref{eq:D2}. After that, we align again the coordinate system with ${\bf y}$ and finally apply Eq. \eqref{eq:D2} again. Since Eq. \eqref{eq:D2} is applied twice, we get the prefactor $\ell^2\left(\ell+1\right)^2$.

\end{widetext}

\bibliography{bibliografia}

\begin{thebibliography}{62}%
\makeatletter
\providecommand \@ifxundefined [1]{%
 \@ifx{#1\undefined}
}%
\providecommand \@ifnum [1]{%
 \ifnum #1\expandafter \@firstoftwo
 \else \expandafter \@secondoftwo
 \fi
}%
\providecommand \@ifx [1]{%
 \ifx #1\expandafter \@firstoftwo
 \else \expandafter \@secondoftwo
 \fi
}%
\providecommand \natexlab [1]{#1}%
\providecommand \enquote  [1]{``#1''}%
\providecommand \bibnamefont  [1]{#1}%
\providecommand \bibfnamefont [1]{#1}%
\providecommand \citenamefont [1]{#1}%
\providecommand \href@noop [0]{\@secondoftwo}%
\providecommand \href [0]{\begingroup \@sanitize@url \@href}%
\providecommand \@href[1]{\@@startlink{#1}\@@href}%
\providecommand \@@href[1]{\endgroup#1\@@endlink}%
\providecommand \@sanitize@url [0]{\catcode `\\12\catcode `\$12\catcode
  `\&12\catcode `\#12\catcode `\^12\catcode `\_12\catcode `\%12\relax}%
\providecommand \@@startlink[1]{}%
\providecommand \@@endlink[0]{}%
\providecommand \url  [0]{\begingroup\@sanitize@url \@url }%
\providecommand \@url [1]{\endgroup\@href {#1}{\urlprefix }}%
\providecommand \urlprefix  [0]{URL }%
\providecommand \Eprint [0]{\href }%
\providecommand \doibase [0]{http://dx.doi.org/}%
\providecommand \selectlanguage [0]{\@gobble}%
\providecommand \bibinfo  [0]{\@secondoftwo}%
\providecommand \bibfield  [0]{\@secondoftwo}%
\providecommand \translation [1]{[#1]}%
\providecommand \BibitemOpen [0]{}%
\providecommand \bibitemStop [0]{}%
\providecommand \bibitemNoStop [0]{.\EOS\space}%
\providecommand \EOS [0]{\spacefactor3000\relax}%
\providecommand \BibitemShut  [1]{\csname bibitem#1\endcsname}%
\let\auto@bib@innerbib\@empty
\bibitem [{\citenamefont {Laureijs}\ \emph {et~al.}(2011)\citenamefont
  {Laureijs} \emph {et~al.}}]{Laureijs:2011gra}%
  \BibitemOpen
  \bibfield  {author} {\bibinfo {author} {\bibfnamefont {R.}~\bibnamefont
  {Laureijs}} \emph {et~al.} (\bibinfo {collaboration} {EUCLID}),\ }\href@noop
  {} {\  (\bibinfo {year} {2011})},\ \Eprint {http://arxiv.org/abs/1110.3193}
  {arXiv:1110.3193 [astro-ph.CO]} \BibitemShut {NoStop}%
\bibitem [{\citenamefont {Abell}\ \emph {et~al.}(2009)\citenamefont {Abell}
  \emph {et~al.}}]{Abell:2009aa}%
  \BibitemOpen
  \bibfield  {author} {\bibinfo {author} {\bibfnamefont {P.~A.}\ \bibnamefont
  {Abell}} \emph {et~al.} (\bibinfo {collaboration} {LSST Science, LSST
  Project}),\ }\href@noop {} {\  (\bibinfo {year} {2009})},\ \Eprint
  {http://arxiv.org/abs/0912.0201} {arXiv:0912.0201 [astro-ph.IM]} \BibitemShut
  {NoStop}%
\bibitem [{\citenamefont {Mellema}\ \emph {et~al.}(2013)\citenamefont {Mellema}
  \emph {et~al.}}]{Mellema:2012ht}%
  \BibitemOpen
  \bibfield  {author} {\bibinfo {author} {\bibfnamefont {G.}~\bibnamefont
  {Mellema}} \emph {et~al.},\ }\href {\doibase 10.1007/s10686-013-9334-5}
  {\bibfield  {journal} {\bibinfo  {journal} {Exper. Astron.}\ }\textbf
  {\bibinfo {volume} {36}},\ \bibinfo {pages} {235} (\bibinfo {year} {2013})},\
  \Eprint {http://arxiv.org/abs/1210.0197} {arXiv:1210.0197 [astro-ph.CO]}
  \BibitemShut {NoStop}%
\bibitem [{\citenamefont {DeBoer}\ \emph {et~al.}(2017)\citenamefont {DeBoer}
  \emph {et~al.}}]{DeBoer:2016tnn}%
  \BibitemOpen
  \bibfield  {author} {\bibinfo {author} {\bibfnamefont {D.~R.}\ \bibnamefont
  {DeBoer}} \emph {et~al.},\ }\href {\doibase 10.1088/1538-3873/129/974/045001}
  {\bibfield  {journal} {\bibinfo  {journal} {Publ. Astron. Soc. Pac.}\
  }\textbf {\bibinfo {volume} {129}},\ \bibinfo {pages} {045001} (\bibinfo
  {year} {2017})},\ \Eprint {http://arxiv.org/abs/1606.07473} {arXiv:1606.07473
  [astro-ph.IM]} \BibitemShut {NoStop}%
\bibitem [{\citenamefont {Marozzi}\ \emph {et~al.}(2016)\citenamefont
  {Marozzi}, \citenamefont {Fanizza}, \citenamefont {Di~Dio},\ and\
  \citenamefont {Durrer}}]{Marozzi:2016uob}%
  \BibitemOpen
  \bibfield  {author} {\bibinfo {author} {\bibfnamefont {G.}~\bibnamefont
  {Marozzi}}, \bibinfo {author} {\bibfnamefont {G.}~\bibnamefont {Fanizza}},
  \bibinfo {author} {\bibfnamefont {E.}~\bibnamefont {Di~Dio}}, \ and\ \bibinfo
  {author} {\bibfnamefont {R.}~\bibnamefont {Durrer}},\ }\href {\doibase
  10.1088/1475-7516/2016/09/028} {\bibfield  {journal} {\bibinfo  {journal}
  {JCAP}\ }\textbf {\bibinfo {volume} {09}},\ \bibinfo {pages} {028} (\bibinfo
  {year} {2016})},\ \Eprint {http://arxiv.org/abs/1605.08761} {arXiv:1605.08761
  [astro-ph.CO]} \BibitemShut {NoStop}%
\bibitem [{\citenamefont {Marozzi}\ \emph {et~al.}(2017)\citenamefont
  {Marozzi}, \citenamefont {Fanizza}, \citenamefont {Di~Dio},\ and\
  \citenamefont {Durrer}}]{Marozzi:2016und}%
  \BibitemOpen
  \bibfield  {author} {\bibinfo {author} {\bibfnamefont {G.}~\bibnamefont
  {Marozzi}}, \bibinfo {author} {\bibfnamefont {G.}~\bibnamefont {Fanizza}},
  \bibinfo {author} {\bibfnamefont {E.}~\bibnamefont {Di~Dio}}, \ and\ \bibinfo
  {author} {\bibfnamefont {R.}~\bibnamefont {Durrer}},\ }\href {\doibase
  10.1103/PhysRevLett.118.211301} {\bibfield  {journal} {\bibinfo  {journal}
  {Phys. Rev. Lett.}\ }\textbf {\bibinfo {volume} {118}},\ \bibinfo {pages}
  {211301} (\bibinfo {year} {2017})},\ \Eprint
  {http://arxiv.org/abs/1612.07650} {arXiv:1612.07650 [astro-ph.CO]}
  \BibitemShut {NoStop}%
\bibitem [{\citenamefont {Marozzi}\ \emph {et~al.}(2018)\citenamefont
  {Marozzi}, \citenamefont {Fanizza}, \citenamefont {Di~Dio},\ and\
  \citenamefont {Durrer}}]{Marozzi:2016qxl}%
  \BibitemOpen
  \bibfield  {author} {\bibinfo {author} {\bibfnamefont {G.}~\bibnamefont
  {Marozzi}}, \bibinfo {author} {\bibfnamefont {G.}~\bibnamefont {Fanizza}},
  \bibinfo {author} {\bibfnamefont {E.}~\bibnamefont {Di~Dio}}, \ and\ \bibinfo
  {author} {\bibfnamefont {R.}~\bibnamefont {Durrer}},\ }\href {\doibase
  10.1103/PhysRevD.98.023535} {\bibfield  {journal} {\bibinfo  {journal} {Phys.
  Rev. D}\ }\textbf {\bibinfo {volume} {98}},\ \bibinfo {pages} {023535}
  (\bibinfo {year} {2018})},\ \Eprint {http://arxiv.org/abs/1612.07263}
  {arXiv:1612.07263 [astro-ph.CO]} \BibitemShut {NoStop}%
\bibitem [{\citenamefont {Di~Dio}\ \emph {et~al.}(2019)\citenamefont {Di~Dio},
  \citenamefont {Durrer}, \citenamefont {Fanizza},\ and\ \citenamefont
  {Marozzi}}]{DiDio:2019rfy}%
  \BibitemOpen
  \bibfield  {author} {\bibinfo {author} {\bibfnamefont {E.}~\bibnamefont
  {Di~Dio}}, \bibinfo {author} {\bibfnamefont {R.}~\bibnamefont {Durrer}},
  \bibinfo {author} {\bibfnamefont {G.}~\bibnamefont {Fanizza}}, \ and\
  \bibinfo {author} {\bibfnamefont {G.}~\bibnamefont {Marozzi}},\ }\href
  {\doibase 10.1103/PhysRevD.100.043508} {\bibfield  {journal} {\bibinfo
  {journal} {Phys. Rev. D}\ }\textbf {\bibinfo {volume} {100}},\ \bibinfo
  {pages} {043508} (\bibinfo {year} {2019})},\ \Eprint
  {http://arxiv.org/abs/1905.12573} {arXiv:1905.12573 [astro-ph.CO]}
  \BibitemShut {NoStop}%
\bibitem [{\citenamefont {Aghanim}\ \emph {et~al.}(2020)\citenamefont {Aghanim}
  \emph {et~al.}}]{Aghanim:2018eyx}%
  \BibitemOpen
  \bibfield  {author} {\bibinfo {author} {\bibfnamefont {N.}~\bibnamefont
  {Aghanim}} \emph {et~al.} (\bibinfo {collaboration} {Planck}),\ }\href
  {\doibase 10.1051/0004-6361/201833910} {\bibfield  {journal} {\bibinfo
  {journal} {Astron. Astrophys.}\ }\textbf {\bibinfo {volume} {641}},\ \bibinfo
  {pages} {A6} (\bibinfo {year} {2020})},\ \Eprint
  {http://arxiv.org/abs/1807.06209} {arXiv:1807.06209 [astro-ph.CO]}
  \BibitemShut {NoStop}%
\bibitem [{\citenamefont {Riess}\ \emph {et~al.}(2021)\citenamefont {Riess},
  \citenamefont {Casertano}, \citenamefont {Yuan}, \citenamefont {Bowers},
  \citenamefont {Macri}, \citenamefont {Zinn},\ and\ \citenamefont
  {Scolnic}}]{Riess:2020fzl}%
  \BibitemOpen
  \bibfield  {author} {\bibinfo {author} {\bibfnamefont {A.~G.}\ \bibnamefont
  {Riess}}, \bibinfo {author} {\bibfnamefont {S.}~\bibnamefont {Casertano}},
  \bibinfo {author} {\bibfnamefont {W.}~\bibnamefont {Yuan}}, \bibinfo {author}
  {\bibfnamefont {J.~B.}\ \bibnamefont {Bowers}}, \bibinfo {author}
  {\bibfnamefont {L.}~\bibnamefont {Macri}}, \bibinfo {author} {\bibfnamefont
  {J.~C.}\ \bibnamefont {Zinn}}, \ and\ \bibinfo {author} {\bibfnamefont
  {D.}~\bibnamefont {Scolnic}},\ }\href {\doibase 10.3847/2041-8213/abdbaf}
  {\bibfield  {journal} {\bibinfo  {journal} {Astrophys. J. Lett.}\ }\textbf
  {\bibinfo {volume} {908}},\ \bibinfo {pages} {L6} (\bibinfo {year} {2021})},\
  \Eprint {http://arxiv.org/abs/2012.08534} {arXiv:2012.08534 [astro-ph.CO]}
  \BibitemShut {NoStop}%
\bibitem [{\citenamefont {Freedman}\ \emph {et~al.}(2019)\citenamefont
  {Freedman}, \citenamefont {Madore}, \citenamefont {Hatt}, \citenamefont
  {Hoyt}, \citenamefont {Jang}, \citenamefont {Beaton}, \citenamefont {Burns},
  \citenamefont {Lee}, \citenamefont {Monson}, \citenamefont {Neeley},\ and\
  \citenamefont {et~al.}}]{Freedman_2019}%
  \BibitemOpen
  \bibfield  {author} {\bibinfo {author} {\bibfnamefont {W.~L.}\ \bibnamefont
  {Freedman}}, \bibinfo {author} {\bibfnamefont {B.~F.}\ \bibnamefont
  {Madore}}, \bibinfo {author} {\bibfnamefont {D.}~\bibnamefont {Hatt}},
  \bibinfo {author} {\bibfnamefont {T.~J.}\ \bibnamefont {Hoyt}}, \bibinfo
  {author} {\bibfnamefont {I.~S.}\ \bibnamefont {Jang}}, \bibinfo {author}
  {\bibfnamefont {R.~L.}\ \bibnamefont {Beaton}}, \bibinfo {author}
  {\bibfnamefont {C.~R.}\ \bibnamefont {Burns}}, \bibinfo {author}
  {\bibfnamefont {M.~G.}\ \bibnamefont {Lee}}, \bibinfo {author} {\bibfnamefont
  {A.~J.}\ \bibnamefont {Monson}}, \bibinfo {author} {\bibfnamefont {J.~R.}\
  \bibnamefont {Neeley}}, \ and\ \bibinfo {author} {\bibnamefont {et~al.}},\
  }\href {\doibase 10.3847/1538-4357/ab2f73} {\bibfield  {journal} {\bibinfo
  {journal} {The Astrophysical Journal}\ }\textbf {\bibinfo {volume} {882}},\
  \bibinfo {pages} {34} (\bibinfo {year} {2019})}\BibitemShut {NoStop}%
\bibitem [{\citenamefont {Freedman}\ \emph {et~al.}(2020)\citenamefont
  {Freedman}, \citenamefont {Madore}, \citenamefont {Hoyt}, \citenamefont
  {Jang}, \citenamefont {Beaton}, \citenamefont {Lee}, \citenamefont {Monson},
  \citenamefont {Neeley},\ and\ \citenamefont {Rich}}]{Freedman_2020}%
  \BibitemOpen
  \bibfield  {author} {\bibinfo {author} {\bibfnamefont {W.~L.}\ \bibnamefont
  {Freedman}}, \bibinfo {author} {\bibfnamefont {B.~F.}\ \bibnamefont
  {Madore}}, \bibinfo {author} {\bibfnamefont {T.}~\bibnamefont {Hoyt}},
  \bibinfo {author} {\bibfnamefont {I.~S.}\ \bibnamefont {Jang}}, \bibinfo
  {author} {\bibfnamefont {R.}~\bibnamefont {Beaton}}, \bibinfo {author}
  {\bibfnamefont {M.~G.}\ \bibnamefont {Lee}}, \bibinfo {author} {\bibfnamefont
  {A.}~\bibnamefont {Monson}}, \bibinfo {author} {\bibfnamefont
  {J.}~\bibnamefont {Neeley}}, \ and\ \bibinfo {author} {\bibfnamefont
  {J.}~\bibnamefont {Rich}},\ }\href {\doibase 10.3847/1538-4357/ab7339}
  {\bibfield  {journal} {\bibinfo  {journal} {The Astrophysical Journal}\
  }\textbf {\bibinfo {volume} {891}},\ \bibinfo {pages} {57} (\bibinfo {year}
  {2020})}\BibitemShut {NoStop}%
\bibitem [{\citenamefont {Wong}\ \emph {et~al.}(2019)\citenamefont {Wong},
  \citenamefont {Suyu}, \citenamefont {Chen}, \citenamefont {Rusu},
  \citenamefont {Millon}, \citenamefont {Sluse}, \citenamefont {Bonvin},
  \citenamefont {Fassnacht}, \citenamefont {Taubenberger}, \citenamefont
  {Auger},\ and\ \citenamefont {et~al.}}]{Wong_2019}%
  \BibitemOpen
  \bibfield  {author} {\bibinfo {author} {\bibfnamefont {K.~C.}\ \bibnamefont
  {Wong}}, \bibinfo {author} {\bibfnamefont {S.~H.}\ \bibnamefont {Suyu}},
  \bibinfo {author} {\bibfnamefont {G.~C.-F.}\ \bibnamefont {Chen}}, \bibinfo
  {author} {\bibfnamefont {C.~E.}\ \bibnamefont {Rusu}}, \bibinfo {author}
  {\bibfnamefont {M.}~\bibnamefont {Millon}}, \bibinfo {author} {\bibfnamefont
  {D.}~\bibnamefont {Sluse}}, \bibinfo {author} {\bibfnamefont
  {V.}~\bibnamefont {Bonvin}}, \bibinfo {author} {\bibfnamefont {C.~D.}\
  \bibnamefont {Fassnacht}}, \bibinfo {author} {\bibfnamefont {S.}~\bibnamefont
  {Taubenberger}}, \bibinfo {author} {\bibfnamefont {M.~W.}\ \bibnamefont
  {Auger}}, \ and\ \bibinfo {author} {\bibnamefont {et~al.}},\ }\href {\doibase
  10.1093/mnras/stz3094} {\bibfield  {journal} {\bibinfo  {journal} {Monthly
  Notices of the Royal Astronomical Society}\ }\textbf {\bibinfo {volume}
  {498}},\ \bibinfo {pages} {1420} (\bibinfo {year} {2019})}\BibitemShut
  {NoStop}%
\bibitem [{\citenamefont {Yuan}\ \emph {et~al.}(2019)\citenamefont {Yuan},
  \citenamefont {Riess}, \citenamefont {Macri}, \citenamefont {Casertano},\
  and\ \citenamefont {Scolnic}}]{Yuan_2019}%
  \BibitemOpen
  \bibfield  {author} {\bibinfo {author} {\bibfnamefont {W.}~\bibnamefont
  {Yuan}}, \bibinfo {author} {\bibfnamefont {A.~G.}\ \bibnamefont {Riess}},
  \bibinfo {author} {\bibfnamefont {L.~M.}\ \bibnamefont {Macri}}, \bibinfo
  {author} {\bibfnamefont {S.}~\bibnamefont {Casertano}}, \ and\ \bibinfo
  {author} {\bibfnamefont {D.~M.}\ \bibnamefont {Scolnic}},\ }\href {\doibase
  10.3847/1538-4357/ab4bc9} {\bibfield  {journal} {\bibinfo  {journal} {The
  Astrophysical Journal}\ }\textbf {\bibinfo {volume} {886}},\ \bibinfo {pages}
  {61} (\bibinfo {year} {2019})}\BibitemShut {NoStop}%
\bibitem [{\citenamefont {Pesce}\ \emph {et~al.}(2020)\citenamefont {Pesce},
  \citenamefont {Braatz}, \citenamefont {Reid}, \citenamefont {Riess},
  \citenamefont {Scolnic}, \citenamefont {Condon}, \citenamefont {Gao},
  \citenamefont {Henkel}, \citenamefont {Impellizzeri}, \citenamefont {Kuo},\
  and\ \citenamefont {et~al.}}]{Pesce_2020}%
  \BibitemOpen
  \bibfield  {author} {\bibinfo {author} {\bibfnamefont {D.~W.}\ \bibnamefont
  {Pesce}}, \bibinfo {author} {\bibfnamefont {J.~A.}\ \bibnamefont {Braatz}},
  \bibinfo {author} {\bibfnamefont {M.~J.}\ \bibnamefont {Reid}}, \bibinfo
  {author} {\bibfnamefont {A.~G.}\ \bibnamefont {Riess}}, \bibinfo {author}
  {\bibfnamefont {D.}~\bibnamefont {Scolnic}}, \bibinfo {author} {\bibfnamefont
  {J.~J.}\ \bibnamefont {Condon}}, \bibinfo {author} {\bibfnamefont
  {F.}~\bibnamefont {Gao}}, \bibinfo {author} {\bibfnamefont {C.}~\bibnamefont
  {Henkel}}, \bibinfo {author} {\bibfnamefont {C.~M.~V.}\ \bibnamefont
  {Impellizzeri}}, \bibinfo {author} {\bibfnamefont {C.~Y.}\ \bibnamefont
  {Kuo}}, \ and\ \bibinfo {author} {\bibnamefont {et~al.}},\ }\href {\doibase
  10.3847/2041-8213/ab75f0} {\bibfield  {journal} {\bibinfo  {journal} {The
  Astrophysical Journal}\ }\textbf {\bibinfo {volume} {891}},\ \bibinfo {pages}
  {L1} (\bibinfo {year} {2020})}\BibitemShut {NoStop}%
\bibitem [{\citenamefont {Verde}\ \emph {et~al.}(2019)\citenamefont {Verde},
  \citenamefont {Treu},\ and\ \citenamefont {Riess}}]{Verde_2019}%
  \BibitemOpen
  \bibfield  {author} {\bibinfo {author} {\bibfnamefont {L.}~\bibnamefont
  {Verde}}, \bibinfo {author} {\bibfnamefont {T.}~\bibnamefont {Treu}}, \ and\
  \bibinfo {author} {\bibfnamefont {A.~G.}\ \bibnamefont {Riess}},\ }\href
  {\doibase 10.1038/s41550-019-0902-0} {\bibfield  {journal} {\bibinfo
  {journal} {Nature Astronomy}\ }\textbf {\bibinfo {volume} {3}},\ \bibinfo
  {pages} {891} (\bibinfo {year} {2019})}\BibitemShut {NoStop}%
\bibitem [{\citenamefont {Bernal}\ \emph {et~al.}(2016)\citenamefont {Bernal},
  \citenamefont {Verde},\ and\ \citenamefont {Riess}}]{Bernal:2016gxb}%
  \BibitemOpen
  \bibfield  {author} {\bibinfo {author} {\bibfnamefont {J.~L.}\ \bibnamefont
  {Bernal}}, \bibinfo {author} {\bibfnamefont {L.}~\bibnamefont {Verde}}, \
  and\ \bibinfo {author} {\bibfnamefont {A.~G.}\ \bibnamefont {Riess}},\ }\href
  {\doibase 10.1088/1475-7516/2016/10/019} {\bibfield  {journal} {\bibinfo
  {journal} {JCAP}\ }\textbf {\bibinfo {volume} {1610}},\ \bibinfo {pages}
  {019} (\bibinfo {year} {2016})},\ \Eprint {http://arxiv.org/abs/1607.05617}
  {arXiv:1607.05617 [astro-ph.CO]} \BibitemShut {NoStop}%
\bibitem [{\citenamefont {Jedamzik}\ \emph {et~al.}(2021)\citenamefont
  {Jedamzik}, \citenamefont {Pogosian},\ and\ \citenamefont
  {Zhao}}]{Jedamzik:2020zmd}%
  \BibitemOpen
  \bibfield  {author} {\bibinfo {author} {\bibfnamefont {K.}~\bibnamefont
  {Jedamzik}}, \bibinfo {author} {\bibfnamefont {L.}~\bibnamefont {Pogosian}},
  \ and\ \bibinfo {author} {\bibfnamefont {G.-B.}\ \bibnamefont {Zhao}},\
  }\href {\doibase 10.1038/s42005-021-00628-x} {\bibfield  {journal} {\bibinfo
  {journal} {Commun. in Phys.}\ }\textbf {\bibinfo {volume} {4}},\ \bibinfo
  {pages} {123} (\bibinfo {year} {2021})},\ \Eprint
  {http://arxiv.org/abs/2010.04158} {arXiv:2010.04158 [astro-ph.CO]}
  \BibitemShut {NoStop}%
\bibitem [{\citenamefont {Beenakker}\ and\ \citenamefont
  {Venhoek}(2021)}]{Beenakker:2021vff}%
  \BibitemOpen
  \bibfield  {author} {\bibinfo {author} {\bibfnamefont {W.}~\bibnamefont
  {Beenakker}}\ and\ \bibinfo {author} {\bibfnamefont {D.}~\bibnamefont
  {Venhoek}},\ }\href@noop {} {\  (\bibinfo {year} {2021})},\ \Eprint
  {http://arxiv.org/abs/2101.01372} {arXiv:2101.01372 [astro-ph.CO]}
  \BibitemShut {NoStop}%
\bibitem [{\citenamefont {Knox}\ and\ \citenamefont
  {Millea}(2020)}]{Knox:2019rjx}%
  \BibitemOpen
  \bibfield  {author} {\bibinfo {author} {\bibfnamefont {L.}~\bibnamefont
  {Knox}}\ and\ \bibinfo {author} {\bibfnamefont {M.}~\bibnamefont {Millea}},\
  }\href {\doibase 10.1103/PhysRevD.101.043533} {\bibfield  {journal} {\bibinfo
   {journal} {Phys. Rev. D}\ }\textbf {\bibinfo {volume} {101}},\ \bibinfo
  {pages} {043533} (\bibinfo {year} {2020})},\ \Eprint
  {http://arxiv.org/abs/1908.03663} {arXiv:1908.03663 [astro-ph.CO]}
  \BibitemShut {NoStop}%
\bibitem [{\citenamefont {Di~Valentino}\ \emph {et~al.}(2021)\citenamefont
  {Di~Valentino} \emph {et~al.}}]{DiValentino:2020zio}%
  \BibitemOpen
  \bibfield  {author} {\bibinfo {author} {\bibfnamefont {E.}~\bibnamefont
  {Di~Valentino}} \emph {et~al.},\ }\href {\doibase
  10.1016/j.astropartphys.2021.102605} {\bibfield  {journal} {\bibinfo
  {journal} {Astropart. Phys.}\ }\textbf {\bibinfo {volume} {131}},\ \bibinfo
  {pages} {102605} (\bibinfo {year} {2021})},\ \Eprint
  {http://arxiv.org/abs/2008.11284} {arXiv:2008.11284 [astro-ph.CO]}
  \BibitemShut {NoStop}%
\bibitem [{\citenamefont {Di~Valentino}\ \emph {et~al.}(2016)\citenamefont
  {Di~Valentino}, \citenamefont {Melchiorri},\ and\ \citenamefont
  {Silk}}]{DiValentino:2016hlg}%
  \BibitemOpen
  \bibfield  {author} {\bibinfo {author} {\bibfnamefont {E.}~\bibnamefont
  {Di~Valentino}}, \bibinfo {author} {\bibfnamefont {A.}~\bibnamefont
  {Melchiorri}}, \ and\ \bibinfo {author} {\bibfnamefont {J.}~\bibnamefont
  {Silk}},\ }\href {\doibase 10.1016/j.physletb.2016.08.043} {\bibfield
  {journal} {\bibinfo  {journal} {Phys. Lett. B}\ }\textbf {\bibinfo {volume}
  {761}},\ \bibinfo {pages} {242} (\bibinfo {year} {2016})},\ \Eprint
  {http://arxiv.org/abs/1606.00634} {arXiv:1606.00634 [astro-ph.CO]}
  \BibitemShut {NoStop}%
\bibitem [{\citenamefont {Di~Valentino}\ \emph {et~al.}(2017)\citenamefont
  {Di~Valentino}, \citenamefont {Melchiorri},\ and\ \citenamefont
  {Mena}}]{DiValentino:2017iww}%
  \BibitemOpen
  \bibfield  {author} {\bibinfo {author} {\bibfnamefont {E.}~\bibnamefont
  {Di~Valentino}}, \bibinfo {author} {\bibfnamefont {A.}~\bibnamefont
  {Melchiorri}}, \ and\ \bibinfo {author} {\bibfnamefont {O.}~\bibnamefont
  {Mena}},\ }\href {\doibase 10.1103/PhysRevD.96.043503} {\bibfield  {journal}
  {\bibinfo  {journal} {Phys. Rev. D}\ }\textbf {\bibinfo {volume} {96}},\
  \bibinfo {pages} {043503} (\bibinfo {year} {2017})},\ \Eprint
  {http://arxiv.org/abs/1704.08342} {arXiv:1704.08342 [astro-ph.CO]}
  \BibitemShut {NoStop}%
\bibitem [{\citenamefont {Vagnozzi}(2020)}]{Vagnozzi:2019ezj}%
  \BibitemOpen
  \bibfield  {author} {\bibinfo {author} {\bibfnamefont {S.}~\bibnamefont
  {Vagnozzi}},\ }\href {\doibase 10.1103/PhysRevD.102.023518} {\bibfield
  {journal} {\bibinfo  {journal} {Phys. Rev. D}\ }\textbf {\bibinfo {volume}
  {102}},\ \bibinfo {pages} {023518} (\bibinfo {year} {2020})},\ \Eprint
  {http://arxiv.org/abs/1907.07569} {arXiv:1907.07569 [astro-ph.CO]}
  \BibitemShut {NoStop}%
\bibitem [{\citenamefont {Ballardini}\ \emph {et~al.}(2020)\citenamefont
  {Ballardini}, \citenamefont {Braglia}, \citenamefont {Finelli}, \citenamefont
  {Paoletti}, \citenamefont {Starobinsky},\ and\ \citenamefont
  {Umilt\`a}}]{Ballardini:2020iws}%
  \BibitemOpen
  \bibfield  {author} {\bibinfo {author} {\bibfnamefont {M.}~\bibnamefont
  {Ballardini}}, \bibinfo {author} {\bibfnamefont {M.}~\bibnamefont {Braglia}},
  \bibinfo {author} {\bibfnamefont {F.}~\bibnamefont {Finelli}}, \bibinfo
  {author} {\bibfnamefont {D.}~\bibnamefont {Paoletti}}, \bibinfo {author}
  {\bibfnamefont {A.~A.}\ \bibnamefont {Starobinsky}}, \ and\ \bibinfo {author}
  {\bibfnamefont {C.}~\bibnamefont {Umilt\`a}},\ }\href {\doibase
  10.1088/1475-7516/2020/10/044} {\bibfield  {journal} {\bibinfo  {journal}
  {JCAP}\ }\textbf {\bibinfo {volume} {10}},\ \bibinfo {pages} {044} (\bibinfo
  {year} {2020})},\ \Eprint {http://arxiv.org/abs/2004.14349} {arXiv:2004.14349
  [astro-ph.CO]} \BibitemShut {NoStop}%
\bibitem [{\citenamefont {Ye}\ \emph {et~al.}(2021)\citenamefont {Ye},
  \citenamefont {Hu},\ and\ \citenamefont {Piao}}]{Ye:2021nej}%
  \BibitemOpen
  \bibfield  {author} {\bibinfo {author} {\bibfnamefont {G.}~\bibnamefont
  {Ye}}, \bibinfo {author} {\bibfnamefont {B.}~\bibnamefont {Hu}}, \ and\
  \bibinfo {author} {\bibfnamefont {Y.-S.}\ \bibnamefont {Piao}},\ }\href
  {\doibase 10.1103/PhysRevD.104.063510} {\bibfield  {journal} {\bibinfo
  {journal} {Phys. Rev. D}\ }\textbf {\bibinfo {volume} {104}},\ \bibinfo
  {pages} {063510} (\bibinfo {year} {2021})},\ \Eprint
  {http://arxiv.org/abs/2103.09729} {arXiv:2103.09729 [astro-ph.CO]}
  \BibitemShut {NoStop}%
\bibitem [{\citenamefont {Baxter}\ and\ \citenamefont
  {Sherwin}(2020)}]{Baxter_2020}%
  \BibitemOpen
  \bibfield  {author} {\bibinfo {author} {\bibfnamefont {E.~J.}\ \bibnamefont
  {Baxter}}\ and\ \bibinfo {author} {\bibfnamefont {B.~D.}\ \bibnamefont
  {Sherwin}},\ }\href {\doibase 10.1093/mnras/staa3706} {\bibfield  {journal}
  {\bibinfo  {journal} {Monthly Notices of the Royal Astronomical Society}\
  }\textbf {\bibinfo {volume} {501}},\ \bibinfo {pages} {1823} (\bibinfo {year}
  {2020})}\BibitemShut {NoStop}%
\bibitem [{\citenamefont {Perivolaropoulos}\ and\ \citenamefont
  {Skara}(2021)}]{Perivolaropoulos:2021jda}%
  \BibitemOpen
  \bibfield  {author} {\bibinfo {author} {\bibfnamefont {L.}~\bibnamefont
  {Perivolaropoulos}}\ and\ \bibinfo {author} {\bibfnamefont {F.}~\bibnamefont
  {Skara}},\ }\href@noop {} {\  (\bibinfo {year} {2021})},\ \Eprint
  {http://arxiv.org/abs/2105.05208} {arXiv:2105.05208 [astro-ph.CO]}
  \BibitemShut {NoStop}%
\bibitem [{\citenamefont {Ben-Dayan}\ \emph {et~al.}(2014)\citenamefont
  {Ben-Dayan}, \citenamefont {Durrer}, \citenamefont {Marozzi},\ and\
  \citenamefont {Schwarz}}]{Ben-Dayan:2014swa}%
  \BibitemOpen
  \bibfield  {author} {\bibinfo {author} {\bibfnamefont {I.}~\bibnamefont
  {Ben-Dayan}}, \bibinfo {author} {\bibfnamefont {R.}~\bibnamefont {Durrer}},
  \bibinfo {author} {\bibfnamefont {G.}~\bibnamefont {Marozzi}}, \ and\
  \bibinfo {author} {\bibfnamefont {D.~J.}\ \bibnamefont {Schwarz}},\ }\href
  {\doibase 10.1103/PhysRevLett.112.221301} {\bibfield  {journal} {\bibinfo
  {journal} {Phys. Rev. Lett.}\ }\textbf {\bibinfo {volume} {112}},\ \bibinfo
  {pages} {221301} (\bibinfo {year} {2014})},\ \Eprint
  {http://arxiv.org/abs/1401.7973} {arXiv:1401.7973 [astro-ph.CO]} \BibitemShut
  {NoStop}%
\bibitem [{\citenamefont {Inserra}\ \emph {et~al.}(2021)\citenamefont {Inserra}
  \emph {et~al.}}]{Inserra:2020uki}%
  \BibitemOpen
  \bibfield  {author} {\bibinfo {author} {\bibfnamefont {C.}~\bibnamefont
  {Inserra}} \emph {et~al.} (\bibinfo {collaboration} {DES}),\ }\href {\doibase
  10.1093/mnras/stab978} {\bibfield  {journal} {\bibinfo  {journal} {Mon. Not.
  Roy. Astron. Soc.}\ }\textbf {\bibinfo {volume} {504}},\ \bibinfo {pages}
  {2535} (\bibinfo {year} {2021})},\ \Eprint {http://arxiv.org/abs/2004.12218}
  {arXiv:2004.12218 [astro-ph.CO]} \BibitemShut {NoStop}%
\bibitem [{\citenamefont {Abazajian}\ \emph {et~al.}(2019)\citenamefont
  {Abazajian} \emph {et~al.}}]{Abazajian:2019eic}%
  \BibitemOpen
  \bibfield  {author} {\bibinfo {author} {\bibfnamefont {K.}~\bibnamefont
  {Abazajian}} \emph {et~al.},\ }\href@noop {} {\  (\bibinfo {year} {2019})},\
  \Eprint {http://arxiv.org/abs/1907.04473} {arXiv:1907.04473 [astro-ph.IM]}
  \BibitemShut {NoStop}%
\bibitem [{\citenamefont {Gasperini}\ \emph {et~al.}(2011)\citenamefont
  {Gasperini}, \citenamefont {Marozzi}, \citenamefont {Nugier},\ and\
  \citenamefont {Veneziano}}]{Gasperini:2011us}%
  \BibitemOpen
  \bibfield  {author} {\bibinfo {author} {\bibfnamefont {M.}~\bibnamefont
  {Gasperini}}, \bibinfo {author} {\bibfnamefont {G.}~\bibnamefont {Marozzi}},
  \bibinfo {author} {\bibfnamefont {F.}~\bibnamefont {Nugier}}, \ and\ \bibinfo
  {author} {\bibfnamefont {G.}~\bibnamefont {Veneziano}},\ }\href {\doibase
  10.1088/1475-7516/2011/07/008} {\bibfield  {journal} {\bibinfo  {journal}
  {JCAP}\ }\textbf {\bibinfo {volume} {07}},\ \bibinfo {pages} {008} (\bibinfo
  {year} {2011})},\ \Eprint {http://arxiv.org/abs/1104.1167} {arXiv:1104.1167
  [astro-ph.CO]} \BibitemShut {NoStop}%
\bibitem [{\citenamefont {Bonvin}\ \emph {et~al.}(2015)\citenamefont {Bonvin},
  \citenamefont {Clarkson}, \citenamefont {Durrer}, \citenamefont {Maartens},\
  and\ \citenamefont {Umeh}}]{Bonvin:2015kea}%
  \BibitemOpen
  \bibfield  {author} {\bibinfo {author} {\bibfnamefont {C.}~\bibnamefont
  {Bonvin}}, \bibinfo {author} {\bibfnamefont {C.}~\bibnamefont {Clarkson}},
  \bibinfo {author} {\bibfnamefont {R.}~\bibnamefont {Durrer}}, \bibinfo
  {author} {\bibfnamefont {R.}~\bibnamefont {Maartens}}, \ and\ \bibinfo
  {author} {\bibfnamefont {O.}~\bibnamefont {Umeh}},\ }\href {\doibase
  10.1088/1475-7516/2015/07/040} {\bibfield  {journal} {\bibinfo  {journal}
  {JCAP}\ }\textbf {\bibinfo {volume} {07}},\ \bibinfo {pages} {040} (\bibinfo
  {year} {2015})},\ \Eprint {http://arxiv.org/abs/1504.01676} {arXiv:1504.01676
  [astro-ph.CO]} \BibitemShut {NoStop}%
\bibitem [{\citenamefont {Heinesen}\ \emph {et~al.}(2019)\citenamefont
  {Heinesen}, \citenamefont {Mourier},\ and\ \citenamefont
  {Buchert}}]{Heinesen:2018vjp}%
  \BibitemOpen
  \bibfield  {author} {\bibinfo {author} {\bibfnamefont {A.}~\bibnamefont
  {Heinesen}}, \bibinfo {author} {\bibfnamefont {P.}~\bibnamefont {Mourier}}, \
  and\ \bibinfo {author} {\bibfnamefont {T.}~\bibnamefont {Buchert}},\ }\href
  {\doibase 10.1088/1361-6382/ab0618} {\bibfield  {journal} {\bibinfo
  {journal} {Class. Quant. Grav.}\ }\textbf {\bibinfo {volume} {36}},\ \bibinfo
  {pages} {075001} (\bibinfo {year} {2019})},\ \Eprint
  {http://arxiv.org/abs/1811.01374} {arXiv:1811.01374 [gr-qc]} \BibitemShut
  {NoStop}%
\bibitem [{\citenamefont {Fanizza}\ \emph {et~al.}(2020)\citenamefont
  {Fanizza}, \citenamefont {Gasperini}, \citenamefont {Marozzi},\ and\
  \citenamefont {Veneziano}}]{Fanizza:2019pfp}%
  \BibitemOpen
  \bibfield  {author} {\bibinfo {author} {\bibfnamefont {G.}~\bibnamefont
  {Fanizza}}, \bibinfo {author} {\bibfnamefont {M.}~\bibnamefont {Gasperini}},
  \bibinfo {author} {\bibfnamefont {G.}~\bibnamefont {Marozzi}}, \ and\
  \bibinfo {author} {\bibfnamefont {G.}~\bibnamefont {Veneziano}},\ }\href
  {\doibase 10.1088/1475-7516/2020/02/017} {\bibfield  {journal} {\bibinfo
  {journal} {JCAP}\ }\textbf {\bibinfo {volume} {02}},\ \bibinfo {pages} {017}
  (\bibinfo {year} {2020})},\ \Eprint {http://arxiv.org/abs/1911.09469}
  {arXiv:1911.09469 [gr-qc]} \BibitemShut {NoStop}%
\bibitem [{\citenamefont {Ben-Dayan}\ \emph
  {et~al.}(2012{\natexlab{a}})\citenamefont {Ben-Dayan}, \citenamefont
  {Gasperini}, \citenamefont {Marozzi}, \citenamefont {Nugier},\ and\
  \citenamefont {Veneziano}}]{BenDayan:2012pp}%
  \BibitemOpen
  \bibfield  {author} {\bibinfo {author} {\bibfnamefont {I.}~\bibnamefont
  {Ben-Dayan}}, \bibinfo {author} {\bibfnamefont {M.}~\bibnamefont
  {Gasperini}}, \bibinfo {author} {\bibfnamefont {G.}~\bibnamefont {Marozzi}},
  \bibinfo {author} {\bibfnamefont {F.}~\bibnamefont {Nugier}}, \ and\ \bibinfo
  {author} {\bibfnamefont {G.}~\bibnamefont {Veneziano}},\ }\href {\doibase
  10.1088/1475-7516/2012/04/036} {\bibfield  {journal} {\bibinfo  {journal}
  {JCAP}\ }\textbf {\bibinfo {volume} {04}},\ \bibinfo {pages} {036} (\bibinfo
  {year} {2012}{\natexlab{a}})},\ \Eprint {http://arxiv.org/abs/1202.1247}
  {arXiv:1202.1247 [astro-ph.CO]} \BibitemShut {NoStop}%
\bibitem [{\citenamefont {Fleury}\ \emph {et~al.}(2017)\citenamefont {Fleury},
  \citenamefont {Clarkson},\ and\ \citenamefont {Maartens}}]{Fleury:2016fda}%
  \BibitemOpen
  \bibfield  {author} {\bibinfo {author} {\bibfnamefont {P.}~\bibnamefont
  {Fleury}}, \bibinfo {author} {\bibfnamefont {C.}~\bibnamefont {Clarkson}}, \
  and\ \bibinfo {author} {\bibfnamefont {R.}~\bibnamefont {Maartens}},\ }\href
  {\doibase 10.1088/1475-7516/2017/03/062} {\bibfield  {journal} {\bibinfo
  {journal} {JCAP}\ }\textbf {\bibinfo {volume} {03}},\ \bibinfo {pages} {062}
  (\bibinfo {year} {2017})},\ \Eprint {http://arxiv.org/abs/1612.03726}
  {arXiv:1612.03726 [astro-ph.CO]} \BibitemShut {NoStop}%
\bibitem [{\citenamefont {Yoo}\ and\ \citenamefont
  {Durrer}(2017)}]{Yoo:2017svj}%
  \BibitemOpen
  \bibfield  {author} {\bibinfo {author} {\bibfnamefont {J.}~\bibnamefont
  {Yoo}}\ and\ \bibinfo {author} {\bibfnamefont {R.}~\bibnamefont {Durrer}},\
  }\href {\doibase 10.1088/1475-7516/2017/09/016} {\bibfield  {journal}
  {\bibinfo  {journal} {JCAP}\ }\textbf {\bibinfo {volume} {1709}},\ \bibinfo
  {pages} {016} (\bibinfo {year} {2017})},\ \Eprint
  {http://arxiv.org/abs/1705.05839} {arXiv:1705.05839 [astro-ph.CO]}
  \BibitemShut {NoStop}%
\bibitem [{\citenamefont {Fanizza}\ \emph {et~al.}(2016)\citenamefont
  {Fanizza}, \citenamefont {Gasperini}, \citenamefont {Marozzi},\ and\
  \citenamefont {Veneziano}}]{Fanizza:2015gdn}%
  \BibitemOpen
  \bibfield  {author} {\bibinfo {author} {\bibfnamefont {G.}~\bibnamefont
  {Fanizza}}, \bibinfo {author} {\bibfnamefont {M.}~\bibnamefont {Gasperini}},
  \bibinfo {author} {\bibfnamefont {G.}~\bibnamefont {Marozzi}}, \ and\
  \bibinfo {author} {\bibfnamefont {G.}~\bibnamefont {Veneziano}},\ }\href
  {\doibase 10.1016/j.physletb.2016.04.032} {\bibfield  {journal} {\bibinfo
  {journal} {Phys. Lett. B}\ }\textbf {\bibinfo {volume} {757}},\ \bibinfo
  {pages} {505} (\bibinfo {year} {2016})},\ \Eprint
  {http://arxiv.org/abs/1512.08489} {arXiv:1512.08489 [astro-ph.CO]}
  \BibitemShut {NoStop}%
\bibitem [{\citenamefont {Marozzi}(2015)}]{Marozzi:2014kua}%
  \BibitemOpen
  \bibfield  {author} {\bibinfo {author} {\bibfnamefont {G.}~\bibnamefont
  {Marozzi}},\ }\href {\doibase 10.1088/0264-9381/32/4/045004} {\bibfield
  {journal} {\bibinfo  {journal} {Class. Quant. Grav.}\ }\textbf {\bibinfo
  {volume} {32}},\ \bibinfo {pages} {045004} (\bibinfo {year} {2015})},\
  \bibinfo {note} {[Erratum: Class.Quant.Grav. 32, 179501 (2015)]},\ \Eprint
  {http://arxiv.org/abs/1406.1135} {arXiv:1406.1135 [astro-ph.CO]} \BibitemShut
  {NoStop}%
\bibitem [{\citenamefont {Fanizza}\ \emph {et~al.}(2018)\citenamefont
  {Fanizza}, \citenamefont {Yoo},\ and\ \citenamefont
  {Biern}}]{Fanizza:2018qux}%
  \BibitemOpen
  \bibfield  {author} {\bibinfo {author} {\bibfnamefont {G.}~\bibnamefont
  {Fanizza}}, \bibinfo {author} {\bibfnamefont {J.}~\bibnamefont {Yoo}}, \ and\
  \bibinfo {author} {\bibfnamefont {S.~G.}\ \bibnamefont {Biern}},\ }\href
  {\doibase 10.1088/1475-7516/2018/09/037} {\bibfield  {journal} {\bibinfo
  {journal} {JCAP}\ }\textbf {\bibinfo {volume} {09}},\ \bibinfo {pages} {037}
  (\bibinfo {year} {2018})},\ \Eprint {http://arxiv.org/abs/1805.05959}
  {arXiv:1805.05959 [gr-qc]} \BibitemShut {NoStop}%
\bibitem [{\citenamefont {Bonvin}\ \emph {et~al.}(2006)\citenamefont {Bonvin},
  \citenamefont {Durrer},\ and\ \citenamefont {Gasparini}}]{Bonvin:2005ps}%
  \BibitemOpen
  \bibfield  {author} {\bibinfo {author} {\bibfnamefont {C.}~\bibnamefont
  {Bonvin}}, \bibinfo {author} {\bibfnamefont {R.}~\bibnamefont {Durrer}}, \
  and\ \bibinfo {author} {\bibfnamefont {M.}~\bibnamefont {Gasparini}},\ }\href
  {\doibase 10.1103/PhysRevD.85.029901} {\bibfield  {journal} {\bibinfo
  {journal} {Phys. Rev. D}\ }\textbf {\bibinfo {volume} {73}},\ \bibinfo
  {pages} {023523} (\bibinfo {year} {2006})},\ \bibinfo {note} {[Erratum:
  Phys.Rev.D 85, 029901 (2012)]},\ \Eprint
  {http://arxiv.org/abs/astro-ph/0511183} {arXiv:astro-ph/0511183} \BibitemShut
  {NoStop}%
\bibitem [{\citenamefont {Ben-Dayan}\ \emph
  {et~al.}(2012{\natexlab{b}})\citenamefont {Ben-Dayan}, \citenamefont
  {Marozzi}, \citenamefont {Nugier},\ and\ \citenamefont
  {Veneziano}}]{BenDayan:2012wi}%
  \BibitemOpen
  \bibfield  {author} {\bibinfo {author} {\bibfnamefont {I.}~\bibnamefont
  {Ben-Dayan}}, \bibinfo {author} {\bibfnamefont {G.}~\bibnamefont {Marozzi}},
  \bibinfo {author} {\bibfnamefont {F.}~\bibnamefont {Nugier}}, \ and\ \bibinfo
  {author} {\bibfnamefont {G.}~\bibnamefont {Veneziano}},\ }\href {\doibase
  10.1088/1475-7516/2012/11/045} {\bibfield  {journal} {\bibinfo  {journal}
  {JCAP}\ }\textbf {\bibinfo {volume} {11}},\ \bibinfo {pages} {045} (\bibinfo
  {year} {2012}{\natexlab{b}})},\ \Eprint {http://arxiv.org/abs/1209.4326}
  {arXiv:1209.4326 [astro-ph.CO]} \BibitemShut {NoStop}%
\bibitem [{\citenamefont {Fanizza}\ \emph {et~al.}(2015)\citenamefont
  {Fanizza}, \citenamefont {Gasperini}, \citenamefont {Marozzi},\ and\
  \citenamefont {Veneziano}}]{Fanizza:2015swa}%
  \BibitemOpen
  \bibfield  {author} {\bibinfo {author} {\bibfnamefont {G.}~\bibnamefont
  {Fanizza}}, \bibinfo {author} {\bibfnamefont {M.}~\bibnamefont {Gasperini}},
  \bibinfo {author} {\bibfnamefont {G.}~\bibnamefont {Marozzi}}, \ and\
  \bibinfo {author} {\bibfnamefont {G.}~\bibnamefont {Veneziano}},\ }\href
  {\doibase 10.1088/1475-7516/2015/08/020} {\bibfield  {journal} {\bibinfo
  {journal} {JCAP}\ }\textbf {\bibinfo {volume} {08}},\ \bibinfo {pages} {020}
  (\bibinfo {year} {2015})},\ \Eprint {http://arxiv.org/abs/1506.02003}
  {arXiv:1506.02003 [astro-ph.CO]} \BibitemShut {NoStop}%
\bibitem [{\citenamefont {Umeh}\ \emph {et~al.}(2014)\citenamefont {Umeh},
  \citenamefont {Clarkson},\ and\ \citenamefont {Maartens}}]{Umeh:2014ana}%
  \BibitemOpen
  \bibfield  {author} {\bibinfo {author} {\bibfnamefont {O.}~\bibnamefont
  {Umeh}}, \bibinfo {author} {\bibfnamefont {C.}~\bibnamefont {Clarkson}}, \
  and\ \bibinfo {author} {\bibfnamefont {R.}~\bibnamefont {Maartens}},\ }\href
  {\doibase 10.1088/0264-9381/31/20/205001} {\bibfield  {journal} {\bibinfo
  {journal} {Class. Quant. Grav.}\ }\textbf {\bibinfo {volume} {31}},\ \bibinfo
  {pages} {205001} (\bibinfo {year} {2014})},\ \Eprint
  {http://arxiv.org/abs/1402.1933} {arXiv:1402.1933 [astro-ph.CO]} \BibitemShut
  {NoStop}%
\bibitem [{\citenamefont {Peter}\ and\ \citenamefont
  {Uzan}(2013)}]{Peter:2013avv}%
  \BibitemOpen
  \bibfield  {author} {\bibinfo {author} {\bibfnamefont {P.}~\bibnamefont
  {Peter}}\ and\ \bibinfo {author} {\bibfnamefont {J.-P.}\ \bibnamefont
  {Uzan}},\ }\href@noop {} {\emph {\bibinfo {title} {{Primordial
  Cosmology}}}},\ Oxford Graduate Texts\ (\bibinfo  {publisher} {Oxford
  University Press},\ \bibinfo {year} {2013})\BibitemShut {NoStop}%
\bibitem [{\citenamefont {Fiorini}(2018)}]{Fiorini2018}%
  \BibitemOpen
  \bibfield  {author} {\bibinfo {author} {\bibfnamefont {B.}~\bibnamefont
  {Fiorini}},\ }\href@noop {} {\bibfield  {journal} {\bibinfo  {journal}
  {Master Thesis}\ } (\bibinfo {year} {2018})}\BibitemShut {NoStop}%
\bibitem [{\citenamefont {Scaccabarozzi}\ \emph {et~al.}(2018)\citenamefont
  {Scaccabarozzi}, \citenamefont {Yoo},\ and\ \citenamefont
  {Biern}}]{Scaccabarozzi:2018vux}%
  \BibitemOpen
  \bibfield  {author} {\bibinfo {author} {\bibfnamefont {F.}~\bibnamefont
  {Scaccabarozzi}}, \bibinfo {author} {\bibfnamefont {J.}~\bibnamefont {Yoo}},
  \ and\ \bibinfo {author} {\bibfnamefont {S.~G.}\ \bibnamefont {Biern}},\
  }\href {\doibase 10.1088/1475-7516/2018/10/024} {\bibfield  {journal}
  {\bibinfo  {journal} {JCAP}\ }\textbf {\bibinfo {volume} {10}},\ \bibinfo
  {pages} {024} (\bibinfo {year} {2018})},\ \Eprint
  {http://arxiv.org/abs/1807.09796} {arXiv:1807.09796 [astro-ph.CO]}
  \BibitemShut {NoStop}%
\bibitem [{\citenamefont {Tansella}\ \emph {et~al.}(2018)\citenamefont
  {Tansella}, \citenamefont {Jelic-Cizmek}, \citenamefont {Bonvin},\ and\
  \citenamefont {Durrer}}]{Tansella:2018sld}%
  \BibitemOpen
  \bibfield  {author} {\bibinfo {author} {\bibfnamefont {V.}~\bibnamefont
  {Tansella}}, \bibinfo {author} {\bibfnamefont {G.}~\bibnamefont
  {Jelic-Cizmek}}, \bibinfo {author} {\bibfnamefont {C.}~\bibnamefont
  {Bonvin}}, \ and\ \bibinfo {author} {\bibfnamefont {R.}~\bibnamefont
  {Durrer}},\ }\href {\doibase 10.1088/1475-7516/2018/10/032} {\bibfield
  {journal} {\bibinfo  {journal} {JCAP}\ }\textbf {\bibinfo {volume} {10}},\
  \bibinfo {pages} {032} (\bibinfo {year} {2018})},\ \Eprint
  {http://arxiv.org/abs/1806.11090} {arXiv:1806.11090 [astro-ph.CO]}
  \BibitemShut {NoStop}%
\bibitem [{\citenamefont {Di~Dio}\ and\ \citenamefont
  {Seljak}(2019)}]{DiDio:2018zmk}%
  \BibitemOpen
  \bibfield  {author} {\bibinfo {author} {\bibfnamefont {E.}~\bibnamefont
  {Di~Dio}}\ and\ \bibinfo {author} {\bibfnamefont {U.}~\bibnamefont
  {Seljak}},\ }\href {\doibase 10.1088/1475-7516/2019/04/050} {\bibfield
  {journal} {\bibinfo  {journal} {JCAP}\ }\textbf {\bibinfo {volume} {04}},\
  \bibinfo {pages} {050} (\bibinfo {year} {2019})},\ \Eprint
  {http://arxiv.org/abs/1811.03054} {arXiv:1811.03054 [astro-ph.CO]}
  \BibitemShut {NoStop}%
\bibitem [{\citenamefont {Beutler}\ and\ \citenamefont
  {Di~Dio}(2020)}]{Beutler:2020evf}%
  \BibitemOpen
  \bibfield  {author} {\bibinfo {author} {\bibfnamefont {F.}~\bibnamefont
  {Beutler}}\ and\ \bibinfo {author} {\bibfnamefont {E.}~\bibnamefont
  {Di~Dio}},\ }\href {\doibase 10.1088/1475-7516/2020/07/048} {\bibfield
  {journal} {\bibinfo  {journal} {JCAP}\ }\textbf {\bibinfo {volume} {07}},\
  \bibinfo {pages} {048} (\bibinfo {year} {2020})},\ \Eprint
  {http://arxiv.org/abs/2004.08014} {arXiv:2004.08014 [astro-ph.CO]}
  \BibitemShut {NoStop}%
\bibitem [{\citenamefont {Eisenstein}\ and\ \citenamefont
  {Hu}(1998)}]{Eisenstein:1997ik}%
  \BibitemOpen
  \bibfield  {author} {\bibinfo {author} {\bibfnamefont {D.~J.}\ \bibnamefont
  {Eisenstein}}\ and\ \bibinfo {author} {\bibfnamefont {W.}~\bibnamefont
  {Hu}},\ }\href {\doibase 10.1086/305424} {\bibfield  {journal} {\bibinfo
  {journal} {Astrophys. J.}\ }\textbf {\bibinfo {volume} {496}},\ \bibinfo
  {pages} {605} (\bibinfo {year} {1998})},\ \Eprint
  {http://arxiv.org/abs/astro-ph/9709112} {arXiv:astro-ph/9709112} \BibitemShut
  {NoStop}%
\bibitem [{\citenamefont {Yoo}(2020)}]{Yoo:2019skw}%
  \BibitemOpen
  \bibfield  {author} {\bibinfo {author} {\bibfnamefont {J.}~\bibnamefont
  {Yoo}},\ }\href {\doibase 10.1103/PhysRevD.101.043507} {\bibfield  {journal}
  {\bibinfo  {journal} {Phys. Rev. D}\ }\textbf {\bibinfo {volume} {101}},\
  \bibinfo {pages} {043507} (\bibinfo {year} {2020})},\ \Eprint
  {http://arxiv.org/abs/1911.07869} {arXiv:1911.07869 [astro-ph.CO]}
  \BibitemShut {NoStop}%
\bibitem [{\citenamefont {Ben-Dayan}\ \emph {et~al.}(2013)\citenamefont
  {Ben-Dayan}, \citenamefont {Gasperini}, \citenamefont {Marozzi},
  \citenamefont {Nugier},\ and\ \citenamefont {Veneziano}}]{BenDayan:2013gc}%
  \BibitemOpen
  \bibfield  {author} {\bibinfo {author} {\bibfnamefont {I.}~\bibnamefont
  {Ben-Dayan}}, \bibinfo {author} {\bibfnamefont {M.}~\bibnamefont
  {Gasperini}}, \bibinfo {author} {\bibfnamefont {G.}~\bibnamefont {Marozzi}},
  \bibinfo {author} {\bibfnamefont {F.}~\bibnamefont {Nugier}}, \ and\ \bibinfo
  {author} {\bibfnamefont {G.}~\bibnamefont {Veneziano}},\ }\href {\doibase
  10.1088/1475-7516/2013/06/002} {\bibfield  {journal} {\bibinfo  {journal}
  {JCAP}\ }\textbf {\bibinfo {volume} {1306}},\ \bibinfo {pages} {002}
  (\bibinfo {year} {2013})},\ \Eprint {http://arxiv.org/abs/1302.0740}
  {arXiv:1302.0740 [astro-ph.CO]} \BibitemShut {NoStop}%
\bibitem [{\citenamefont {Smith}\ \emph {et~al.}(2003)\citenamefont {Smith},
  \citenamefont {Peacock}, \citenamefont {Jenkins}, \citenamefont {White},
  \citenamefont {Frenk}, \citenamefont {Pearce}, \citenamefont {Thomas},
  \citenamefont {Efstathiou},\ and\ \citenamefont {Couchman}}]{Smith_2003}%
  \BibitemOpen
  \bibfield  {author} {\bibinfo {author} {\bibfnamefont {R.~E.}\ \bibnamefont
  {Smith}}, \bibinfo {author} {\bibfnamefont {J.~A.}\ \bibnamefont {Peacock}},
  \bibinfo {author} {\bibfnamefont {A.}~\bibnamefont {Jenkins}}, \bibinfo
  {author} {\bibfnamefont {S.~D.~M.}\ \bibnamefont {White}}, \bibinfo {author}
  {\bibfnamefont {C.~S.}\ \bibnamefont {Frenk}}, \bibinfo {author}
  {\bibfnamefont {F.~R.}\ \bibnamefont {Pearce}}, \bibinfo {author}
  {\bibfnamefont {P.~A.}\ \bibnamefont {Thomas}}, \bibinfo {author}
  {\bibfnamefont {G.}~\bibnamefont {Efstathiou}}, \ and\ \bibinfo {author}
  {\bibfnamefont {H.~M.~P.}\ \bibnamefont {Couchman}},\ }\href {\doibase
  10.1046/j.1365-8711.2003.06503.x} {\bibfield  {journal} {\bibinfo  {journal}
  {Monthly Notices of the Royal Astronomical Society}\ }\textbf {\bibinfo
  {volume} {341}},\ \bibinfo {pages} {1311} (\bibinfo {year}
  {2003})}\BibitemShut {NoStop}%
\bibitem [{\citenamefont {Takahashi}\ \emph {et~al.}(2012)\citenamefont
  {Takahashi}, \citenamefont {Sato}, \citenamefont {Nishimichi}, \citenamefont
  {Taruya},\ and\ \citenamefont {Oguri}}]{Takahashi_2012}%
  \BibitemOpen
  \bibfield  {author} {\bibinfo {author} {\bibfnamefont {R.}~\bibnamefont
  {Takahashi}}, \bibinfo {author} {\bibfnamefont {M.}~\bibnamefont {Sato}},
  \bibinfo {author} {\bibfnamefont {T.}~\bibnamefont {Nishimichi}}, \bibinfo
  {author} {\bibfnamefont {A.}~\bibnamefont {Taruya}}, \ and\ \bibinfo {author}
  {\bibfnamefont {M.}~\bibnamefont {Oguri}},\ }\href {\doibase
  10.1088/0004-637x/761/2/152} {\bibfield  {journal} {\bibinfo  {journal} {The
  Astrophysical Journal}\ }\textbf {\bibinfo {volume} {761}},\ \bibinfo {pages}
  {152} (\bibinfo {year} {2012})}\BibitemShut {NoStop}%
\bibitem [{\citenamefont {Cosmai}\ \emph {et~al.}(2013)\citenamefont {Cosmai},
  \citenamefont {Fanizza}, \citenamefont {Gasperini},\ and\ \citenamefont
  {Tedesco}}]{Cosmai:2013iga}%
  \BibitemOpen
  \bibfield  {author} {\bibinfo {author} {\bibfnamefont {L.}~\bibnamefont
  {Cosmai}}, \bibinfo {author} {\bibfnamefont {G.}~\bibnamefont {Fanizza}},
  \bibinfo {author} {\bibfnamefont {M.}~\bibnamefont {Gasperini}}, \ and\
  \bibinfo {author} {\bibfnamefont {L.}~\bibnamefont {Tedesco}},\ }\href
  {\doibase 10.1088/0264-9381/30/9/095011} {\bibfield  {journal} {\bibinfo
  {journal} {Class. Quant. Grav.}\ }\textbf {\bibinfo {volume} {30}},\ \bibinfo
  {pages} {095011} (\bibinfo {year} {2013})},\ \Eprint
  {http://arxiv.org/abs/1303.5484} {arXiv:1303.5484 [gr-qc]} \BibitemShut
  {NoStop}%
\bibitem [{\citenamefont {Romano}(2018)}]{Romano:2016utn}%
  \BibitemOpen
  \bibfield  {author} {\bibinfo {author} {\bibfnamefont {A.~E.}\ \bibnamefont
  {Romano}},\ }\href {\doibase 10.1142/S021827181850102X} {\bibfield  {journal}
  {\bibinfo  {journal} {Int. J. Mod. Phys. D}\ }\textbf {\bibinfo {volume}
  {27}},\ \bibinfo {pages} {1850102} (\bibinfo {year} {2018})},\ \Eprint
  {http://arxiv.org/abs/1609.04081} {arXiv:1609.04081 [astro-ph.CO]}
  \BibitemShut {NoStop}%
\bibitem [{\citenamefont {Cosmai}\ \emph {et~al.}(2019)\citenamefont {Cosmai},
  \citenamefont {Fanizza}, \citenamefont {Sylos~Labini}, \citenamefont
  {Pietronero},\ and\ \citenamefont {Tedesco}}]{Cosmai:2018nvx}%
  \BibitemOpen
  \bibfield  {author} {\bibinfo {author} {\bibfnamefont {L.}~\bibnamefont
  {Cosmai}}, \bibinfo {author} {\bibfnamefont {G.}~\bibnamefont {Fanizza}},
  \bibinfo {author} {\bibfnamefont {F.}~\bibnamefont {Sylos~Labini}}, \bibinfo
  {author} {\bibfnamefont {L.}~\bibnamefont {Pietronero}}, \ and\ \bibinfo
  {author} {\bibfnamefont {L.}~\bibnamefont {Tedesco}},\ }\href {\doibase
  10.1088/1361-6382/aae8f7} {\bibfield  {journal} {\bibinfo  {journal} {Class.
  Quant. Grav.}\ }\textbf {\bibinfo {volume} {36}},\ \bibinfo {pages} {045007}
  (\bibinfo {year} {2019})},\ \Eprint {http://arxiv.org/abs/1810.06318}
  {arXiv:1810.06318 [astro-ph.CO]} \BibitemShut {NoStop}%
\bibitem [{\citenamefont {Vallejo-Pe\~na}\ and\ \citenamefont
  {Romano}(2020)}]{Vallejo-Pena:2019agp}%
  \BibitemOpen
  \bibfield  {author} {\bibinfo {author} {\bibfnamefont {S.~A.}\ \bibnamefont
  {Vallejo-Pe\~na}}\ and\ \bibinfo {author} {\bibfnamefont {A.~E.}\
  \bibnamefont {Romano}},\ }\href {\doibase 10.1088/1475-7516/2020/03/023}
  {\bibfield  {journal} {\bibinfo  {journal} {JCAP}\ }\textbf {\bibinfo
  {volume} {03}},\ \bibinfo {pages} {023} (\bibinfo {year} {2020})},\ \Eprint
  {http://arxiv.org/abs/1906.04946} {arXiv:1906.04946 [astro-ph.CO]}
  \BibitemShut {NoStop}%
\bibitem [{\citenamefont {Cai}\ \emph {et~al.}(2021)\citenamefont {Cai},
  \citenamefont {Guo}, \citenamefont {Li}, \citenamefont {Wang},\ and\
  \citenamefont {Yu}}]{Cai:2021wgv}%
  \BibitemOpen
  \bibfield  {author} {\bibinfo {author} {\bibfnamefont {R.-G.}\ \bibnamefont
  {Cai}}, \bibinfo {author} {\bibfnamefont {Z.-K.}\ \bibnamefont {Guo}},
  \bibinfo {author} {\bibfnamefont {L.}~\bibnamefont {Li}}, \bibinfo {author}
  {\bibfnamefont {S.-J.}\ \bibnamefont {Wang}}, \ and\ \bibinfo {author}
  {\bibfnamefont {W.-W.}\ \bibnamefont {Yu}},\ }\href {\doibase
  10.1103/PhysRevD.103.L121302} {\bibfield  {journal} {\bibinfo  {journal}
  {Phys. Rev. D}\ }\textbf {\bibinfo {volume} {103}},\ \bibinfo {pages}
  {121302} (\bibinfo {year} {2021})},\ \Eprint
  {http://arxiv.org/abs/2102.02020} {arXiv:2102.02020 [astro-ph.CO]}
  \BibitemShut {NoStop}%
\bibitem [{\citenamefont {Dainotti}\ \emph {et~al.}(2021)\citenamefont
  {Dainotti}, \citenamefont {De~Simone}, \citenamefont {Schiavone},
  \citenamefont {Montani}, \citenamefont {Rinaldi},\ and\ \citenamefont
  {Lambiase}}]{Dainotti:2021pqg}%
  \BibitemOpen
  \bibfield  {author} {\bibinfo {author} {\bibfnamefont {M.~G.}\ \bibnamefont
  {Dainotti}}, \bibinfo {author} {\bibfnamefont {B.}~\bibnamefont {De~Simone}},
  \bibinfo {author} {\bibfnamefont {T.}~\bibnamefont {Schiavone}}, \bibinfo
  {author} {\bibfnamefont {G.}~\bibnamefont {Montani}}, \bibinfo {author}
  {\bibfnamefont {E.}~\bibnamefont {Rinaldi}}, \ and\ \bibinfo {author}
  {\bibfnamefont {G.}~\bibnamefont {Lambiase}},\ }\href {\doibase
  10.3847/1538-4357/abeb73} {\bibfield  {journal} {\bibinfo  {journal}
  {Astrophys. J.}\ }\textbf {\bibinfo {volume} {912}},\ \bibinfo {pages} {150}
  (\bibinfo {year} {2021})},\ \Eprint {http://arxiv.org/abs/2103.02117}
  {arXiv:2103.02117 [astro-ph.CO]} \BibitemShut {NoStop}%
\end{thebibliography}%

\end{document}